\documentclass[14pt]{article}

\usepackage[utf8]{inputenc}
\usepackage[T1]{fontenc}
\usepackage{threeparttable}
\usepackage{authblk}

\usepackage{geometry}
\usepackage{amsmath, amssymb}
\usepackage{graphicx}
\usepackage{float}
\usepackage{booktabs}
\usepackage{eurosym}
\usepackage{setspace}
\usepackage{tikz}
\usepackage{natbib}

\usepackage{hyperref}
\hypersetup{
  colorlinks=true,
  linkcolor=black,
  citecolor=black,
  urlcolor=black
}

\newcounter{hp}
\renewcommand{\thehp}{H\arabic{hp}}

\newenvironment{hp}[1][]{
  \refstepcounter{hp}
  \par\medskip
  \noindent\textbf{\thehp\ifx#1\empty\else\ (#1)\fi.}\ \itshape
}{
  \par\medskip
}

\begin{document}

\title{When Is Prosocial Behavior Intuitive?\\
The Moderating Role of Empathic Perspective Taking%
\thanks{Funding: S.R. acknowledges funding for the experiments from the VERA project, Università Ca' Foscari Venezia. F.P. and S.R. were funded by FAR2024 UNIMORE (CUP E83D2400063000), and E.D. and B.L. were funded by FAR2025 DEMB UNIMORE. Data availability: The data and analysis code will be made publicly available in an open repository before acceptance.}}

\author[1]{Enrico D'Ecclesiis [\textbf{Corresponding author}]}
\author[2]{Francesca Pancotto}
\author[3]{Barbara Luppi}
\author[3]{Simone Righi}

\affil[1]{Department of Economics Marco Biagi, University of Modena and Reggio Emilia, Via Jacopo Berengario 51, 41121 Modena, Italy. Email: enricoangeloraffaele.decclesiis@unimore.it}

\affil[2]{Department of Linguistic and Cultural Studies, University of Modena and Reggio Emilia, Largo Sant'Eufemia 19, 41121 Modena, Italy}

\affil[3]{Department of Economics Marco Biagi, University of Modena and Reggio Emilia, Via Jacopo Berengario 51, 41121 Modena, Italy}

\date{}

\maketitle

\begin{abstract}
Why is prosocial behavior intuitive for some individuals but deliberative for others? We examine whether empathic perspective taking moderates the association between reflective processing and prosocial choice. We run a laboratory experiment with 652 participants, where we measure prosociality using a four-player Distribution Game that holds aggregate payoffs constant and orders allocations from self-regarding to egalitarian. We combine participants’ choices with the Perspective Taking subscale of the Interpersonal Reactivity Index, a five-item Cognitive Reflection Test, and choice-stage response time in the allocation decision. 
Perspective taking significantly moderates the association between response time and prosociality. At high levels of perspective taking, longer response times are associated with modestly lower prosociality; at low levels, the association is positive but not statistically significant. No comparable moderation emerges for Cognitive Reflection Test performance. These results are consistent with the possibility that, when others’ outcomes are readily represented (high perspective taking), prosocial choices may arise relatively quickly, while additional decision-specific processing may be associated with greater attention to their private cost. When others’ outcomes are less salient (low perspective taking), additional processing may instead help bring them into consideration. The findings highlight the importance of distinguishing processing proximal to the specific decision from domain-general analytic reflection when explaining whether prosocial behavior appears intuitive.

\end{abstract}


\vspace{1.5cm}

\noindent \textbf{Declaration of competing interest.} Simone Righi reports financial support from Ca’ Foscari University of Venice. The other authors declare that they have no known competing financial interests or personal relationships that could have appeared to influence the work reported in this paper.

\newpage

\section{Introduction}

Prosocial behavior---the willingness to incur personal costs in order to benefit others---is a central feature of human social life and a long-standing puzzle across the behavioral sciences \citep{AxelrodHamilton1981,Nowak2006,BowlesGintis2011}. Human beings frequently cooperate, help, share, and redistribute resources even when material incentives alone would favor more self-regarding behavior. In economics, this puzzle is closely related to the study of social preferences: individuals often deviate from narrowly selfish choices in anonymous and non-repeated interactions, and these deviations have been interpreted in terms of inequality aversion, reciprocity, altruism, maximin preferences, and concern for social efficiency \citep{FehrFischbacher2003,FehrSchmidt1999,BoltonOckenfels2000,charness2002understanding,andreoni2002giving,fisman2007individual}. In psychology and cognitive neuroscience, the same broad class of behaviors has been studied through the lenses of empathy, moral cognition, social learning, and the neural mechanisms of other-regarding choice \citep{Eisenbergetal1987,batson1991altruism,SingerFehr2005,bellucci2020neural}. A central question cutting across these literatures is which cognitive processes make prosocial behavior more or less likely.

Dual-process theories provide a natural starting point for addressing this question. These theories broadly distinguish between fast, automatic, associative, and intuitive processes and slower, controlled, reflective processes that involve explicit reasoning and cognitive control \citep{WasonEvans1974,Sloman1996,StanovichWest2000,Evans2003,Kahneman2003,Alos-FerrerStrack2014}. Applied to social behavior, this distinction has generated a major debate on whether prosocial choices are primarily intuitive or deliberative. One family of accounts, commonly described as the intuitive-selfishness hypothesis, treats self-interest as the default response and views prosocial behavior as requiring reflective control \citep{StevensHauser2004,MooreLoewenstein2004}. In this view, deliberation allows individuals to inhibit an initially selfish response and act more prosocially, a prediction consistent with findings from some studies using response times, time pressure, and cognitive depletion \citep{PiovesanWengstrom2009,Achtzigeretal2015,CapraroCococcioni2016}. A second approach instead argues that prosociality itself can be intuitive. The Social Heuristics Hypothesis proposes that cooperative behavior may become internalized as a heuristic because cooperation is often advantageous in repeated everyday social interactions; individuals can then generalize such heuristics to anonymous one-shot environments, where cooperation is no longer strategically profitable \citep{Randetal2012,Randetal2014SocialHeuristics,Rand2016,PeysakhovichRand2016}.

The empirical evidence does not support a simple one-to-one mapping between intuition and generosity, or between reflection and self-interest. Some studies find that faster or more intuitive decisions are associated with greater cooperation or giving \citep{Randetal2012,Rand2016,Cappelenetal2014}, while others show that intuition can favor self-regarding behavior or that time-related effects are fragile and sensitive to the decision environment \citep{Tinghogetal2013,Bouwmeesteretal2017,CapraroCococcioni2016,Fromelletal2020,Andersenetal2018}. Evidence based on cognitive reflection is similarly mixed. Cognitive reflection can override antisocial motives, but does not necessarily promote prosocial behavior in all contexts \citep{Corgnetetal2015a,Corgnetetal2015b}; moreover, it might also promote social efficiency by causing individuals to disregard relative shares \citep{CapraroCorgnetEtAl2017}. Meta-analytic evidence and recent work document a mixed pattern of findings across domains of social behavior, experimental manipulations, incentive structures, and motivational contexts \citep{Kvarvenetal2020,capraro2024dualprocess,KvarvenStromlandTorsvik2026}. At the individual level, related evidence also shows that the association between decision time and prosocial behavior can vary with pre-existing social preferences. Longer decision times have been associated with lower cooperation among prosocially oriented individuals \citep{MischkowskiGlockner2016}, while other evidence shows opposite response-time patterns among prosocial and proself individuals \citep{Yamagishietal2017}. Using experimentally incentivized time-pressure and time-delay manipulations
in public-good games, \citet{AlosFerrerGaragnani2020} similarly find no
average effect of time pressure on cooperation, but provide suggestive
evidence that its relationship with cooperative behavior varies with
individual social preferences and attitudes toward strategic uncertainty. These findings suggest that the relevant question is no longer whether intuition or reflection is prosocial in general. Rather, the key issue is when prosociality becomes intuitive, for whom, and through which psychological mechanism.

This shift from average effects to heterogeneity is central to our paper, which elaborates on the approach of \citet{PancottoRighi2021}, who study the relationship between reflectivity (measured with a Cognitive Reflection Test), strategic reasoning (measured with a guessing game), and prosociality (measured with a Distribution Game). Participants are classified as naïve or strategic based on their guesses, while their written comments are used to identify responses explicitly reported as random. Their main result is that reflectivity relates to prosociality in opposite ways across these groups. Among naïve participants, more reflective individuals are more prosocial. Among strategic participants, more reflective individuals are less prosocial. This finding is important because it shows that prosocial behavior may be intuitive for some individuals but deliberative for others.

The present paper builds directly on this insight. If prosociality is intuitive only for some individuals, a natural next question is which psychological dispositions make other-regarding considerations available early in the decision process. \citet{PancottoRighi2021} suggest that the intuitive prosociality observed among strategic participants may be related to the ability to perceive the consequences of one's actions for others, and point to empathy as a possible explanation. More specifically, empathy may underlie both strategic reasoning and prosociality because it helps individuals represent others' motives and states, as well as the consequences of one's actions for them. Within this broader construct, they point to cognitive empathy as particularly relevant because, unlike affective sharing, it involves perspective taking and mental-state inference. We develop this conjecture by focusing on empathic perspective taking, defined as the dispositional tendency to adopt another person's point of view and to represent how situations and outcomes appear from that perspective.

Perspective taking is therefore particularly relevant to our argument because it captures a cognitive component of empathy that is distinct from more affective or self-oriented responses, such as empathic concern and personal distress \citep{Davis1983,DuanHill1996,DecetyJackson2004,Reniers2011,SingerLamm2009}. While the broader relationship between empathy and generosity is well established in the empathy-altruism and prosocial-behavior literature \citep{Eisenbergetal1987,batson1991altruism,deWaal2008}, our analysis more specifically examines whether individual differences in empathic perspective taking help explain why prosocial behavior appears intuitive for some individuals and deliberative for others.

One interpretation motivating our analysis is as follows. Individuals who are high in perspective taking may be more likely to spontaneously represent how their choices affect others. For them, the consequences of a self-regarding allocation for other group members may already be salient relatively early in the decision process, making prosocial choices more readily available. Additional choice-stage processing may then be associated with greater attention to the private cost of the prosocial allocation or to self--other trade-offs. Conversely, when perspective taking is low, the consequences for others may be less salient initially, and additional processing may be associated with greater consideration of other-regarding consequences. Under this interpretation, the association between processing and prosociality need not have a fixed direction; it may depend on which considerations are already salient when the choice is made.

This argument connects the dual-process literature on prosocial behavior with research on empathy, mentalizing, and social cognition. The empathy-altruism tradition emphasizes that empathic responses can increase the motivational weight attached to others' welfare \citep{batson1991altruism}. Research in social neuroscience and psychology further distinguishes affective sharing, empathic concern, perspective taking, and mental-state inference as related but separable components of social cognition \citep{DecetyJackson2004,SingerLamm2009,Kanskeetal2016}. Experimental and neurobehavioral evidence further shows that empathy, perspective taking, and attentional shifts can make distinct contributions to prosocial behavior, with their relative influence varying across individuals \citep{LammBatsonDecety2007,Tuscheetal2016,lehmann2022empathy}. Our contribution is to bring this distinction into the dual-process debate. Perspective taking may not only increase other-regarding concern on average; it may also help shape whether prosociality is expressed relatively intuitively or is more closely associated with additional processing.

Our analysis considers reflective processing from two complementary perspectives. One is domain-general analytic reflection: the ability or disposition to override intuitive but incorrect answers in abstract reasoning tasks. This is the construct typically measured by the Cognitive Reflection Test (CRT) \citep{Frederick2005,Primietal2016,BranasGarzaetal2019CRT}. The other concerns processing associated with the specific decision at hand. Choice-stage response time provides an imperfect but informative behavioral proxy for decision-specific processing. Unlike CRT performance, which is measured in a separate reasoning task, response time is observed at the stage at which participants submit the distributive choice and is therefore more proximal to the specific decision process. This distinction matters because the mechanism we propose concerns how processing unfolds within a specific distributive choice, whereas CRT performance allows us to assess whether the same moderation pattern extends to domain-general analytic reflection.

Recent work on response times and time pressure reinforces this point while also cautioning against interpreting decision time as a one-to-one indicator of intuition or deliberation. Response times can vary with conflict, difficulty, strength of preference, information search, and the weighting of different attributes of a choice \citep{Evansetal2015,Fiedleretal2013}. In social decisions, \citet{KrajbichEtAl2015} show that apparent associations between response time and selfish or prosocial choice can depend on the discriminability of the available options, while \citet{ChenKrajbich2018} show that time pressure can amplify pre-existing choice biases within a sequential-sampling framework. Related evidence indicates that time pressure can alter which attributes receive greater attention during altruistic choice \citep{TeohEtAl2020}, while empathy traits can moderate the effect of time pressure on prosocial decision-making \citep{liu2023prosocial}. These findings do not make response time uninformative; rather, they motivate treating it as a behavioral measure of processing at the decision stage rather than as a direct readout of an underlying cognitive state. In our setting, this makes it particularly informative to ask whether the predicted moderation pattern appears in this decision-proximal measure, in domain-general cognitive reflection, or in both.

We test these ideas using experimental data from a four-player Distribution Game adapted from \citet{EngelmannStrobel2004}, where the aggregate payoff is held constant across allocation options to avoid the confounding role of efficiency concerns and to focus on distributional motives. Participants choose among allocations that differ in the decision-maker's payoff and in the distribution of payoffs among the other group members. These allocations represent ordered levels of prosociality, ranging from the option that maximizes the decision-maker's payoff to the option that is most equal and maximizes the average payoff of the other group members.

The same participants also completed an extended CRT and the Interpersonal Reactivity Index (IRI), from which we use the Perspective Taking subscale---the tendency to adopt others' viewpoints---as our main empathy measure. In addition, we observe choice-stage response times in the Distribution Game, allowing us to compare domain-general analytic reflection with a behavioral proxy for decision-specific processing. This design therefore combines a non-strategic distributive-choice task with psychological and cognitive measures that allow us to examine whether the relationship between cognitive processing and prosocial behavior differs across individuals.

The results show that empathic perspective taking moderates the association between choice-stage response time and prosocial behavior. At lower levels of perspective taking, the estimated association between response time and prosocial choice is positive but not statistically distinguishable from zero. At higher levels of perspective taking, the estimated association is negative and statistically significant, a pattern consistent with the possibility that prosociality is relatively more intuitive in this group and that additional decision-specific processing may increase attention to its private cost. The same moderation pattern is not detected for CRT performance, our task-based measure of domain-general analytic reflection. An analysis of Empathic Concern, the affective dimension of empathy, reveals an interaction with response time in the opposite direction.

Our paper yields four main contributions.  First, it contributes to the literature on dual-process prosocial behavior by showing that the association between cognitive processing and prosociality depends on how processing is operationalized and, in the case of choice-stage response time, on the individual's level of empathic perspective taking. Second, it extends \citet{PancottoRighi2021} by examining empathic perspective taking as a plausible psychological source of a related form of heterogeneity: prosociality may be intuitive when others' outcomes are already represented in the decision process, and deliberative when they are not. Third, it clarifies the difference between CRT-based analytic reflection and choice-stage response time as a behavioral proxy for decision-specific processing. The former captures a domain-general cognitive disposition, whereas the latter is closer to the processing of the specific social trade-off. Fourth, it contributes to the literature on distributional preferences by studying these mechanisms in a Distribution Game in which the aggregate payoff is held constant, thereby isolating other-regarding allocation motives from efficiency concerns.

\section{Methods}

\subsection{Distribution Game (DG) and prosociality measure}

Prosocial behavior is measured using a four-player Distribution Game (DG), adapted from \citet{EngelmannStrobel2004}. In this task, each participant chooses how to allocate a fixed amount of experimental points among themselves and three other anonymous group members. Participants are informed that, within each group, one individual will be randomly selected as the decision-maker, and that only the allocation chosen by the selected decision-maker will determine payoffs for all group members. The task is therefore payoff-relevant: participants' choices can directly determine both their own payoff and the payoffs of the other members of their group.

\begin{table}[H]
\centering
\footnotesize
\caption{Parameters of the Distribution Game (DG)}
\label{table:DG_parameters}

\begin{threeparttable}
\begin{tabular}{lccc}
\toprule
Player & Choice A & Choice B & Choice C \\
\midrule
Person 1 & 51 & 45 & 42 \\
Decision-maker & 30 & 27 & 24 \\
Person 3 & 9 & 15 & 18 \\
Person 4 & 6 & 9 & 12 \\
\midrule
Aggregate payoff & 96 & 96 & 96 \\
\addlinespace
\textit{Criteria} & & & \\
\midrule
Group variance & 328.5 & 189 & 126 \\
Bolton--Ockenfels (ERC) & $-6.25$ & $-3.13$ & $0$ \\
Fehr--Schmidt strict inequality (FSS) & $-22$ & $-16$ & $-12$ \\
Minimum payoff & 6 & 9 & 12 \\
Average payoff of other group members & 22 & 23 & 24 \\
\bottomrule
\end{tabular}

\begin{tablenotes}[flushleft]
\scriptsize
\item \textit{Note.} Group variance is the variance of individual payoffs under each allocation.
The Bolton--Ockenfels (ERC) index is reported as
\( ERC = -100\left| \frac{\text{Decision-maker payoff}}{96} - \frac{1}{4} \right| \).
Following \citet{EngelmannStrobel2004}, the Fehr--Schmidt strict inequality measure (FSS) assigns equal weight to advantageous and disadvantageous payoff differences and is calculated as
\( -\frac{1}{3} \sum_{i \neq d} \lvert \text{Payoff}_i - \text{Payoff}_d \rvert \),
where \(d\) denotes the decision-maker.
Minimum payoff denotes the lowest individual payoff among the four group members.
Average payoff of other group members refers to the mean payoff of the three non-decision-makers.
\end{tablenotes}

\end{threeparttable}
\end{table}

Each participant chooses among three allocation options, reported in Table~\ref{table:DG_parameters}. In line with \citet{PancottoRighi2021}, the aggregate payoff is constant across choices and always equal to 96 points. The task therefore differs from distribution problems in which a more other-regarding allocation may also increase total surplus. Here, moving from Choice A to Choice C necessarily requires the decision-maker to give up part of his or her own payoff while increasing the average payoff of the other group members and reducing payoff inequality. Experimental points were converted at a rate of 100 points = EUR 3. Thus, conditional on being selected as the decision-maker, choosing Choice C rather than Choice A reduced the decision-maker's payoff by 6 points, equivalent to EUR 0.18. Because aggregate payoffs are held constant, this ordering cannot be driven by a preference for greater social efficiency. Moreover, the allocation decision does not require forming beliefs about other participants' behavior. The game therefore provides a particularly clean measure of costly other-regarding allocation choices, separating them from both efficiency concerns and strategic expectations.

Choice A is the least prosocial option: it maximizes the decision-maker's payoff and generates the highest inequality across group members. Choice B represents an intermediate allocation, while Choice C is the most prosocial option according to the criteria reported in Table~\ref{table:DG_parameters}: it reduces payoff inequality, increases the minimum payoff, and maximizes the average payoff of the other group members. This ordering is consistent with standard measures of social and distributional preferences \citep{FehrSchmidt1999,BoltonOckenfels2000,charness2002understanding,andreoni2002giving,fisman2007individual}. Because concern for others' payoffs and preference for greater equality move in the same direction across the three options, however, the task does not allow us to distinguish which of these distributional motives underlies a more prosocial choice.

\subsection{Cognitive Reflection Test (CRT)}

Cognitive reflection is measured using a five-item version of the Cognitive Reflection Test (CRT). The CRT was originally introduced by \citet{Frederick2005} and is designed to capture individual differences in the tendency to override salient but incorrect intuitive responses in favor of more reflective reasoning. We use items from the extended CRT developed by \citet{Primietal2016}, which provides a broader and less familiar set of questions than the original three-item version.

Each item elicits an intuitive answer that is incorrect, while the correct answer requires analytical reasoning and cognitive control. Responses are coded as binary indicators equal to 1 when the participant provides the reflective/correct answer and 0 otherwise. The main CRT measure is the sum of correct answers across the five items, ranging from 0 to 5, with higher values indicating greater cognitive reflection.

The use of items from the extended CRT developed by \citet{Primietal2016} offers two advantages for our purposes. First, the extended scale was developed to improve measurement across a broader range of cognitive-reflection ability and to reduce the floor effects associated with the original three-item CRT. Second, using alternative items reduces reliance on the best-known original CRT questions, whose widespread familiarity may increase the risk that correct answers reflect prior exposure. At the same time, the items remain numerical reasoning problems, so CRT performance may partly reflect numeracy and calculation ability in addition to the tendency to override an initially compelling response \citep{SinayevPeters2015}.

The task is individual and non-strategic. Instructions remain visible throughout the task, and no feedback is provided before the end of the experiment. Correct answers are monetarily incentivized: participants earn 7 experimental points, equivalent to \euro{}0.21, for each correct answer. The full list of CRT items and correct answers is reported in Table S12 in the Online Appendix.

In the present study, CRT performance provides a task-based measure of domain-general analytic reflection. This measure complements choice-stage response time in the Distribution Game, which we use as a behavioral proxy for decision-specific processing. Comparing the two allows us to examine whether the predicted moderation pattern emerges in performance on a general reflection task, in processing proximal to the distributive choice, or in both.

\subsection{Measures of empathy}

Empathic dispositions are measured using the Interpersonal Reactivity Index (IRI) \citep{Davis1983}. Participants indicate how well each statement describes them on a five-point Likert scale ranging from 1 (`never true') to 5 (`always true'). The task is individual and non-incentivized. As with the other non-incentivized components of the design, the IRI is intended to capture pre-existing psychological characteristics rather than responses induced by monetary incentives or strategic considerations.

The IRI comprises four theoretically distinct subscales capturing complementary dimensions of empathy: Perspective Taking, Empathic Concern, Fantasy, and Personal Distress. While all four dimensions were collected to provide a comprehensive characterization of empathic traits, the primary focus of the present study is on \textit{Perspective Taking} (PT). PT captures the cognitive component of empathy, defined as the tendency to adopt others' viewpoints and represent situations from their perspective.

This focus follows directly from our theoretical argument. The mechanism proposed in the paper concerns whether individuals spontaneously represent the consequences of their choices for others during the distributive decision. PT is therefore the most relevant IRI dimension for examining our hypothesis that other-regarding considerations may be relatively salient early in the decision process.

We also consider \textit{Empathic Concern} (EC), both separately and jointly with PT, and report additional specifications in the Online Appendix. EC captures a more affective dimension of empathy, related to feelings of sympathy, compassion, and concern for others. Examining EC separately and jointly with PT allows us to assess whether the moderating role of empathy differs between its cognitive and affective components.

The remaining IRI subscales capture further dimensions of empathy. The Fantasy Scale measures imaginative identification with fictional characters, while Personal Distress captures self-oriented emotional discomfort in response to others' suffering. These subscales are reported descriptively in the Online Appendix but are not analyzed further because they are not central to the research question.

For the empirical analysis, we construct a continuous PT score by averaging responses across the items of the PT subscale, with higher values indicating greater perspective taking. The full list of items included in each IRI subscale is reported in Table S16 in the Online Appendix. Descriptive statistics for all study variables are also reported in the Online Appendix.

\subsection{Experimental procedures}

The experiment consisted of a fixed sequence of tasks completed by all participants. Participants first took part in a bribery game, whose data are analyzed separately in \citet{Pancottoetal2026}. We also show in the Online Appendix that our main result remains robust when controlling for the treatment and number of rounds of this preceding game. They then completed the Distribution Game, used here to capture prosocial allocation choices, followed by the Cognitive Reflection Test, the Interpersonal Reactivity Index, and a post-experimental survey collecting socio-demographic and background information. No feedback on outcomes was provided between tasks. The full set of instructions is reported in the Online Appendix.

Participants were randomly and anonymously matched in groups of four for the bribery game and the Distribution Game and were rematched between the two tasks. The Cognitive Reflection Test, the Interpersonal Reactivity Index, and the post-experimental survey were completed individually. Prior to each incentivized experimental task, participants answered comprehension questions designed to ensure understanding of the instructions. Participants were not allowed to proceed until the correct answers were provided.

In the Distribution Game, participants were shown the allocation options and their associated payoffs in the task instructions and completed the comprehension questions before reaching the final choice page. Response time is measured as the time spent on this final page, where participants submitted Choice A, B, or C. It therefore captures choice-stage response time rather than the full amount of time during which participants could consider the problem.

The experiment was implemented in Python using oTree, the web-based platform developed by \citet{Chen2016}, and conducted at the VERA Laboratory for Experiments in Social Sciences (VERALabEx) at Ca' Foscari University of Venice. The study followed the laboratory's Regulations and Guidelines and was conducted under the oversight of the VERALabEx Steering Committee. The experimental application, including the description of the experiment and its procedures, was reviewed by the Steering Committee, which identified no critical aspects concerning ethics or privacy and did not require further review by an Ethics Committee.\footnote{The Executive Board of the Department of Economics at Ca' Foscari University of Venice reviewed the application on 5 September 2023, approved partial financing of the experiment, and endorsed the project. The decision was subsequently ratified by the Department Council on 14 September 2023 (decision 299/2023).}

Upon arrival, participants were registered, assigned to individual cubicles, and provided with informed consent and privacy information. Visual and verbal communication among participants was not permitted during the sessions. All task instructions were displayed on participants' screens and read aloud by the same experimenter across sessions; written instructions remained accessible throughout each task.

Experimental sessions lasted approximately 90 minutes. Payoffs were expressed in experimental points and converted to euros at a fixed exchange rate. Participants earned an average payment of approximately \euro{}15.

Each participant took part in only one session. Because the data were collected as part of a broader experimental project, the sample size was not determined specifically for the present analysis. We use the full sample of 652 participants recruited across two data-collection waves, with 39.3\% recruited in April 2023 and 60.7\% recruited in October--November 2023. The composition of the final sample is reported in detail in Tables S1--S10 and Figure S1 in the Online Appendix.

The analysis plan for the relationships among Distribution Game choices, empathy, and cognitive reflection was preregistered on the Open Science Framework on January 23, 2024, after data collection had been completed but before any human observation of the data (OSF: 10.17605/OSF.IO/K8C29). The preregistration specified analyses of fairness in the Distribution Game, overall empathy and its IRI subscales, CRT performance, and multivariate specifications including an interaction between empathy and reflectivity. Choice-stage response time was not included in the preregistration and is examined in the present study as a theory-driven extension of the preregistered analysis.

\subsection{Hypotheses}

The theoretical argument developed above implies a general moderation pattern. As discussed in the literature review, reflection does not have a uniform association with prosocial choice. Rather, its association with prosocial behavior may depend on which considerations are already salient when additional processing occurs \citep{PancottoRighi2021,capraro2024dualprocess}. Empathic perspective taking is expected to be associated with greater initial salience of other-regarding consequences, because it captures the dispositional tendency to represent others' perspectives and the consequences of one's choices for them \citep{Davis1983,DecetyJackson2004,lehmann2022empathy}.

For individuals high in perspective taking, other-regarding consequences may already be relatively salient at an early stage of the decision process. In this case, additional processing is expected to be associated with greater attention to the private cost of the prosocial allocation or to the self--other trade-off, and therefore with less prosocial choice. For individuals low in perspective taking, by contrast, others' outcomes may be less salient initially. In this case, additional processing is expected to be associated with greater consideration of other-regarding consequences and therefore with more prosocial choice.

The preregistered analysis plan included a specification in which fairness in the Distribution Game was regressed on empathy, cognitive reflection, and their interaction, considering both overall empathy and the individual IRI subscales. The preregistration therefore anticipated testing whether the association between CRT-based reflectivity and fairness varies with empathy, including Perspective Taking, although it did not specify the directional conditional pattern developed below. Choice-stage response time was not included in the preregistration. Building on the same theoretical argument, we formulate two complementary hypotheses. H1 specifies the directional pattern for the preregistered empathy $\times$ CRT analysis using Perspective Taking, whereas H2 applies the same moderation logic to choice-stage response time.

We first consider CRT performance, which provides the closest link to \citet{PancottoRighi2021}. If empathic perspective taking captures a psychological source of heterogeneity related to that documented across naïve and strategic participants, the association between domain-general analytic reflection and prosocial behavior should vary across levels of perspective taking.

\begin{hp}[Perspective taking and cognitive reflection]
Empathic perspective taking moderates the association between Cognitive Reflection Test performance and prosocial behavior. Higher CRT performance is associated with more prosocial behavior among individuals with low levels of perspective taking and with less prosocial behavior among individuals with high levels of perspective taking.
\end{hp}

We then consider choice-stage response time in the Distribution Game. Because the proposed mechanism concerns how processing unfolds within the distributive choice itself, choice-stage response time provides a behavioral proxy that is more proximal to the specific decision than CRT performance. If perspective taking shapes which considerations are initially salient during that decision, the association between response time and prosocial behavior should display the same conditional pattern.

\begin{hp}[Perspective taking and choice-stage response time]
Empathic perspective taking moderates the association between choice-stage response time in the Distribution Game and prosocial behavior. Longer response times are associated with more prosocial behavior among individuals with low levels of perspective taking and with less prosocial behavior among individuals with high levels of perspective taking.
\end{hp}

Both hypotheses are formulated as conditional associations rather than as causal effects, since perspective taking, CRT performance, and response time are not experimentally manipulated. Examining the same conditional pattern across the two indicators allows us to assess whether the proposed moderation extends to domain-general analytic reflection, emerges in choice-stage response time as a behavioral proxy for decision-specific processing, or is observed in both.

\subsection{Empirical Strategy}

To evaluate these hypotheses, we estimate regression models of prosocial choice that include reflective processing, empathic perspective taking, and their interaction. This empirical strategy allows us to assess whether the association between reflection and prosocial behavior depends on individual differences in perspective taking. The key empirical object is the conditional association between reflective processing and prosocial behavior at different levels of perspective taking.

For the empirical analysis, individual prosocial behavior is operationalized using participants' choices in the DG. We code Choice A, Choice B, and Choice C as increasing levels of prosocial behavior, taking values 1, 2, and 3, respectively. This coding reflects the natural ordering of the allocation options from least to most prosocial. Our main specifications are estimated using OLS, which provides a transparent interpretation of interaction terms and conditional slopes. Because the outcome is ordinal and the distances between adjacent categories need not be identical, we also estimate ordered logit models as a robustness check. These models yield the same qualitative interaction pattern and are reported in the Online Appendix.

We examine reflective processing using two complementary empirical approaches. First, cognitive reflection is measured by participants' total number of correct answers in the Cognitive Reflection Test. We denote this measure by \(\mathrm{CRT}_i\). The variable ranges from 0 to 5, with higher values indicating greater cognitive reflection. Second, choice-stage response time in the Distribution Game, denoted by \(\mathrm{RT}_i\), is used as a behavioral proxy for decision-specific processing. Thus, whereas \(\mathrm{CRT}_i\) captures performance on a domain-general reflection task, \(\mathrm{RT}_i\) provides information about processing proximal to the distributive choice itself.

Choice-stage response time was unavailable in five experimental sessions. The response-time specifications therefore use the subsample for which response time and the other variables required for estimation are available (\(N=525\)). The CRT specifications use the larger available sample (\(N=649\)). We additionally report common-sample and joint specifications to assess whether the comparison between CRT and response time is affected by the difference in estimation samples. Additional checks reported in the Online Appendix show no statistically significant differences between participants with and without response-time information on the main observed characteristics used in the analysis.

We also construct a continuous PT score, denoted by \(\mathrm{PTmean}_i\), by averaging responses across the items of the PT subscale, with higher values indicating greater perspective taking. This continuous measure is used in the main specifications to capture variation in empathic disposition across individuals.

For regression estimation, Response Time is expressed in units of 10 seconds, so that the estimated coefficients and conditional slopes refer to a 10-second increase in decision time. Before estimation, the continuous variables entering each interaction---either CRT score or choice-stage response time, together with perspective taking---are mean-centered within the relevant estimation sample. For notational simplicity, we retain the notation \(R_i\) and \(\mathrm{PTmean}_i\) for the centered variables in the equations below. Centering makes the lower-order coefficients interpretable at the sample mean of the other variable and is standard practice in the interpretation of interaction models \citep{AikenWest1991}. It does not alter the interaction coefficient, fitted values, or overall model fit; it mainly changes the interpretation of the lower-order terms and may reduce nonessential collinearity between the interaction term and its constituent variables \citep{DalalZickar2012}.

Let \(Y_i\) denote individual \(i\)'s Distribution Game choice, coded as an ordered prosociality score from 1 to 3, where higher values indicate more prosocial allocations. Let \(R_i\) denote the cognitive-processing indicator used in a given specification, either the CRT score \((\mathrm{CRT}_i)\) or choice-stage response time \((\mathrm{RT}_i)\). We estimate the following specification:

\begin{equation}\label{eq:main_interaction}
Y_i = \alpha
+ \beta_1 R_i
+ \beta_2 \mathrm{PTmean}_i
+ \beta_3 (R_i \times \mathrm{PTmean}_i)
+ \gamma X_i
+ \varepsilon_i ,
\end{equation}

where \(X_i\) is a set of demographic and background controls---age, gender, generalized trust, association membership, voting behavior, and blood-donation status---and \(\varepsilon_i\) is an error term. The model is estimated separately using \(\mathrm{CRT}_i\) and \(\mathrm{RT}_i\) as alternative measures of \(R_i\), both without and with the control variables. Because \(R_i\) and \(\mathrm{PTmean}_i\) are mean-centered, \(\beta_1\) represents the association between reflective processing and prosocial behavior at the average level of perspective taking, while \(\beta_2\) represents the association between perspective taking and prosocial behavior at the average level of reflective processing.

From Eq.~\ref{eq:main_interaction}, the marginal effect of reflective processing on prosocial behavior is:

\begin{equation}\label{eq:marginal_reflection}
\frac{\partial Y_i}{\partial R_i}
= \beta_1 + \beta_3 \mathrm{PTmean}_i .
\end{equation}

This expression shows that the association between reflective processing and prosocial behavior varies linearly with perspective taking. A positive marginal effect indicates that higher reflective processing is associated with more prosocial behavior, whereas a negative marginal effect indicates that higher reflective processing is associated with less prosocial behavior.

Both hypotheses predict opposite conditional associations at low and high levels of perspective taking. At relatively low levels of perspective taking, reflective processing is expected to be positively associated with prosocial behavior:

\begin{equation}\label{eq:low_pt_prediction}
\beta_1 + \beta_3 \mathrm{PTmean}_{L} > 0,
\end{equation}

where \(\mathrm{PTmean}_{L}\) denotes a low value of perspective taking. At relatively high levels of perspective taking, reflective processing is expected to be negatively associated with prosocial behavior:

\begin{equation}\label{eq:high_pt_prediction}
\beta_1 + \beta_3 \mathrm{PTmean}_{H} < 0,
\end{equation}

where \(\mathrm{PTmean}_{H}\) denotes a high value of perspective taking. Empirically, we evaluate these conditional associations by computing marginal effects of reflective processing at one standard deviation below the mean, at the mean, and at one standard deviation above the mean of perspective taking.

Estimating the model using both CRT scores and choice-stage response times allows us to assess whether the moderating role of empathic perspective taking emerges across these complementary indicators of cognitive processing. CRT performance captures domain-general analytic reflection, while choice-stage response time provides a behavioral proxy for processing proximal to the allocation decision. Additional robustness checks and extended specifications are reported in the Online Appendix. 
To compare the cognitive and affective dimensions of empathy, we also estimate analogous specifications using Empathic Concern and a joint model including both empathy dimensions and their interactions with response time.

\section{Results}
\subsection{Regression Results}

We estimate linear interaction models to examine whether empathic perspective taking moderates the association between reflective processing and prosocial behavior. Table~\ref{tab:reflection_prosocial} reports the main specifications. The interaction with perspective taking is negative and statistically significant for response time, whereas the corresponding interaction for CRT performance is not statistically significant. In columns (1)--(2), domain-general analytic reflection is measured using the CRT score. In columns (3)--(4), choice-stage response time in the Distribution Game is used as a behavioral proxy for decision-specific processing. For each indicator, we report specifications without and with demographic and social controls.

\begin{table}[htbp]
\centering
\caption{OLS estimates for prosocial choice in the Distribution Game}
\label{tab:reflection_prosocial}
\begin{threeparttable}
\begin{tabular}{lcccc}
\toprule
& (1) & (2) & (3) & (4) \\
\midrule

Perspective Taking (PT) & 0.092 & 0.092 & 0.097 & 0.093 \\
                        & (0.056) & (0.057) & (0.061) & (0.063) \\

Cognitive Reflection (CRT) & -0.018 & -0.019 &  &  \\
                           & (0.028) & (0.029) &  &  \\

Cognitive Reflection (CRT) $\times$ PT & 0.010 & 0.015 &  &  \\
                & (0.045) & (0.046) &  &  \\

Response Time (DG) &  &  & 0.000 & 0.004 \\
                   &  &  & (0.026) & (0.026) \\

Response Time (DG) $\times$ PT &  &  & -0.064* & -0.072** \\
                               &  &  & (0.027) & (0.028) \\

Constant & 2.074*** & 2.422*** & 2.049*** & 2.533*** \\
         & (0.033) & (0.236) & (0.037) & (0.249) \\
         
\midrule

Controls & No & Yes & No & Yes \\

Observations & 649 & 649 & 525 & 525 \\

$R^2$ & 0.005 & 0.021 & 0.012 & 0.034 \\

\bottomrule
\end{tabular}

\begin{tablenotes}[flushleft]
\footnotesize
\item Notes: The dependent variable is the ordered prosociality score in the Distribution Game, ranging from 1 to 3. Models (1)--(2) use the CRT score as a measure of domain-general analytic reflection. Models (3)--(4) use choice-stage response time in the Distribution Game, expressed in units of 10 seconds, as a behavioral proxy for decision-specific processing. The continuous variables entering each interaction term are mean-centered within the relevant estimation sample. Robust standard errors are reported in parentheses. Coefficients are rounded to three decimal places. *** $p < .001$, ** $p < .01$, * $p < .05$.
\end{tablenotes}

\end{threeparttable}
\end{table}

\subsubsection{Cognitive Reflection Test}

Columns (1) and (2) report the results using scores on the Cognitive Reflection Test (CRT) as a measure of reflective processing. Across both specifications, neither the main effect of CRT nor the interaction between CRT and perspective taking is statistically significant. The inclusion of demographic and social controls in column (2) does not materially alter these results.

These estimates do not provide evidence that perspective taking moderates the association between CRT performance and prosocial allocations in the Distribution Game. Nor do we detect a statistically significant main association between CRT performance and prosocial choice. These null estimates should not be interpreted as showing that domain-general analytic reflection is irrelevant to distributive behavior; rather, the moderation pattern predicted in H1 is not detected in this sample. This is consistent with evidence that the relationship between cognitive reflection and social behavior varies substantially across decision environments and domains of sociality \citep{capraro2024dualprocess}.

\subsubsection{Response Time and Perspective Taking}

Columns (3) and (4) report the results using choice-stage response time in the Distribution Game as a behavioral proxy for decision-specific processing. In contrast to the CRT specifications, the interaction between response time and perspective taking is negative and statistically significant across specifications. This result remains robust after including demographic and social controls.

\begin{figure}[H]
\centering

\includegraphics[width=0.85\textwidth]{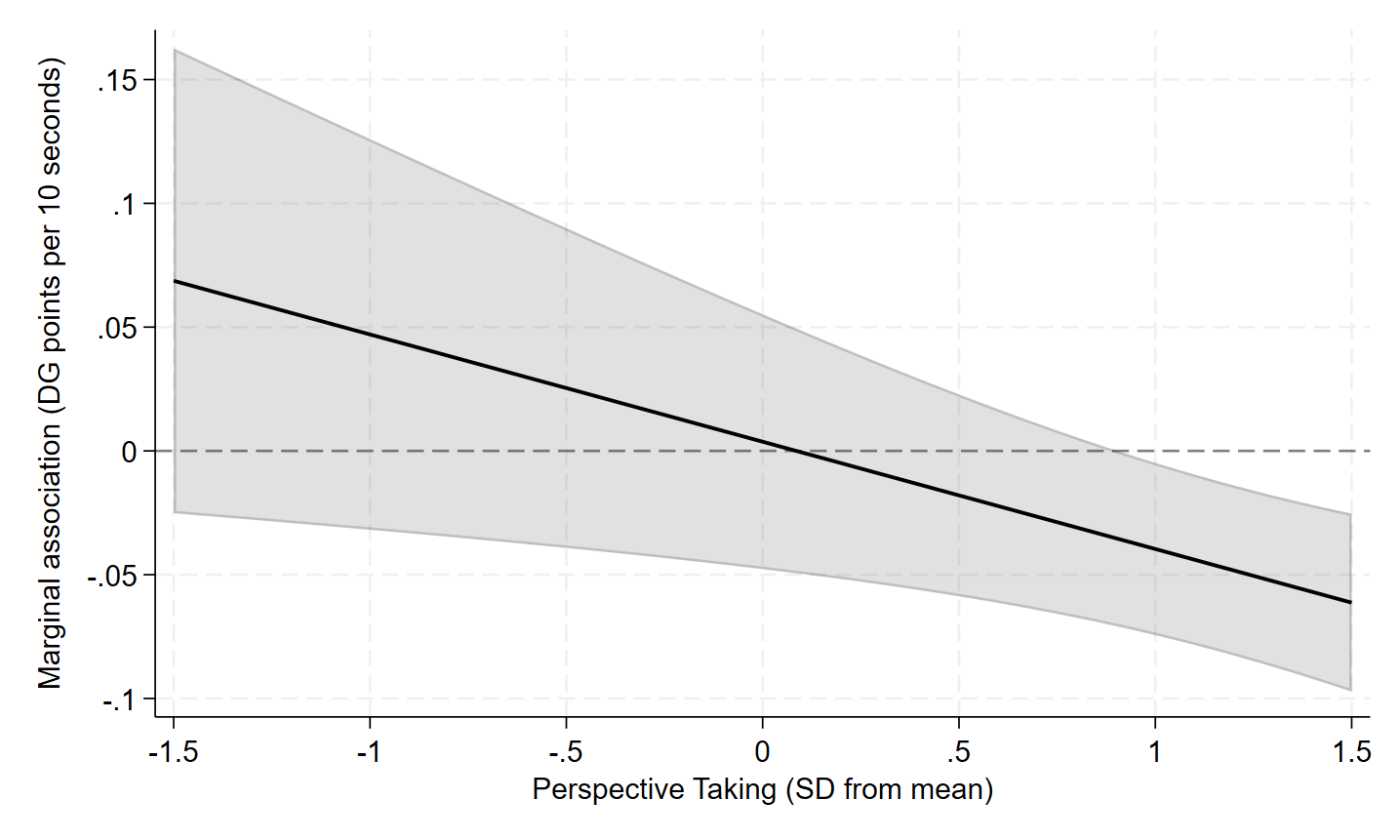}

\caption{Marginal association of response time across perspective taking}
\label{fig:rt_pt_marginal_main}

\begin{minipage}{0.85\textwidth}
\footnotesize
\textit{Notes.} The figure reports the estimated marginal association
between choice-stage response time and the Distribution Game
prosociality score across levels of Perspective Taking.
Estimates are obtained from the controlled OLS specification
($N=525$), with robust standard errors.
Perspective Taking is expressed in standard deviations from
its sample mean, over the range $-1.5$ to $+1.5$ SD.
The vertical axis reports the estimated change in the
prosociality score associated with 10 additional seconds
of response time.
The shaded area represents the 95\% confidence interval.
The horizontal dashed line indicates zero.
\end{minipage}

\end{figure}

The negative interaction implies that the association between choice-stage response time and prosocial behavior varies with the individual's level of empathic perspective taking. To illustrate this relationship, we compute the marginal effect of response time at different levels of perspective taking. The estimated marginal effects are reported in Table~\ref{tab:marginal_rt_pt}, while Figure~\ref{fig:rt_pt_marginal_main} illustrates how the marginal association varies across levels of perspective taking.

\begin{table}[htbp]
\centering
\begin{threeparttable}
\caption{Marginal effect of response time at different levels of perspective taking}
\label{tab:marginal_rt_pt}
\begin{tabular}{lcc}
\toprule
 & Without controls & With controls \\
\midrule
PT = $-1$ SD & 0.039 & 0.047 \\
    & (0.040) & (0.040) \\
PT = mean    & 0.000 & 0.004 \\
    & (0.026) & (0.026) \\
PT = $+1$ SD & -0.038* & -0.040* \\
    & (0.018) & (0.018) \\
\bottomrule
\end{tabular}
\begin{tablenotes}[flushleft]
\footnotesize
\item Notes: The table reports conditional marginal effects of response time in the centered Response Time $\times$ Perspective Taking models. Marginal effects refer to a 10-second increase in response time. Perspective Taking is evaluated at one standard deviation below the mean, at the mean, and at one standard deviation above the mean. Standard errors, reported in parentheses, are obtained by the delta method. Values are rounded to three decimal places. *** $p < .001$, ** $p < .01$, * $p < .05$.
\end{tablenotes}
\end{threeparttable}
\end{table}

Figure~\ref{fig:rt_pt_marginal_main} shows how the relationship between decision time and prosocial choice differs with Perspective Taking. Among participants with higher Perspective Taking, those who take longer to decide tend to choose less prosocial allocations, and this association is statistically significant. At lower levels of Perspective Taking, the estimated relationship is positive but not statistically significant, while around the mean it is close to zero.

The magnitude of this conditional association should be interpreted in relation to both the scale of the dependent variable and the observed variation in response time. The dependent variable ranges from 1 to 3, and the estimated marginal effect at high levels of perspective taking is approximately $-0.040$ for a 10-second increase in response time. This corresponds to a relatively small change in the prosociality score. However, response times vary substantially in the data: the average response time is about 18.5 seconds, with a standard deviation of about 16.3 seconds. A one-standard-deviation increase in response time therefore corresponds to an expected change of approximately $-0.065$ points in the prosociality score at high levels of perspective taking. The substantive magnitude is therefore modest, but the estimate indicates that the conditional association between response time and prosocial choice varies meaningfully across levels of perspective taking.

This pattern should not be interpreted as showing that longer response time is inherently prosocial or self-interested. Rather, it is consistent with the possibility that additional decision-specific processing changes the relative weight assigned to different considerations in the allocation decision \citep{TeohEtAl2020,teoh2022games}. For individuals scoring relatively low on perspective taking---that is, individuals less prone to spontaneously adopt others' viewpoints---additional processing may increase attention to the consequences of one's choice for others. For individuals scoring relatively high on perspective taking, those consequences may already be salient, and additional processing may instead increase attention to the private cost of generosity. These interpretations concern a possible mechanism underlying the observed conditional association rather than a directly identified cognitive process.

Importantly, the evidence for moderation does not depend on the statistical significance of any single conditional slope. Difference-in-slopes tests reported in the Online Appendix show that the estimated association between response time and prosocial choice differs significantly between low and high levels of perspective taking. The estimated slope crosses zero close to the mean of the perspective-taking distribution, indicating that the change in the conditional association is not confined to extreme levels of perspective taking.

Additional sensitivity analyses reported in the Online Appendix assess the stability of the response-time interaction to extreme response-time values and session heterogeneity. The Response Time $\times$ PT interaction remains negative and statistically significant when response time is winsorized at the 1st and 99th percentiles and when session fixed effects are included. Ordered logit specifications also yield the same qualitative interaction pattern.

\subsubsection{Comparing the Two Indicators of Cognitive Processing}

Taken together, the main specifications do not support H1 for CRT performance and support H2 for choice-stage response time. The Response Time $\times$ PT interaction is negative and statistically significant, and the difference between the conditional slopes at low and high levels of perspective taking is statistically significant. Consistent with the predicted directional pattern, the estimated response-time slope is positive at low levels of perspective taking and negative at high levels, although only the latter conditional slope is individually statistically significant.

Because the CRT and response-time specifications use different estimation samples, we conduct two additional comparisons on the common response-time sample (\(N=525\)). First, re-estimating the CRT interaction model on this restricted sample leaves the CRT $\times$ PT interaction close to zero and statistically insignificant both without controls (\(\beta=-0.010\), SE \(=0.052\)) and with controls (\(\beta=0.006\), SE \(=0.053\)). Second, we estimate a joint specification including CRT performance, choice-stage response time, perspective taking, and both interaction terms. In the controlled specification, the Response Time $\times$ PT interaction remains negative and statistically significant (\(\beta=-0.075\), SE \(=0.028\), \(p=.008\)), whereas the CRT $\times$ PT interaction remains statistically insignificant (\(\beta=0.020\), SE \(=0.054\), \(p=.705\)). The same qualitative pattern holds without controls.

Response time and CRT scores are positively but modestly correlated in the data (\(r=0.108\), \(p=.013\)), indicating limited overlap between the two indicators. Taken together, the separate, common-sample, and joint specifications show that the moderation pattern is detected for choice-stage response time but not for CRT performance. This contrast is consistent with the distinction developed above between domain-general analytic reflection and processing proximal to the specific distributive choice, while not implying that either measure uniquely identifies the underlying cognitive process.

\subsubsection{Cognitive versus Affective Empathy: Opposite Response-Time Moderation}

While our main analysis focuses on Perspective Taking (PT), we use Empathic Concern (EC) as a natural benchmark against which to interpret it. Both are subscales of the same instrument and are estimated on the same sample and with the same specifications, but they capture different aspects of empathy: PT reflects the cognitive tendency to adopt others' perspectives, whereas EC reflects affective concern for others. Comparing how response time moderates each dimension therefore helps clarify what our PT measure captures.

Columns (1) and (2) of Table~\ref{tab:rt_pt_ec_comparison} show the separate models for PT and EC, estimated on the same 525 observations with the same controls. The interaction with Response Time is negative for PT and positive for EC, and both are statistically significant in the separate models. The interaction between CRT performance and EC is not statistically significant, either with or without controls (Section S4 of the Online Appendix).

Column (3) includes both empathy measures and their interactions with Response Time. The interaction coefficients retain opposite signs. Neither is individually significant ($p=.118$ for PT and $p=.147$ for EC), but the joint test rejects that both coefficients are zero ($p=.027$), and the test of equality rejects that they are the same ($p=.010$). The estimates and tests are similar when CRT performance is added as a control (Section S4.4 of the Online Appendix).

\begin{table}[htbp]
\centering
\caption{Response Time interactions with Perspective Taking and Empathic Concern}
\label{tab:rt_pt_ec_comparison}
\begin{threeparttable}
\begin{tabular}{lccc}
\toprule
& (1) & (2) & (3) \\
\midrule
Response Time $\times$ PT & -0.072** & & -0.053 \\
& (0.028) & & (0.034) \\
\addlinespace
Response Time $\times$ EC & & 0.042* & 0.032 \\
& & (0.017) & (0.022) \\
\midrule
Controls & Yes & Yes & Yes \\
Observations & 525 & 525 & 525 \\
$R^2$ & 0.034 & 0.032 & 0.038 \\
\midrule
Joint significance ($p$-value) & & & .027 \\
Equality of interaction coefficients ($p$-value) & & & .010 \\
\bottomrule
\end{tabular}
\begin{tablenotes}[flushleft]
\footnotesize
\item Notes: The dependent variable is the Distribution Game prosociality score, ranging from 1 to 3. The table reports selected OLS coefficients with robust standard errors in parentheses. Response Time is expressed in units of 10 seconds. Response Time and the empathy scores are mean-centered. All models include the main effects corresponding to their interaction terms and the same demographic and social controls. Column (1) reproduces column (4) of Table~\ref{tab:reflection_prosocial}. Column (2) uses Empathic Concern in place of Perspective Taking. Column (3) includes both empathy measures and their interactions with Response Time. The last two rows report robust Wald tests for column (3). Full results are reported in Section S4 of the Online Appendix. *** $p<.001$, ** $p<.01$, * $p<.05$.
\end{tablenotes}
\end{threeparttable}
\end{table}

\begin{figure}[H]
\centering

\includegraphics[width=\textwidth]{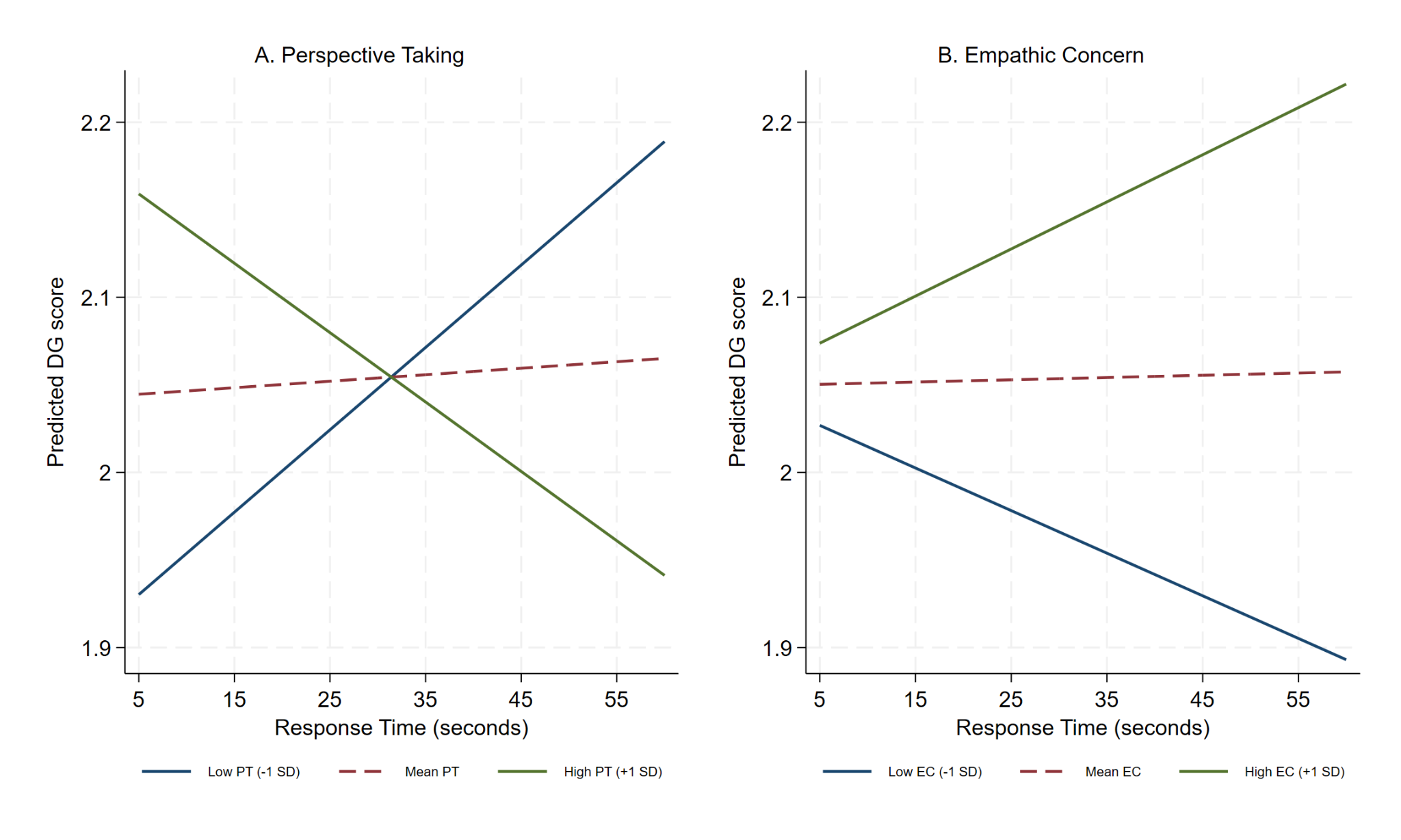}

\caption{Predicted prosocial choice across response time and empathy dimensions}
\label{fig:rt_pt_ec_predicted}

\begin{minipage}{0.95\textwidth}
\footnotesize
\textit{Notes.} The figure shows predicted Distribution Game
prosociality scores across choice-stage response times for
different levels of Perspective Taking (Panel A) and
Empathic Concern (Panel B).
Low and high empathy correspond to one standard deviation
below and above the respective sample mean.
Predictions are obtained from separate controlled OLS
specifications, using the same estimation sample ($N=525$)
and demographic and social controls.
The lines represent point predictions.
Confidence intervals are omitted for visual clarity.
The corresponding regression estimates and additional
marginal-effect analyses are reported in the Online Appendix.
\end{minipage}

\end{figure}

Figure~\ref{fig:rt_pt_ec_predicted} plots predicted Distribution Game scores across response times for different levels of PT and EC, using the separate controlled specifications. The two dimensions display mirror-image patterns. For PT, predicted scores of high-PT participants decline with response time while those of low-PT participants rise, so that the prosocial advantage associated with PT is concentrated among fast decisions: at the 25th percentile of response time (10 seconds), the predicted difference between high and low PT is 0.186 points ($p=.030$), and it shrinks to 0.082 points ($p=.276$) at the 75th percentile (22 seconds). For EC the pattern is reversed: predicted scores of high-EC participants increase with response time while those of low-EC participants decline, so that the gap is small among fast decisions (0.073 points, $p=.396$) and widens among slower ones (0.134 points, $p=.086$).

One interpretation consistent with these patterns is that our PT measure captures a spontaneous, intuitive component of other-regard that is distinct from affective concern: high-PT individuals are more prosocial when they decide quickly, and this advantage is attenuated among longer decisions. EC does not display this intuitive signature; its association with prosocial choice, if anything, emerges in slower decisions. This result is in line with \citet{PancottoRighi2021}, where prosociality was intuitive for strategically sophisticated participants, who became less prosocial with greater reflectivity, whereas non-strategic participants were intuitively selfish and became more prosocial with reflection. The authors conjecture that the common root of strategic reasoning and intuitive prosociality lies in the cognitive, rather than affective, component of empathy, a conjecture their design could not test directly. Our results are consistent with it: the intuitive-prosociality pattern appears for perspective taking, the cognitive dimension of empathy, but not for empathic concern, its affective counterpart. One difference should be noted: whereas \citet{PancottoRighi2021} measure reflectivity through the CRT, in our data the moderation emerges with decision-level response time and not with CRT performance, suggesting that the relevant heterogeneity is more clearly reflected in processing proximal to the specific decision than in a general disposition toward reflection.

Two caveats apply. First, response time is observed rather than exogenously manipulated, so these patterns are correlational and should not be read as the causal effect of deliberation. They may also reflect selection into fast versus slow responding or differences in decision conflict. Second, the PT lines cross at around 31 seconds. The predicted gap narrows as response time approaches the crossing, consistent with the negative interaction between Response Time and PT, but is estimated less precisely beyond that point. This may reflect the smaller number of observations at longer response times, with 60 of the 525 observations in the estimation sample, or 11.4\%, above 30 seconds.

\section{Discussion}

The results show that empathic perspective taking moderates the association between choice-stage response time and prosocial behavior, whereas no comparable moderation is detected for CRT performance. This contrast suggests that the predicted heterogeneity is more clearly detected in processing proximal to the specific distributive choice than in performance on a domain-general reflection task.

The response-time results are consistent with the idea that prosocial behavior can emerge through different cognitive routes. Among individuals with relatively low levels of perspective taking, the estimated association between response time and prosocial choice is positive, although not statistically significant. Among individuals with relatively high levels of perspective taking, the association is negative and statistically significant: longer choice-stage response times are associated with less prosocial behavior. The magnitude of this conditional association is modest, but the variation in the response-time slope across the perspective-taking distribution is theoretically informative. It is consistent with the possibility that additional decision-specific processing operates differently depending on which considerations are already salient at the time of choice.

One interpretation consistent with our theoretical framework is that empathic perspective taking makes others' outcomes salient relatively early in the decision process. When this occurs, prosocial allocations may require less reflective correction because the consequences of one's choices for other group members are already represented. Additional processing may then increase attention to the private cost of generosity or to the self--other trade-off embedded in the allocation decision. Conversely, when perspective taking is lower, others' outcomes may be less salient initially, and additional processing may help bring them into consideration. Because response time does not uniquely identify the underlying cognitive process, this mechanism should be understood as an interpretation of the observed moderation pattern rather than as a directly measured sequence of mental operations.

The contrast with CRT performance is also informative, although the CRT null result should not be interpreted as evidence that domain-general analytic reflection is irrelevant to distributive behavior. CRT performance captures success on a separate reflection task and may also partly reflect numeracy and calculation ability, whereas choice-stage response time provides information about processing proximal to the allocation decision itself. Moreover, response time can reflect not only additional processing but also decision conflict, difficulty, preference strength, and information search \citep{KrajbichEtAl2015,ChenKrajbich2018,TeohEtAl2020}. The evidence therefore supports treating the two indicators as conceptually distinct sources of information about cognitive processing, without implying that either provides a pure measure of the underlying cognitive process.

These results fit a broader pattern in the dual-process literature: the association between cognitive processing and prosocial behavior varies across individuals and decision environments \citep{Randetal2012,Tinghogetal2013,Bouwmeesteretal2017,MischkowskiGlockner2016,Yamagishietal2017,capraro2024dualprocess}. In our data, this heterogeneity is related to perspective taking: the association between additional processing and prosocial choice becomes more negative as perspective taking increases. This pattern is consistent with the idea that the role of additional processing depends partly on which considerations are already salient when the choice is made.

The results also extend the interpretation proposed by \citet{PancottoRighi2021}. In their study, reflectivity is positively associated with prosociality among naïve participants but negatively associated with prosociality among strategically sophisticated participants. Our findings point to empathic perspective taking as a plausible psychological source of a related form of heterogeneity. Strategic reasoning and perspective taking are conceptually distinct: strategic reasoning involves analytic inference about others' behavior and possible reactions, whereas perspective taking reflects a dispositional tendency to adopt others' viewpoints and represent how outcomes affect them. Recent evidence similarly shows that the relationship between social-cognitive ability and cooperation can depend strongly on the strategic context \citep{RidingerMcBride2025}. However, both processes require representing others' perspectives or outcomes. This shared element may help explain why prosociality can appear more intuitive among both strategically sophisticated individuals and individuals high in perspective taking.

The analysis of Empathic Concern reveals a contrasting pattern to that observed for Perspective Taking. Perspective Taking reflects the tendency to adopt others' viewpoints, whereas Empathic Concern captures sympathy, compassion, and concern for others and is therefore more closely connected to the affective-motivational component of empathy emphasized in the empathy--altruism literature \citep{batson1991altruism,Eisenbergetal1987}. The two dimensions are positively but only moderately correlated in our data (\(r=0.309\)), consistent with their being related but distinct components of empathy. In line with this distinction, the interaction between response time and Empathic Concern is positive, whereas the corresponding interaction with Perspective Taking is negative. When PT and EC enter the same model, their interactions with response time retain opposite signs, and the difference between the two coefficients is statistically significant. In the fitted profiles, individuals with higher PT have higher predicted prosociality at short response times, but this difference narrows as response time increases. For EC, by contrast, the predicted difference between high and low levels of empathy increases with response time. These contrasting patterns are consistent with the possibility that cognitive perspective taking and affective concern relate differently to the salience and weighting of other-regarding considerations during choice \citep{Tuscheetal2016,TeohEtAl2020,liu2023prosocial}. However, they do not identify a temporal sequence of psychological processes: response times are observed across individuals rather than as within-person processing trajectories. The two empathy dimensions should therefore not be treated as interchangeable measures of empathy.

The results also suggest a practical implication: encouraging people to spend more time on a social decision may not have the same consequences for everyone. If the association between processing and prosocial choice depends on whether others' outcomes are already salient, additional decision-stage processing may be associated with different choice patterns across individuals. A useful next step is to test this idea directly by combining time-based interventions with perspective-taking cues. Because neither processing time nor perspective taking is experimentally manipulated here, this remains a hypothesis for future causal work rather than a direct policy prescription.

Some features of the design nevertheless qualify the interpretation of these findings. Choice-stage response time provides a behavioral proxy for decision-specific processing, but it does not identify the particular cognitive process underlying longer choices. Because both the CRT and the IRI were administered after the Distribution Game, task-order or carryover effects on these post-task measures cannot be ruled out. In addition, because concern for others' payoffs and preference for greater equality move together across the Distribution Game options, the task does not separately identify the distributional motive underlying a more prosocial choice. Future research could test these mechanisms more directly through experimental manipulations of processing conditions and distributive tasks that more sharply separate competing social-preference motives.

Our main conclusion is that the association between decision time and prosocial choice is not uniform across individuals. At higher levels of perspective taking, longer choice-stage response times are associated with less prosocial choice, whereas at lower levels the estimated association is positive but imprecise. This pattern is consistent with our central argument that additional processing may play a different role depending on which social considerations are already salient when the choice is made.

\section*{Declaration of generative AI and AI-assisted technologies in the manuscript preparation process}

During the preparation of this work, the authors used ChatGPT to support language editing, text organization, and consistency checks. After using this tool, the authors reviewed and edited the content as needed and take full responsibility for the content of the article.

\bibliographystyle{plainnat}
\bibliography{biblio}

\end{document}


\begin{center}
{\Large \textbf{Online Appendix:}} {\large \textbf{When Is Prosocial Behavior Intuitive? \\The Moderating Role of Empathic Perspective Taking}}\\[10pt]

{\normalsize
Enrico D’Ecclesiis\textsuperscript{1} [\textbf{Corresponding author}],
Francesca Pancotto\textsuperscript{2},
Barbara Luppi\textsuperscript{3} and
Simone Righi\textsuperscript{3}
}\\[6pt]
\end{center}
{\small
\textsuperscript{1} Department of Economics Marco Biagi, University of Modena and Reggio Emilia, Via Jacopo Berengario 51, 41121 Modena, Italy. Email: enricoangeloraffaele.decclesiis@unimore.it\\
\textsuperscript{2} Department of Linguistic and Cultural Studies, University of Modena and Reggio Emilia, Largo Sant'Eufemia 19, 41121 Modena, Italy
\textsuperscript{3} Department of Economics Marco Biagi, University of Modena and Reggio Emilia, Via Jacopo Berengario 51, 41121 Modena, Italy
}

\vspace{10pt}

\noindent
This Online Appendix reports descriptive statistics of the variables in the dataset, details on measures and coding, additional analyses supporting the results presented in the main text, and the full experimental instructions.

\tableofcontents
\clearpage

\section{Sample and Descriptive Statistics}

The data analyzed in this study were collected across two distinct waves, yielding a total sample of 652 participants. The first wave took place in April 2023 and accounts for 39.3\% of the observations, while the second wave was conducted between October and November 2023 and accounts for the remaining 60.7\%. The Distribution Game and the subsequent CRT, IRI, and survey components analyzed in the present study followed the same procedures across the two waves.

This section reports descriptive statistics for sample composition, Distribution Game choices, CRT performance, choice-stage response time, the IRI subscales, and the civic-capital and corruption measures. It also reports pairwise correlations among the main variables used in the analysis.

\subsection{Sample composition}

The final sample consists of 652 participants and is predominantly composed of young individuals, with an average age of 22.1 years (SD = 4.1), reflecting the student-centered nature of the recruitment. Students represent the large majority of respondents (89.1\%), while the remaining participants are mainly employed or inactive. Women account for 63.2\% of the sample, men for 35.0\%, with a small residual share reporting other or missing gender categories. Participants were mainly recruited in Northern Italy, with a strong concentration in Veneto in terms of current residence, place of birth, and region of primary education. Overall, these characteristics indicate a relatively homogeneous socio-demographic background, consistent with the student-centered recruitment of a university laboratory experiment.

\begin{figure}[H]
\centering
\caption{Age distribution}
\includegraphics[width=0.85\textwidth]{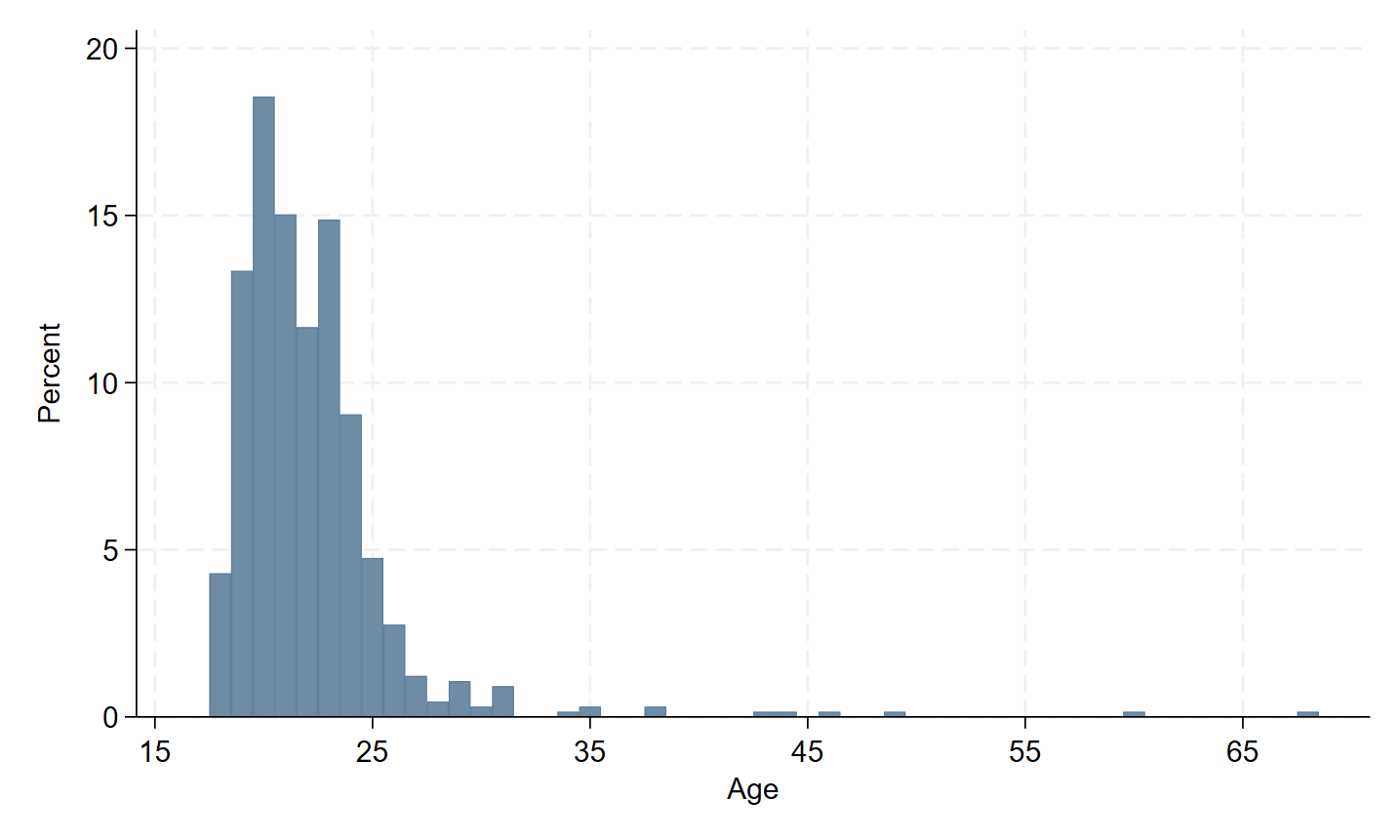}
\label{fig:age_hist}
\end{figure}

A complete list and descriptive statistics for socio-demographic and background characteristics are reported in Tables~\ref{list_covariates}--\ref{descriptive_dep_cat4} and Figure~\ref{fig:age_hist}.

\begin{center}
\begin{longtable}{P{4.0cm} P{11cm}}
\caption{List of socio-demographic and background characteristics}
\label{list_covariates}\\
\toprule
Variable & Description \\
\midrule
\endfirsthead

\toprule
Variable & Description \\
\midrule
\endhead

Gender &
Self-reported gender. Response categories include: Female, Male, Prefer not to answer, and Not identified. \\

Age &
Age of the participant, reported in years. \\

Region of residence &
Italian region of current residence. Responses are coded according to the 20 Italian administrative regions; an additional category identifies participants residing abroad. \\

Region of birth &
Italian region of birth, coded according to the 20 Italian administrative regions; participants born abroad are recorded in a separate category. \\

Region of primary education &
Italian region where the participant completed primary education. Responses are coded according to the 20 Italian administrative regions; primary education completed abroad is recorded separately. \\

Education level &
Highest educational attainment. Categories include: High school diploma, Bachelor’s degree, and Master’s degree. \\

Main occupation &
Self-reported main occupational status. Categories include: Employed, Student, Unemployed, Retired, and Not looking for a job. \\

Department of affiliation &
Department of academic affiliation at Ca’ Foscari University of Venice. Categories include: Economics, Management, Philosophy and Cultural Heritage, Environmental Sciences, Computer Science and Statistics, Molecular and Nanosystems Sciences, Linguistic and Contemporary Cultural Studies, Humanities, and Asian and Mediterranean African Studies. Additional categories identify non--Ca’ Foscari students and non-students. \\

Department area &
Recoded version of the department of affiliation, grouped into four macro-areas: STEM, Economic--Social Sciences, Humanities, and Other. \\

\bottomrule
\end{longtable}
\end{center}

\begin{center}
\begin{table}[H]
\centering
\caption{Descriptive statistics of socio-demographic and background characteristics}
\label{tab:descriptive_background}
\begin{threeparttable}
\begin{tabular}{P{4.5cm} C{1.5cm} C{1.5cm} C{1.5cm} C{1.5cm} C{1.5cm}}
\toprule
Variable & N & Mean & SD & Min & Max \\
\midrule
Gender & 652 & -- & -- & 1 & 4 \\
Age & 652 & 22.141 & 4.063 & 18 & 68 \\
Region of residence & 652 & -- & -- & 1 & 20 \\
Region of birth & 652 & -- & -- & 1 & 21 \\
Region of primary educ. & 652 & -- & -- & 1 & 21 \\
Education level & 652 & 1.768 & 0.597 & 1 & 3 \\
Main occupation & 652 & -- & -- & 1 & 5 \\
Department of affiliation & 652 & -- & -- & 1 & 10 \\
Department area & 652 & -- & -- & 1 & 4 \\
\bottomrule
\end{tabular}

\begin{tablenotes}[flushleft]\footnotesize
\item \textit{Notes.} Mean and standard deviation are not reported for categorical variables, for which such summary statistics are not informative. Full distributional statistics for these variables are reported in the subsequent tables.
\end{tablenotes}
\end{threeparttable}
\end{table}
\end{center}

\begin{longtable}{lccc}
\caption{Descriptive statistics for gender}\label{descriptive_gender}\\
\hline
Gender & Freq. & Percent & Cum. \\
\hline
Female & 412 & 63.19 & 63.19 \\
Male & 228 & 34.97 & 98.16 \\
Prefer not to answer & 8 & 1.23 & 99.39 \\
Not identified & 4 & 0.61 & 100.00 \\
\hline
Total & 652 & 100.00 & 100.00 \\
\hline
\end{longtable}

\begin{longtable}{lccc}
\caption{Descriptive statistics for region of residence}\label{descriptive_region_res}\\
\hline
Region of residence & Freq. & Percent & Cum. \\
\hline
Abruzzo & 3 & 0.46 & 0.46 \\
Basilicata & 1 & 0.15 & 0.61 \\
Calabria & 1 & 0.15 & 0.77 \\
Campania & 5 & 0.77 & 1.53 \\
Emilia-Romagna & 24 & 3.68 & 5.21 \\
Friuli-Venezia Giulia & 33 & 5.06 & 10.28 \\
Lazio & 14 & 2.15 & 12.42 \\
Liguria & 9 & 1.38 & 13.80 \\
Lombardia & 45 & 6.90 & 20.71 \\
Marche & 14 & 2.15 & 22.85 \\
Molise & 1 & 0.15 & 23.01 \\
Piemonte & 16 & 2.45 & 25.46 \\
Puglia & 4 & 0.61 & 26.07 \\
Sardegna & 5 & 0.77 & 26.84 \\
Sicilia & 13 & 1.99 & 28.83 \\
Toscana & 11 & 1.69 & 30.52 \\
Trentino-Alto Adige & 3 & 0.46 & 30.98 \\
Umbria & 5 & 0.77 & 31.75 \\
Valle d'Aosta & 1 & 0.15 & 31.90 \\
Veneto & 444 & 68.10 & 100.00 \\
\hline
Total & 652 & 100.00 & 100.00 \\
\hline
\end{longtable}

\begin{longtable}{lccc}
\caption{Descriptive statistics for region of birth}\label{descriptive_region_birt}\\
\hline
Region of birth & Freq. & Percent & Cum. \\
\hline
Abruzzo & 6 & 0.92 & 0.92 \\
Basilicata & 1 & 0.15 & 1.07 \\
Calabria & 3 & 0.46 & 1.53 \\
Campania & 8 & 1.23 & 2.76 \\
Emilia-Romagna & 28 & 4.29 & 7.06 \\
Friuli-Venezia Giulia & 33 & 5.06 & 12.12 \\
Lazio & 16 & 2.45 & 14.57 \\
Liguria & 9 & 1.38 & 15.95 \\
Lombardia & 52 & 7.98 & 23.93 \\
Marche & 14 & 2.15 & 26.07 \\
Molise & 2 & 0.31 & 26.38 \\
Piemonte & 21 & 3.22 & 29.60 \\
Puglia & 8 & 1.23 & 30.83 \\
Sardegna & 9 & 1.38 & 32.21 \\
Sicilia & 22 & 3.37 & 35.58 \\
Toscana & 11 & 1.69 & 37.27 \\
Trentino-Alto Adige & 4 & 0.61 & 37.88 \\
Umbria & 5 & 0.77 & 38.65 \\
Valle d'Aosta & 2 & 0.31 & 38.96 \\
Veneto & 390 & 59.82 & 98.77 \\
Abroad / Other & 8 & 1.23 & 100.00 \\
\hline
Total & 652 & 100.00 & 100.00 \\
\hline
\end{longtable}

\begin{longtable}{lccc}
\caption{Descriptive statistics for region of primary education}\label{descriptive_region_ele}\\
\hline
Region of primary education & Freq. & Percent & Cum. \\
\hline
Abruzzo & 6 & 0.92 & 0.92 \\
Basilicata & 1 & 0.15 & 1.07 \\
Calabria & 3 & 0.46 & 1.53 \\
Campania & 8 & 1.23 & 2.76 \\
Emilia-Romagna & 28 & 4.29 & 7.06 \\
Friuli-Venezia Giulia & 33 & 5.06 & 12.12 \\
Lazio & 16 & 2.45 & 14.57 \\
Liguria & 9 & 1.38 & 15.95 \\
Lombardia & 52 & 7.98 & 23.93 \\
Marche & 14 & 2.15 & 26.07 \\
Molise & 2 & 0.31 & 26.38 \\
Piemonte & 21 & 3.22 & 29.60 \\
Puglia & 8 & 1.23 & 30.83 \\
Sardegna & 9 & 1.38 & 32.21 \\
Sicilia & 22 & 3.37 & 35.58 \\
Toscana & 11 & 1.69 & 37.27 \\
Trentino-Alto Adige & 4 & 0.61 & 37.88 \\
Umbria & 5 & 0.77 & 38.65 \\
Valle d'Aosta & 2 & 0.31 & 38.96 \\
Veneto & 390 & 59.82 & 98.77 \\
Abroad / Other & 8 & 1.23 & 100.00 \\
\hline
Total & 652 & 100.00 & 100.00 \\
\hline
\end{longtable}

\begin{longtable}{lccc}
\caption{Descriptive statistics for education level}\label{descriptive_education_level}\\
\hline
Education level & Freq. & Percent & Cum. \\
\hline
Bachelor's degree & 209 & 32.06 & 32.06 \\
High school diploma & 385 & 59.05 & 91.10 \\
Master's degree & 58 & 8.90 & 100.00 \\
\hline
Total & 652 & 100.00 & 100.00 \\
\hline
\end{longtable}

\begin{longtable}{lccc}
\caption{Descriptive statistics for main occupation}\label{descriptive_occupation}\\
\hline
Main occupation & Freq. & Percent & Cum. \\
\hline
Employed & 55 & 8.44 & 8.44 \\
Not looking for a job & 7 & 1.07 & 9.51 \\
Retired & 1 & 0.15 & 9.66 \\
Student & 581 & 89.11 & 98.77 \\
Unemployed & 8 & 1.23 & 100.00 \\
\hline
Total & 652 & 100.00 & 100.00 \\
\hline
\end{longtable}

\begin{longtable}{lccc}
\caption{Descriptive statistics for department of affiliation}\label{descriptive_department_aff}\\
\hline
Department of affiliation & Freq. & Percent & Cum. \\
\hline
Economics & 222 & 34.05 & 34.05 \\
Philosophy and Cultural Heritage & 63 & 9.66 & 43.71 \\
Management & 142 & 21.78 & 65.49 \\
Environmental Sciences, Computer Science and Statistics & 7 & 1.07 & 66.56 \\
Molecular and Nanosystems Sciences & 6 & 0.92 & 67.48 \\
Linguistic and Contemporary Cultural Studies & 39 & 5.98 & 73.47 \\
Humanities & 36 & 5.52 & 78.99 \\
Asian and Mediterranean African Studies & 111 & 17.02 & 96.01 \\
Not a student & 12 & 1.84 & 97.85 \\
Not a Ca’ Foscari student & 14 & 2.15 & 100.00 \\
\hline
Total & 652 & 100.00 & 100.00 \\
\hline
\end{longtable}

\begin{longtable}{lccc}
\caption{Descriptive statistics for department area}\label{descriptive_dep_cat4}\\
\hline
Department area & Freq. & Percent & Cum. \\
\hline
STEM & 13 & 1.99 & 1.99 \\
Economic--Social Sciences & 364 & 55.83 & 57.82 \\
Humanities & 249 & 38.19 & 96.01 \\
Other & 26 & 3.99 & 100.00 \\
\hline
Total & 652 & 100.00 & 100.00 \\
\hline
\end{longtable}

\subsection{Distribution Game (DG)}

\begin{figure}[H]
\centering
\caption{Distribution Game -- decision types}
\includegraphics[width=0.85\textwidth]{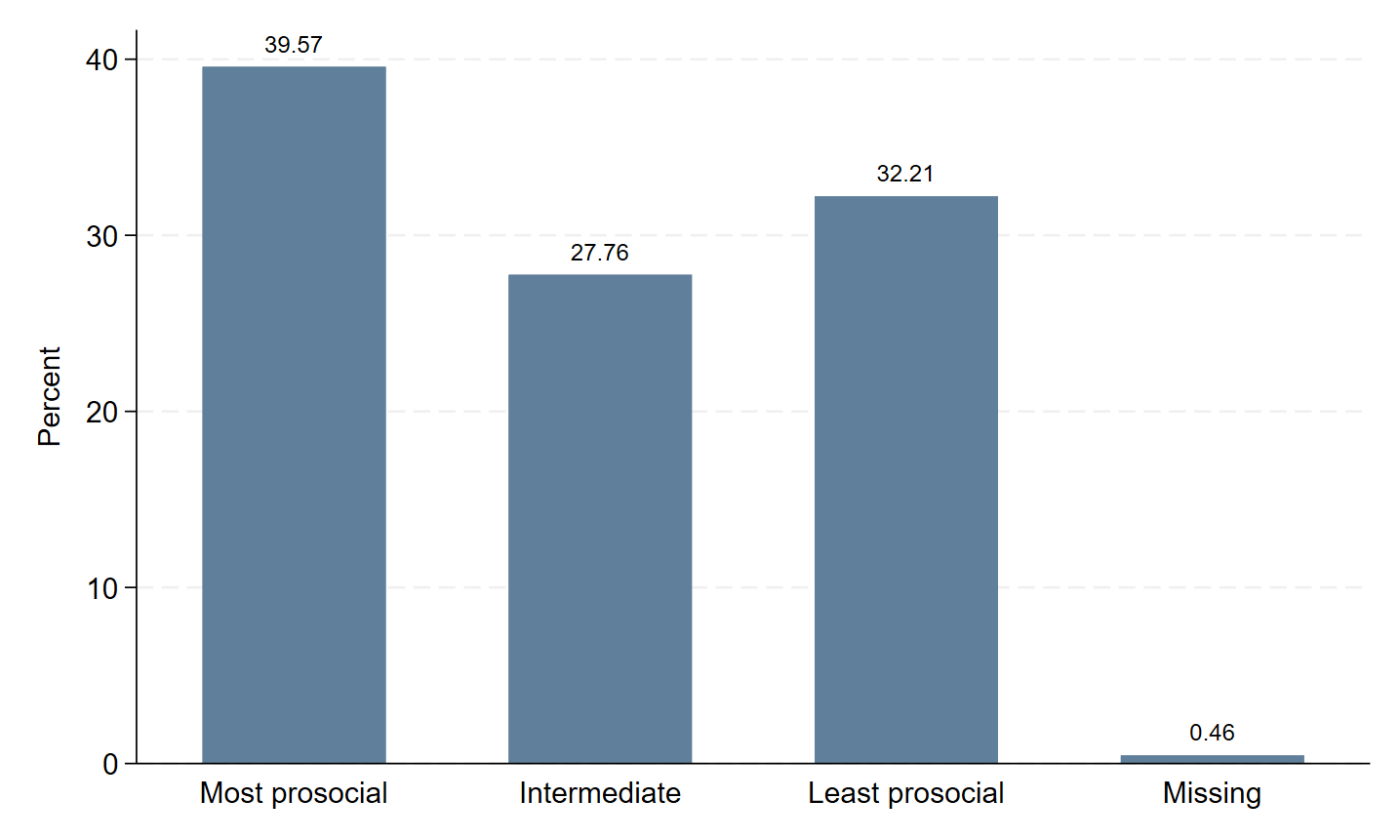}
\label{fig:dg_bar}
\end{figure}

Distributive preferences are measured using a Distribution Game (DG), which provides a behavioral measure of individual distributive orientation ranging from more prosocial to more self-regarding choices. The game presents participants with a set of alternative payoff allocations, allowing us to classify observed decisions into most prosocial, intermediate, and least prosocial types. Looking at the distribution of decision types, most prosocial choices represent the most frequent outcome, accounting for 39.6\% of observations, followed by least prosocial decisions (32.2\%) and intermediate choices (27.8\%). Missing decisions are negligible (0.5\%). Overall, the distribution suggests substantial heterogeneity in distributive preferences, with no single decision type clearly dominating the others. While most prosocial behavior emerges as the modal choice, a sizeable share of participants opts for either intermediate or fully least prosocial allocations, highlighting the coexistence of prosocial and self-interested motivations within the sample. Complete statistics and graphical representations are reported in Table~\ref{descriptive_dist_game} and Figure~\ref{fig:dg_bar}.

\begin{longtable}{lccc}
\caption{Descriptive statistics for Distribution Game decision types}\label{descriptive_dist_game}\\
\hline
Distribution Game decision & Freq. & Percent & Cum. \\
\hline
Most prosocial & 258 & 39.57 & 39.57 \\
Intermediate & 181 & 27.76 & 67.33 \\
Least prosocial & 210 & 32.21 & 99.54 \\
Missing & 3 & 0.46 & 100.00 \\
\hline
Total & 652 & 100.00 & 100.00 \\
\hline
\end{longtable}

\subsection{Cognitive Reflection Test (CRT)}

Cognitive reflection is assessed through a five-item Cognitive Reflection Test (CRT), capturing the propensity to override intuitive responses in favor of reflective reasoning. Table~\ref{tab:final_vars1} reports the five CRT items included in the questionnaire, along with the corresponding reflective/correct answers.

\begin{center}
\begin{longtable}{|P{2.0cm}|p{9.5cm}|P{3.0cm}|}
    \caption{Summary of Cognitive Reflection Test (CRT) variables}\label{tab:final_vars1}\\
 \hline
\textbf{Item} & \textbf{Question} & \textbf{Reflective answer} \\ \hline \hline
    \hline
    CRT item 1 (Percentage)               
        & If you flipped a fair coin three times, what is the probability that it would land ‘Heads’ at least once? Answer expressed as a PERCENTAGE & 87.5 percent \\ \hline
    CRT item 2 (Workers) 
        & If three workers can wrap three toys in one hour, how many workers are needed to wrap six toys in two hours? Answer expressed as WORKERS & 3 workers \\ \hline
    CRT item 3 (Laps)           
        & Giulia and Sara are running around a track. They run equally fast but Giulia started later. When Giulia has run 5 laps, Sara has run 15 laps. When Giulia has run 30 laps, how many laps has Sara run? Answer expressed as LAPS & 40 laps \\ \hline
    CRT item 4 (Students)    
        & Marco received both the 15th highest and the 15th lowest mark in the class. How many students are in the class? Answer expressed as STUDENTS & 29 students \\ \hline
    CRT item 5 (Medals)                
        & In an athletics team, tall members are three times more likely to win a medal than short members. This year, the team has won 60 medals so far. How many of these have been won by short athletes? Answer expressed as MEDALS & 15 medals \\ \hline
    \hline

\end{longtable}
\end{center}

Table~\ref{descriptive_crt_items} reports the share of respondents providing the reflective answer for each item. Performance varies substantially across questions: while a large majority answers correctly the “Workers” item (90.0\%), success rates are considerably lower for the “Percentage” item (8.4\%) and intermediate for the remaining items, ranging between 47.5\% and 60.3\%. This heterogeneity suggests meaningful variation in item difficulty and cognitive demands within the test.

\begin{figure}[H]
\centering
\caption{Cognitive Reflection Test -- total number of reflective/correct answers}
\includegraphics[width=0.85\textwidth]{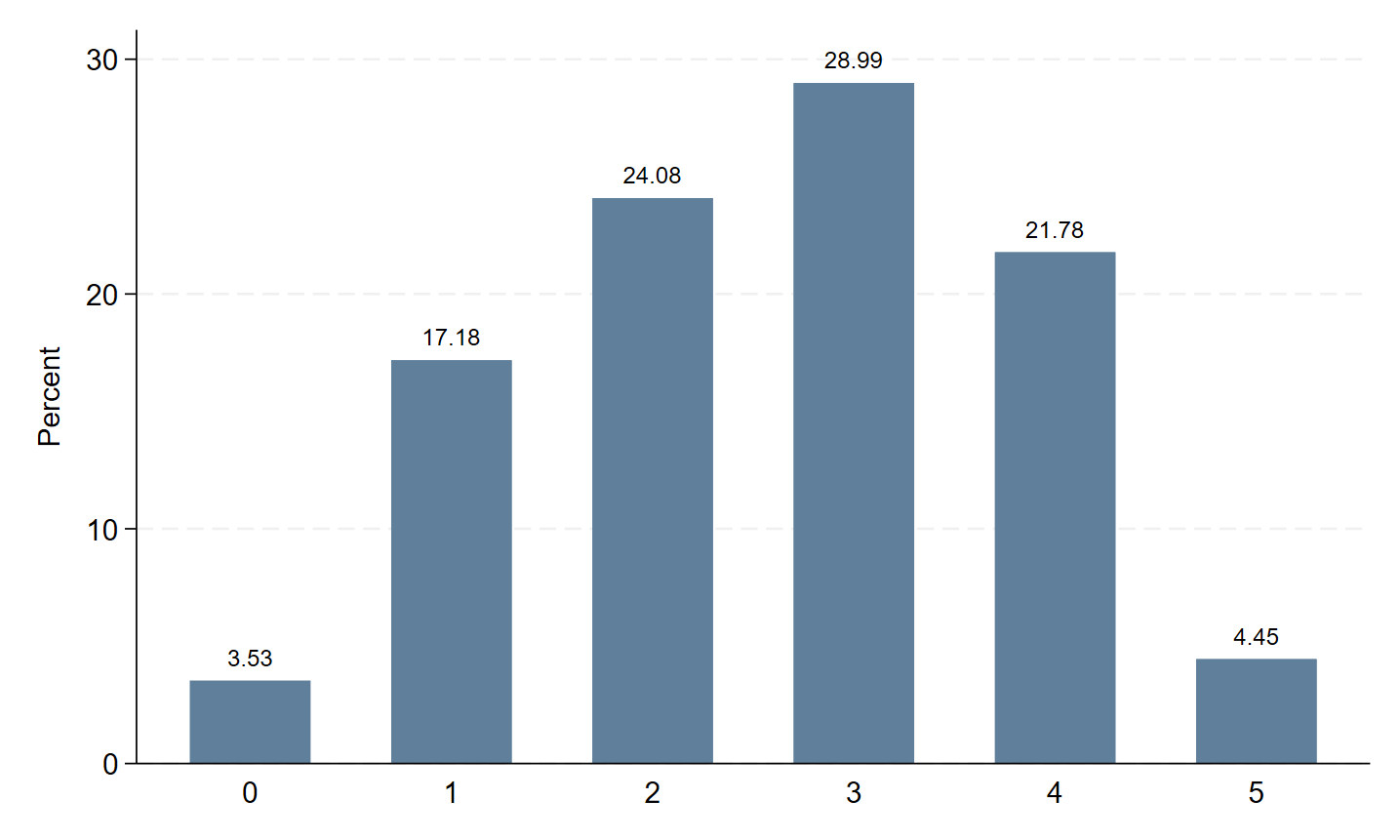}
\label{fig:crt_bar}
\end{figure}

Aggregating across items, the average CRT score is 2.62 (SD = 1.22) on a scale from 0 to 5 (Table~\ref{descriptive_crt_items}). The distribution of total scores is approximately unimodal and centered around intermediate values, as shown in Table~\ref{descriptive_crt_sum_tab} and Figure~\ref{fig:crt_bar}: most respondents provide between two and four reflective answers, with relatively few individuals at the extremes (3.5\% scoring zero and 4.4\% scoring the maximum). Overall, these patterns indicate substantial heterogeneity in cognitive reflection within the sample, with a clear concentration around moderate levels of reflective reasoning.

\begin{center}    
    \begin{longtable}{|P{4.8cm}|C{1cm}|C{1cm}|C{1cm}|C{0.5cm}|C{0.5cm}|}
    \caption{Cognitive Reflection Test (CRT) items (binary) and CRT sum score (0--5)}\label{descriptive_crt_items}\\
    \hline
        Variable  & count & mean & sd & min & max \\ \hline
        CRT item 1 (Percentage) & 652 & 0.084 & 0.278 & 0 & 1 \\ \hline
        CRT item 2 (Workers) & 652 & 0.900 & 0.300 & 0 & 1 \\ \hline
        CRT item 3 (Laps) & 652 & 0.554 & 0.497 & 0 & 1 \\ \hline
        CRT item 4 (Students) & 652 & 0.475 & 0.500 & 0 & 1 \\ \hline
        CRT item 5 (Medals) & 652 & 0.603 & 0.490 & 0 & 1 \\ \hline
        CRT sum score & 652 & 2.617 & 1.223 & 0 & 5 \\ \hline
    \end{longtable}

    \vspace{-0.8cm}
    {\footnotesize \textit{Note:} CRT items are coded as binary indicators (1 = reflective/correct answer; 0 = otherwise), and the means represent the share of respondents providing the reflective answer. The CRT sum score corresponds to the total number of reflective/correct answers across the five CRT items (range 0--5).}
\end{center}

\begin{longtable}{lccc}
\caption{Descriptive statistics for CRT sum score}\label{descriptive_crt_sum_tab}\\
\hline
CRT sum score & Freq. & Percent & Cum. \\
\hline
0 & 23  & 3.53  & 3.53 \\
1 & 112 & 17.18 & 20.71 \\
2 & 157 & 24.08 & 44.79 \\
3 & 189 & 28.99 & 73.77 \\
4 & 142 & 21.78 & 95.55 \\
5 & 29  & 4.45  & 100.00 \\
\hline
Total & 652 & 100.00 & 100.00 \\
\hline
\end{longtable}

\subsection{Response Time (DG)}

In addition to the Cognitive Reflection Test (CRT), the dataset records the time spent on the final choice page of the Distribution Game. In the main analysis, \textit{Response Time (DG)} is therefore interpreted as choice-stage response time and used as a behavioral proxy for decision-specific processing, complementing CRT performance as a task-based measure of domain-general analytic reflection.

Response Time is available for 528 of the 652 participants. Because three participants with recorded response times have a missing Distribution Game choice, the Response Time regression sample consists of 525 observations. Response-time availability and the comparability of the resulting estimation sample are examined further in Section~S3.

\begin{figure}[H]
\centering
\caption{Response Time (DG) (seconds)}
\label{fig:dg_rt_hist}
\includegraphics[width=0.85\textwidth]{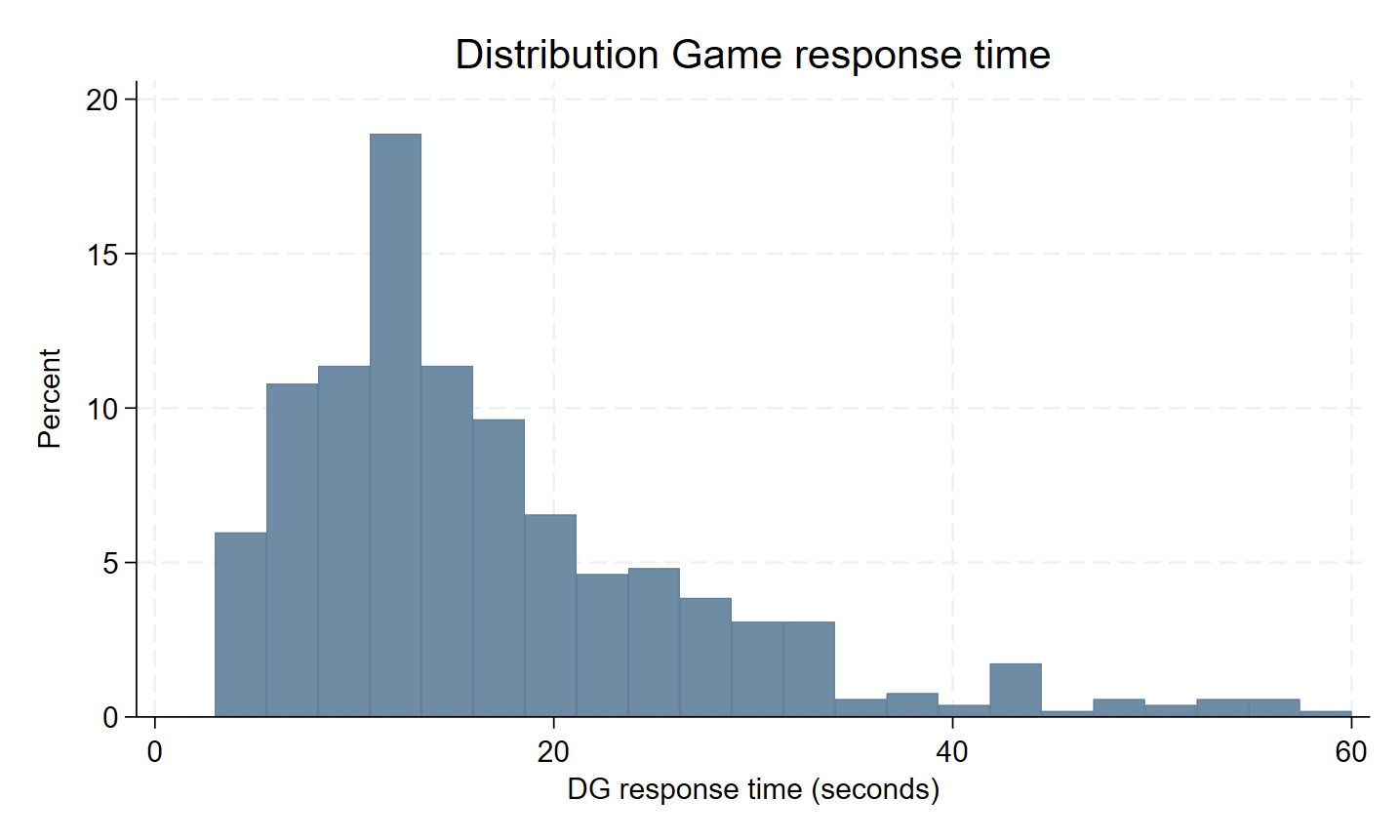}

\begin{threeparttable}
\begin{tablenotes}[flushleft]\footnotesize
\item \textit{Notes.} Histogram of Response Time (DG), restricted to observations with recorded response times and trimmed at 60 seconds for display. Percentages refer to the trimmed sample.
\end{tablenotes}
\end{threeparttable}
\end{figure}

Response Time (DG) displays substantial heterogeneity. The average time is 18.5 seconds (SD = 16.3; $N=528$), with a minimum of 3 seconds and a small number of very long decision times (maximum = 168 seconds). Table~\ref{tab:dg_rt_sum} reports summary statistics, while Figure~\ref{fig:dg_rt_hist} shows the distribution (for ease of visualization, the histogram is trimmed at 60 seconds). As is typical in response-time data, the distribution is right-skewed: most participants decide relatively quickly, while a minority spend considerably longer on the task.

\begin{center}
\begin{table}[H]
\centering
\caption{Descriptive statistics for Response Time (DG) (seconds)}
\label{tab:dg_rt_sum}
\begin{threeparttable}
\begin{tabular}{lccccc}
\toprule
Variable & N & Mean & SD & Min & Max \\
\midrule
Response Time (DG) & 528 & 18.517 & 16.306 & 3 & 168 \\
\bottomrule
\end{tabular}
\end{threeparttable}
\end{table}
\end{center}

\subsection{Interpersonal Reactivity Index (IRI)}

Empathy is captured through the Interpersonal Reactivity Index (IRI), which comprises four distinct but related dimensions: Perspective Taking (PT), Fantasy (FS), Empathic Concern (EC), and Personal Distress (PD). Each dimension is assessed through seven items measured on a five-point Likert scale (from 1 = Never true to 5 = Always true), with negatively worded items reverse-coded prior to aggregation. Table~\ref{tab:final_varsIRI} provides a detailed description of the IRI items, their allocation across the four dimensions, and the response scale.

\begin{center}
\begin{longtable}{|P{2.0cm}|p{9.5cm}|P{3.0cm}|}
    \caption{Summary of Interpersonal Reactivity Index (IRI) variables}\label{tab:final_varsIRI}\\
 \hline
\textbf{Subscale} & \textbf{Item wording} & \textbf{Response scale} \\ \hline \hline
    \hline
    \multicolumn{3}{|l|}{\textbf{Perspective Taking (PT)}}\\
    \hline
    IRI item 1 (PT)               
        & Sometimes I find it difficult to see things from another person’s point of view (Inverted) & From 1 Never true to 5 Always true \\ \hline
    IRI item 2 (PT) 
        & In case of disagreement, I try to take everyone’s point of view into account before making a decision & From 1 Never true to 5 Always true \\ \hline
    IRI item 3 (PT)           
        & Sometimes I try to understand my friends better by imagining how things may appear from their perspective & From 1 Never true to 5 Always true \\ \hline
    IRI item 4 (PT)    
        & If I’m sure I’m right about something, I don’t waste much time listening to other people’s arguments (Inverted) & From 1 Never true to 5 Always true \\ \hline
    IRI item 5 (PT)                
        & I believe that there are two opposite sides to every story and I try to take both into consideration & From 1 Never true to 5 Always true \\ \hline
    IRI item 6 (PT)                
        & When I am in conflict with someone, I usually try to put myself in their shoes for a moment & From 1 Never true to 5 Always true \\ \hline
    IRI item 7 (PT)                
        & Before criticizing someone, I try to imagine how I would feel if I were in their place & From 1 Never true to 5 Always true \\ \hline
    \multicolumn{3}{|l|}{\textbf{Fantasy (FS)}}\\
    \hline
    IRI item 8 (FS)               
        & I daydream and fantasize quite regularly about things that might happen to me & From 1 Never true to 5 Always true \\ \hline
    IRI item 9 (FS) 
        & I get truly involved in the moods of the main characters in a story & From 1 Never true to 5 Always true \\ \hline
    IRI item 10 (FS)           
        & I can usually be objective when watching a movie or a play and I rarely get completely involved (Inverted) & From 1 Never true to 5 Always true \\ \hline
    IRI item 11 (FS)    
        & I rarely get really involved with a good book or a good movie (Inverted) & From 1 Never true to 5 Always true \\ \hline
    IRI item 12 (FS)                
        & After watching a play or a movie, I feel as if I am one of the protagonists myself & From 1 Never true to 5 Always true \\ \hline
    IRI item 13 (FS)                
        & When I watch a good movie, I can very easily put myself in the shoes of a main character & From 1 Never true to 5 Always true \\ \hline
    IRI item 14 (FS)                
        & When I read an interesting story or tale, I imagine how I would feel if those events were happening to me & From 1 Never true to 5 Always true \\ \hline
    \multicolumn{3}{|l|}{\textbf{Empathic Concern (EC)}}\\
    \hline
    IRI item 15 (EC)               
        & I often feel tenderness and concern for people less fortunate than I am & From 1 Never true to 5 Always true \\ \hline
    IRI item 16 (EC) 
        & Sometimes I don’t feel much sorry for other people who have problems (Inverted) & From 1 Never true to 5 Always true \\ \hline
    IRI item 17 (EC)           
        & When I see someone being exploited, I have protective feelings toward them & From 1 Never true to 5 Always true \\ \hline
    IRI item 18 (EC)    
        & Sometimes other people’s misfortunes don’t upset me much (Inverted) & From 1 Never true to 5 Always true \\ \hline
    IRI item 19 (EC)                
        & When I see someone being treated unfairly, I sometimes don’t feel so much pity for them (Inverted) & From 1 Never true to 5 Always true \\ \hline
    IRI item 20 (EC)                
        & Things that happen affect me quite often & From 1 Never true to 5 Always true \\ \hline
    IRI item 21 (EC)                
        & I could describe myself as a person with a rather tender heart & From 1 Never true to 5 Always true \\ \hline
    \multicolumn{3}{|l|}{\textbf{Personal Distress (PD)}}\\
    \hline
    IRI item 22 (PD)               
        & In emergency situations, I feel apprehensive and uncomfortable & From 1 Never true to 5 Always true \\ \hline
    IRI item 23 (PD) 
        & Sometimes I feel helpless when I find myself in very emotionally involving situations & From 1 Never true to 5 Always true \\ \hline
    IRI item 24 (PD)           
        & When I see someone getting hurt, I tend to stay calm (Inverted) & From 1 Never true to 5 Always true \\ \hline
    IRI item 25 (PD)    
        & Finding myself in situations that cause emotional tension scares me & From 1 Never true to 5 Always true \\ \hline
    IRI item 26 (PD)                
        & I am usually quite efficient in dealing with emergency situations (Inverted) & From 1 Never true to 5 Always true \\ \hline
    IRI item 27 (PD)                
        & I tend to lose control in emergency cases & From 1 Never true to 5 Always true \\ \hline
    IRI item 28 (PD)                
        & When I see someone who urgently needs help in an emergency situation, I break down & From 1 Never true to 5 Always true \\ \hline

\end{longtable}

\vspace{-0.8cm}
{\footnotesize \textit{Note:} Items labeled as inverted in the questionnaire were reverse-coded prior to the analysis.}
\end{center}

Here we provide a brief description of each subscale.

\paragraph{Perspective Taking (PT)}

The \textit{Perspective Taking (PT)} subscale measures the cognitive component of empathy, defined as the tendency to adopt the psychological point of view of others. Items in this scale capture respondents' propensity to consider others' perspectives in everyday situations, reflecting a cognitive rather than primarily affective component of empathy. PT is other-oriented and conceptually distinct from affective forms of empathy. The PT scale is computed as the mean of the seven items listed in Table~\ref{tab:final_varsIRI}.

\paragraph{Empathic Concern (EC)}

The \textit{Empathic Concern (EC)} subscale captures affective, other-oriented responses to others’ distress, including feelings of sympathy, compassion, and concern. This dimension reflects emotional engagement with others’ well-being rather than cognitive perspective taking. EC is commonly associated with prosocial motivation and prosocial behavior driven by concern for others. The EC scale is constructed as the mean of the seven items reported in Table~\ref{tab:final_varsIRI}.

\paragraph{Personal Distress (PD)}

The \textit{Personal Distress (PD)} subscale measures self-oriented emotional responses to others’ distress, such as feelings of anxiety, discomfort, or emotional unease in tense interpersonal situations. Although affective in nature, PD differs from Empathic Concern in that the emotional focus is on the self rather than on the other person. High levels of PD may therefore be associated with avoidance or withdrawal rather than prosocial behavior. The PD scale is computed as the mean of the seven items listed in Table~\ref{tab:final_varsIRI}.

\paragraph{Fantasy (FS)}

The \textit{Fantasy (FS)} subscale captures the tendency to imaginatively identify with fictional characters in books, films, or other narratives. This dimension reflects imaginative involvement and narrative engagement rather than direct interpersonal empathy in real-world social interactions. FS involves elements of both cognitive simulation and affective engagement but is less directly linked to observed interpersonal behavior. The FS scale is constructed as the mean of the seven items reported in Table~\ref{tab:final_varsIRI}.

\subsection*{Scale construction}

All items are measured on the same response scale. Reverse-coded items were recoded prior to scale construction. Higher values of each subscale indicate higher levels of the corresponding empathic trait.

\begin{figure}[H]
\centering
\caption{Distributions of Interpersonal Reactivity Index subscale scores}
\includegraphics[width=1.0\textwidth]{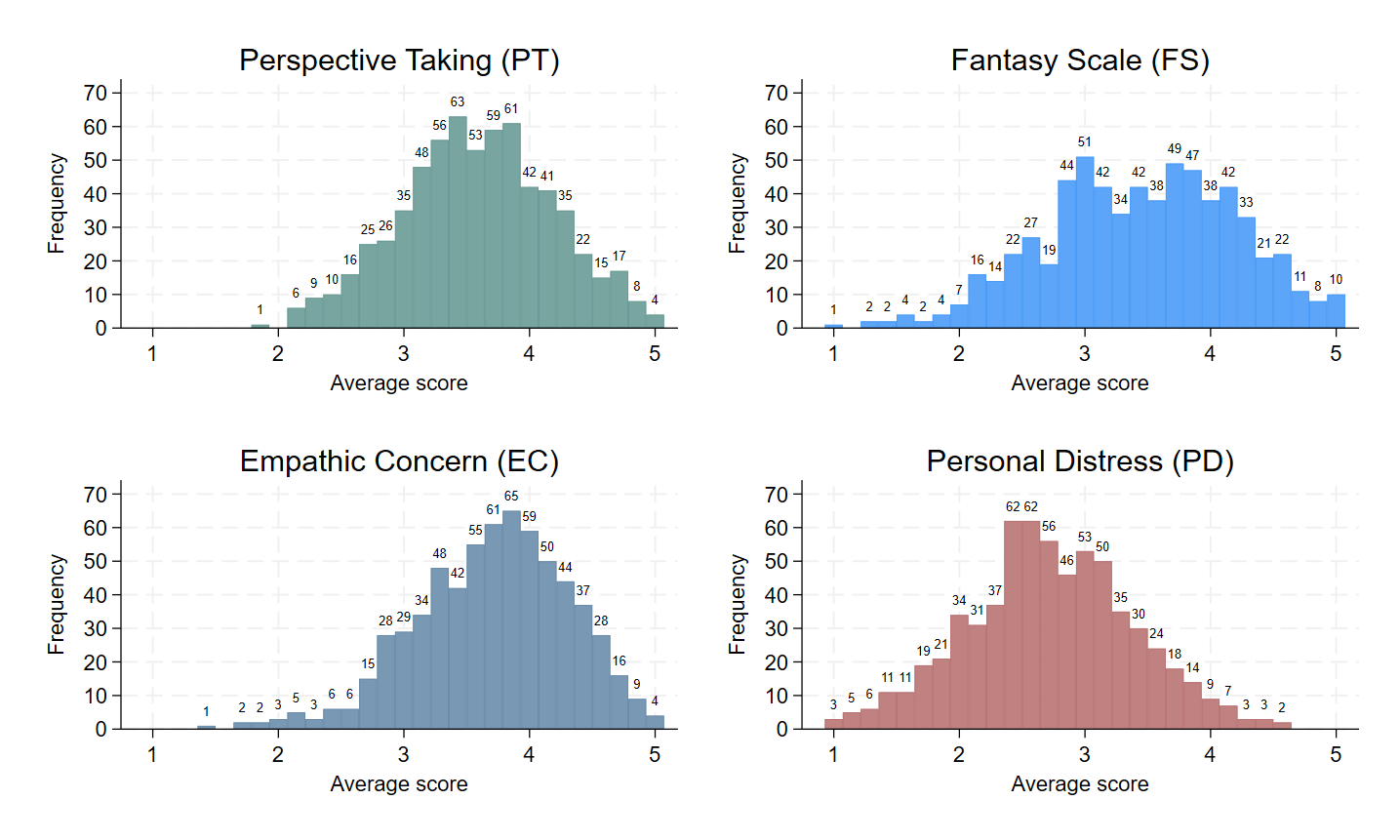}
\label{fig:iri_hist_comb}
\end{figure}

Item-level responses display substantial variation both within and across subscales, with mean values generally concentrated above the midpoint of the scale for PT, FS, and EC items, and lower average scores for PD items. Full descriptive statistics of the individual Interpersonal Reactivity Index items and indices are reported in Tables~\ref{descriptive_iri_indices} and \ref{descriptive_iri_items}, and in Figure~\ref{fig:iri_hist_comb}.

At the index level, respondents score highest on Perspective Taking (mean 3.58, SD 0.60) and Empathic Concern (mean 3.70, SD 0.61), followed by Fantasy (mean 3.44, SD 0.76). Personal Distress exhibits a lower average level (mean 2.72, SD 0.67), suggesting that emotional discomfort in response to others’ distress is less pronounced than cognitive or affective empathic orientations in this sample. The overall IRI mean is 3.36 (SD 0.43), indicating moderate-to-high empathic tendencies on average. Across all indices, the distributions show meaningful dispersion, pointing to substantial heterogeneity in empathic traits among participants.

\begin{center}    
    \begin{longtable}{|P{4.8cm}|C{1cm}|C{1cm}|C{1cm}|C{1.5cm}|C{1.5cm}|}
    \caption{Descriptive statistics of IRI indices (means)}\label{descriptive_iri_indices}\\
    \hline
        Variable  & count & mean & sd & min & max \\ \hline
        Perspective Taking (PT) mean & 652 & 3.582 & 0.603 & 1.857 & 5.000 \\ \hline
        Fantasy (FS) mean            & 652 & 3.444 & 0.755 & 1.000 & 5.000 \\ \hline
        Empathic Concern (EC) mean   & 652 & 3.704 & 0.612 & 1.429 & 5.000 \\ \hline
        Personal Distress (PD) mean  & 652 & 2.721 & 0.666 & 1.000 & 4.571 \\ \hline
        IRI total mean               & 652 & 3.363 & 0.428 & 1.964 & 4.393 \\ \hline
    \end{longtable}

    \vspace{-0.2cm}
    {\footnotesize \textit{Note:} Each index is computed as the mean of its corresponding IRI items after reverse-coding the inverted statements (scale: 1 = Never true; 5 = Always true).}
\end{center}

\begin{center}
\begin{longtable}{|P{4.8cm}|C{1cm}|C{1cm}|C{1cm}|C{0.5cm}|C{0.5cm}|}
\caption{Descriptive statistics of Interpersonal Reactivity Index (IRI) items}
\label{descriptive_iri_items}\\
\hline
Variable & count & mean & sd & min & max \\ 
\hline
\multicolumn{6}{|l|}{\textbf{Perspective Taking (PT)}} \\ \hline
IRI item 1 (PT) & 652 & 3.621 & 0.971 & 1 & 5 \\ \hline
IRI item 2 (PT) & 652 & 3.831 & 0.910 & 1 & 5 \\ \hline
IRI item 3 (PT) & 652 & 3.831 & 0.937 & 1 & 5 \\ \hline
IRI item 4 (PT) & 652 & 3.138 & 1.107 & 1 & 5 \\ \hline
IRI item 5 (PT) & 652 & 3.709 & 0.949 & 1 & 5 \\ \hline
IRI item 6 (PT) & 652 & 3.394 & 0.987 & 1 & 5 \\ \hline
IRI item 7 (PT) & 652 & 3.548 & 0.985 & 1 & 5 \\ \hline

\multicolumn{6}{|l|}{\textbf{Fantasy (FS)}} \\ \hline
IRI item 8 (FS)  & 652 & 3.635 & 1.102 & 1 & 5 \\ \hline
IRI item 9 (FS)  & 652 & 3.337 & 1.121 & 1 & 5 \\ \hline
IRI item 10 (FS) & 652 & 3.425 & 1.094 & 1 & 5 \\ \hline
IRI item 11 (FS) & 652 & 3.810 & 1.140 & 1 & 5 \\ \hline
IRI item 12 (FS) & 652 & 2.952 & 1.239 & 1 & 5 \\ \hline
IRI item 13 (FS) & 652 & 3.420 & 1.080 & 1 & 5 \\ \hline
IRI item 14 (FS) & 652 & 3.528 & 1.123 & 1 & 5 \\ \hline

\multicolumn{6}{|l|}{\textbf{Empathic Concern (EC)}} \\ \hline
IRI item 15 (EC) & 652 & 3.630 & 1.038 & 1 & 5 \\ \hline
IRI item 16 (EC) & 652 & 3.667 & 1.021 & 1 & 5 \\ \hline
IRI item 17 (EC) & 652 & 4.048 & 0.860 & 1 & 5 \\ \hline
IRI item 18 (EC) & 652 & 3.423 & 1.033 & 1 & 5 \\ \hline
IRI item 19 (EC) & 652 & 4.094 & 1.070 & 1 & 5 \\ \hline
IRI item 20 (EC) & 652 & 3.374 & 0.930 & 1 & 5 \\ \hline
IRI item 21 (EC) & 652 & 3.690 & 1.009 & 1 & 5 \\ \hline

\multicolumn{6}{|l|}{\textbf{Personal Distress (PD)}} \\ \hline
IRI item 22 (PD) & 652 & 3.020 & 1.143 & 1 & 5 \\ \hline
IRI item 23 (PD) & 652 & 3.331 & 1.179 & 1 & 5 \\ \hline
IRI item 24 (PD) & 652 & 2.813 & 1.064 & 1 & 5 \\ \hline
IRI item 25 (PD) & 652 & 3.210 & 1.192 & 1 & 5 \\ \hline
IRI item 26 (PD) & 652 & 2.475 & 0.894 & 1 & 5 \\ \hline
IRI item 27 (PD) & 652 & 2.425 & 1.054 & 1 & 5 \\ \hline
IRI item 28 (PD) & 652 & 1.775 & 0.814 & 1 & 5 \\ \hline
\end{longtable}

\vspace{-0.8cm}
{\footnotesize \textit{Note:} All items are measured on a 5-point Likert scale (1 = Never true; 5 = Always true). Items labeled as inverted in the questionnaire were reverse-coded prior to the analysis.}
\end{center}

\subsection{Civic capital and corruption survey}

To account for heterogeneity in civic engagement, prosocial behavior, institutional trust, and norms related to legality and corruption, we draw on a broader set of behavioral and attitudinal measures available in the dataset. These variables were collected as part of the wider research project and were not elicited specifically for the distributive-choice task analyzed here. We report them descriptively for transparency and completeness, while in the main analysis we include only a parsimonious subset of controls that is most relevant for our hypotheses.

A large majority of respondents report having voted in the most recent election (81.6\%), while 10.1\% declare that they did not vote and 8.3\% indicate that they were not eligible at the time. Participation in voluntary or community-based activities is widespread, with 74.7\% reporting membership or involvement in associations. Prosocial behaviors related to donation display more heterogeneous patterns: 16.3\% of respondents report being blood donors, 62.1\% report not donating blood, and 21.6\% indicate that they are not eligible; by contrast, 51.2\% report being registered organ donors, 33.0\% report not being registered, and 15.8\% declare uncertainty about their status.

Attitudinal measures suggest moderate levels of generalized trust, with an average score of 2.99 on a six-point scale. Norms related to legality and corruption are characterized by relatively low average tolerance: the mean scores for the justifiability of welfare fraud (1.81), fee evasion (2.55), tax evasion (1.79), and bribe acceptance (2.13) all lie in the lower half of the respective scales. Finally, perceived corruption diffusion is moderate, with a mean value of 2.76 on a four-point scale. Higher values in these indices indicate greater acceptance, tolerance, or perceived prevalence, in line with the original coding.

Complete descriptive statistics for all items included in the civic capital and corruption survey are reported in Tables~\ref{descriptive_controls}--\ref{descriptive_organ}.

\begin{center}    
    \begin{longtable}{|P{5.0cm}|C{1cm}|C{1cm}|C{1cm}|C{0.5cm}|C{0.5cm}|}
    \caption{Descriptive statistics for behavioral and attitudinal control variables}\label{descriptive_controls}\\
    \hline
        Variable  & count & mean & sd & min & max \\ \hline \hline
        Voting behavior & 652 & -- & -- & 1 & 3 \\ \hline
        Association membership & 652 & -- & -- & 0 & 1 \\ \hline
        Blood donation & 652 & -- & -- & 1 & 3 \\ \hline
        Organ donation & 652 & -- & -- & 1 & 3 \\ \hline
        Trust in people & 652 & 2.986 & 1.139 & 1 & 6 \\ \hline
        Tolerance of welfare fraud & 652 & 1.805 & 1.053 & 1 & 6 \\ \hline
        Tolerance of fee evasion & 652 & 2.551 & 1.309 & 1 & 6 \\ \hline
        Tolerance of tax evasion & 652 & 1.790 & 1.326 & 1 & 6 \\ \hline
        Acceptance of bribes & 652 & 2.132 & 1.367 & 1 & 6 \\ \hline
        Perceived corruption diffusion & 652 & 2.761 & 0.577 & 1 & 4 \\ \hline
    \end{longtable}

    \vspace{-0.8cm}
    {\footnotesize \textit{Note:} Higher values indicate higher intensity, acceptance, trust, or perceived prevalence depending on the variable.}
\end{center}

\begin{longtable}{lccc}
\caption{Descriptive statistics for voting behavior}\label{descriptive_vote}\\
\hline
Voting behavior & Freq. & Percent & Cum. \\
\hline
Voted & 532 & 81.60 & 81.60 \\
Did not vote & 66 & 10.12 & 91.72 \\
Not eligible & 54 & 8.28 & 100.00 \\
\hline
Total & 652 & 100.00 & 100.00 \\
\hline
\end{longtable}

\begin{longtable}{lccc}
\caption{Descriptive statistics for association membership}\label{descriptive_associations}\\
\hline
Association membership & Freq. & Percent & Cum. \\
\hline
Not a member & 165 & 25.31 & 25.31 \\
Member & 487 & 74.69 & 100.00 \\
\hline
Total & 652 & 100.00 & 100.00 \\
\hline
\end{longtable}

\begin{longtable}{lccc}
\caption{Descriptive statistics for blood donation}\label{descriptive_blood}\\
\hline
Blood donation & Freq. & Percent & Cum. \\
\hline
Donor & 106 & 16.26 & 16.26 \\
Non-donor & 405 & 62.12 & 78.37 \\
Not eligible & 141 & 21.63 & 100.00 \\
\hline
Total & 652 & 100.00 & 100.00 \\
\hline
\end{longtable}

\begin{longtable}{lccc}
\caption{Descriptive statistics for organ donation}\label{descriptive_organ}\\
\hline
Organ donation & Freq. & Percent & Cum. \\
\hline
Donor & 334 & 51.23 & 51.23 \\
Non-donor & 215 & 32.98 & 84.20 \\
Do not know & 103 & 15.80 & 100.00 \\
\hline
Total & 652 & 100.00 & 100.00 \\
\hline
\end{longtable}

\subsection*{S1.7 Correlation matrix of the main variables}
\addcontentsline{toc}{subsection}{S1.7 Correlation matrix of the main variables}

Table~\ref{tab:main_corr} reports the pairwise Pearson correlations among the main variables used in the analysis. Overall, the bivariate associations are modest in magnitude. The Distribution Game outcome is only weakly correlated with the Cognitive Reflection Test sum score, response time, Perspective Taking, and Empathic Concern. Perspective Taking and Empathic Concern are positively correlated, while the Cognitive Reflection Test sum score is negatively correlated with Empathic Concern and only weakly related to Perspective Taking and response time. Correlations involving response time are based on a smaller number of observations because valid timing information is not available for all participants.

\begin{table}[H]
\centering
\caption{Pairwise Pearson correlations among the main variables}
\label{tab:main_corr}
\begin{tabular}{lccccc}
\toprule
 & DG & CRT & RT & PT & EC \\
\midrule
DG  & 1.000      &           &           &           &           \\
CRT & -0.029     & 1.000     &           &           &           \\
RT  & -0.034     & 0.108*   & 1.000     &           &           \\
PT  & 0.066     & -0.038    & 0.030     & 1.000     &           \\
EC  & 0.054      & -0.159*** & -0.108*  & 0.309***  & 1.000     \\
\bottomrule
\end{tabular}

\begin{minipage}{0.88\textwidth}
\footnotesize
\textit{Notes:} The table reports pairwise Pearson correlation coefficients computed using pairwise deletion; therefore, the number of observations varies across variable pairs. In particular, correlations involving RT are based on the smaller subsample of participants with non-missing valid timing information. DG = Distribution Game choice; CRT = Cognitive Reflection Test sum score; RT = response time in seconds in the Distribution Game; PT = Perspective Taking mean score; EC = Empathic Concern mean score. Higher values of DG indicate more prosocial choices. *** $p<.001$, ** $p<.01$, * $p<.05$.
\end{minipage}
\end{table}

\section{Measures and Coding}

This section describes how the main variables used in the empirical analysis are constructed and coded. We focus on the operationalization of the dependent variable, the measures of cognitive reflection and response time, the empathy measures, and the control variables included in the baseline specifications. Rather than providing descriptive evidence, which is reported in Section S1, this section documents the coding rules adopted in the regression analysis.

\subsection*{S2.1 Distribution Game (DG)}
\addcontentsline{toc}{subsection}{S2.1 Distribution Game (DG)}

The dependent variable is derived from the choice made in the four-player Distribution Game (DG). Participants choose one of three alternative allocations of a fixed total endowment among four group members, including themselves.

For the analysis, DG choices are coded as an ordinal variable taking values from 1 to 3, where higher values denote more prosocial allocations: Choice A = 1, Choice B = 2, and Choice C = 3. The coding reflects the monotonic ordering of the three allocations in terms of prosociality. Relative to Choice A, moving to Choices B and C implies a lower own payoff for the decision-maker, a higher average payoff for the other group members, and lower payoff inequality. Choice C is therefore the most prosocial allocation, Choice A the least prosocial, and Choice B an intermediate option.

\subsection*{S2.2 Cognitive Reflection Test (CRT)}
\addcontentsline{toc}{subsection}{S2.2 Cognitive Reflection Test (CRT)}

Cognitive reflection is measured using a five-item Cognitive Reflection Test (CRT). For each item, responses are coded as 1 when the participant provides the reflective/correct answer and 0 otherwise. The main CRT measure used in the analysis is the total number of correct answers across the five items. This yields a score ranging from 0 to 5, where higher values denote a greater propensity to override intuitive but incorrect responses in favor of reflective reasoning.

\subsection*{S2.3 Response Time (RT)}
\addcontentsline{toc}{subsection}{S2.3 Response Time}

Choice-stage response time is measured as the number of seconds participants spend on the final Distribution Game choice page, where they submit Choice A, B, or C. Participants had already viewed the allocation options and their associated payoffs and completed the comprehension questions before reaching this page. Response Time therefore measures latency at the choice stage rather than the full amount of time during which participants could consider the allocation problem. In the baseline empirical analysis, Response Time is analyzed on its linear scale, with regression coefficients expressed per 10-second increase. Descriptive statistics and graphical representations of response-time distributions remain expressed in seconds. A sensitivity analysis using response time winsorized at the 1st and 99th percentiles is reported in Section~S3.

\subsection*{S2.4 Perspective Taking (PT)}
\addcontentsline{toc}{subsection}{S2.4 Perspective Taking (PT)}

Perspective Taking (PT) is measured through the seven-item Perspective Taking subscale of the Interpersonal Reactivity Index (IRI). The PT score is constructed as the mean of the seven corresponding items, after reverse-coding negatively worded items where appropriate. Higher values indicate a stronger dispositional tendency to adopt the point of view of others.

Empathic Concern (EC), which is also examined in the main text and in Section S4 below, is constructed analogously as the mean of the seven items belonging to the corresponding IRI subscale, after reverse-coding negatively worded items where required.

\subsection*{S2.5 Control variables}
\addcontentsline{toc}{subsection}{S2.5 Control variables}

The baseline specifications include a parsimonious set of socio-demographic and survey-based controls. Among the socio-demographic variables, Age is included as a continuous regressor, while Gender is entered as a categorical variable. Among the survey-based controls, Trust in people is included as a continuous variable measured on a six-point scale, Association membership is coded as a binary indicator, and Voting behavior and Blood donation are entered as categorical variables.

More specifically, higher values of Trust in people indicate greater generalized trust. Association membership takes value 1 if the respondent reports having participated in an association or community project without deriving any profit from it, and 0 otherwise. Voting behavior distinguishes between respondents who voted in the last election, those who did not vote, and those who were not eligible to vote. Blood donation distinguishes between respondents who report being blood donors, non-donors, and those who declare that they are not eligible to donate blood.

This set of controls is intended to account, in a parsimonious way, for basic demographic heterogeneity and for individual differences in civic engagement and prosocial orientation that may be correlated with distributive choices.

\subsection*{S2.6 Centering in interaction models}
\addcontentsline{toc}{subsection}{S2.6 Centering in interaction models}

For the interaction models, we report specifications both with and without mean-centering of the continuous variables entering the interaction terms. Mean-centering is performed by subtracting from each observation the sample mean of the relevant variable, so that the transformed variable has mean equal to zero. Formally, for a generic variable $X$, the centered version is defined as $X_c = X - \bar{X}$.

This transformation does not change the overall fit of the model, the predicted values, or the coefficient of the interaction term itself. Its main purpose is interpretative. In models with interactions, the coefficient on a lower-order term represents the effect of that variable when the interacting variable is equal to zero. When zero is not substantively meaningful, this interpretation can be uninformative. After centering, the lower-order coefficients are instead interpreted at the sample mean of the interacting variable.

In addition, mean-centering reduces non-essential multicollinearity between the constituent terms and their interaction, which is useful when assessing the stability and interpretability of lower-order coefficients. For completeness, the appendix reports both the raw and the centered interaction models.

\section{Extended Results and Robustness Checks}

This section reports extended specifications and robustness checks for the main interaction results, including centering diagnostics, conditional slopes, common-sample and joint CRT--Response Time models, ordered logit estimates, response-time sensitivity analyses, sample-comparability checks, and detectable-effect calculations.

A broader analysis plan for these data was preregistered on OSF (10.17605/OSF.IO/K8C29). It considered overall empathy and the four IRI subscales and identified Perspective Taking (PT) and Empathic Concern (EC) as the dimensions expected to be most closely related to fair choices. The present paper focuses on PT because it is the dimension most directly linked to the theoretical mechanism developed here, while EC is examined as a conceptually distinct affective comparison. Fantasy and Personal Distress are reported descriptively in Section~S1 but are not pursued further, and the overall IRI score is not used in the interaction analyses because it combines conceptually distinct empathy dimensions. Choice-stage response time was not part of the preregistration and is analyzed here as a theory-driven extension.

\subsection{Extended interaction models for the main analysis}

Table~\ref{tab:s3_main_models_pt} reports the extended versions of the four interaction models presented in the main text. Columns~(1) and~(2) use CRT performance as a task-based measure of domain-general analytic reflection, while columns~(3) and~(4) use choice-stage Response Time in the Distribution Game as a behavioral proxy for decision-specific processing. Across the CRT specifications, neither the main effect of CRT nor the interaction between CRT and Perspective Taking is statistically significant. By contrast, in the Response Time specifications, the interaction between Response Time and Perspective Taking is negative and statistically significant both without and with controls. This confirms the core result of the paper: the association between choice-stage response time and prosociality varies with individual differences in Perspective Taking.

\begin{center}
\begin{longtable}{lcccc}
\caption{Extended uncentered interaction models with Perspective Taking}
\label{tab:s3_main_models_pt}\\
\toprule
& (1) & (2) & (3) & (4) \\
\midrule
\endfirsthead

\toprule
& (1) & (2) & (3) & (4) \\
\midrule
\endhead

\midrule
\multicolumn{5}{r}{\textit{Continued on next page}}\\
\endfoot

\bottomrule
\endlastfoot

Cognitive Reflection Test (CRT)
    & -0.055 & -0.072 &  &  \\
    & (0.162) & (0.166) &  &  \\

Perspective Taking (PT)
    & 0.065 & 0.053 & 0.216* & 0.226** \\
    & (0.129) & (0.131) & (0.085) & (0.087) \\

Response Time (DG)
    &  &  & 0.230 & 0.262* \\
    &  &  & (0.119) & (0.120) \\

Cognitive Reflection (CRT) $\times$ PT
    & 0.010 & 0.015 & & \\
    & (0.045) & (0.046) & & \\

Response Time (DG) $\times$ PT
    & & & -0.064* & -0.072** \\
    & & & (0.027) & (0.028) \\

Age
    &  & -0.010 &  & -0.010 \\
    &  & (0.009) &  & (0.009) \\

Gender: Male
    &  & -0.031 &  & -0.132 \\
    &  & (0.076) &  & (0.082) \\

Gender: Prefer not to answer
    &  & 0.536* &  & 0.351 \\
    &  & (0.253) &  & (0.411) \\

Gender: Not identified
    &  & -0.057 &  & -0.090 \\
    &  & (0.499) &  & (0.476) \\

Trust in people
    &  & 0.017 &  & 0.005 \\
    &  & (0.030) &  & (0.034) \\

Association membership
    &  & -0.041 &  & -0.073 \\
    &  & (0.080) &  & (0.090) \\

Voting behavior: Did not vote
    &  & 0.061 &  & 0.012 \\
    &  & (0.110) &  & (0.124) \\

Voting behavior: Not eligible
    &  & 0.017 &  & 0.116 \\
    &  & (0.129) &  & (0.144) \\

Blood donation: Non-donor
    &  & -0.206* &  & -0.238* \\
    &  & (0.096) &  & (0.104) \\

Blood donation: Not eligible
    &  & -0.136 &  & -0.228 \\
    &  & (0.112) &  & (0.122) \\

Constant
    & 1.890*** & 2.281*** & 1.275*** & 1.716*** \\
    & (0.466) & (0.532) & (0.330) & (0.396) \\

\midrule
Controls & No & Yes & No & Yes \\
Observations & 649 & 649 & 525 & 525 \\
R-squared & 0.005 & 0.021 & 0.012 & 0.034 \\
Log likelihood & -810.638 & -805.364 & -655.441 & -649.590 \\
AIC & 1629.276 & 1638.728 & 1318.881 & 1327.180 \\
BIC & 1647.178 & 1701.384 & 1335.935 & 1386.868 \\

\midrule

\multicolumn{5}{p{0.95\textwidth}}{\footnotesize
\textit{Notes:} The dependent variable is the Distribution Game outcome, coded so that higher values indicate more prosocial choices. Robust standard errors are reported in parentheses. The variables entering the interaction terms are uncentered. Response Time is expressed in units of 10 seconds, while CRT and Perspective Taking retain their original scales. Perspective Taking (PT) is measured as the mean score of the corresponding Interpersonal Reactivity Index subscale. Gender, Voting behavior, and Blood donation are entered as categorical variables; the omitted categories are Female, Voted, and Donor, respectively. Association membership is coded as a binary indicator. Columns (1)--(2) use CRT performance, whereas columns (3)--(4) use choice-stage Response Time. *** $p<.001$, ** $p<.01$, * $p<.05$.
}\\

\end{longtable}
\end{center}

\subsection{Separate associations of Perspective Taking, Cognitive Reflection, and Response Time}

As a complementary specification, Table~\ref{tab:separate_associations} reports the association of Perspective Taking, Cognitive Reflection, and Response Time with Distribution Game choices when each focal variable is entered separately, both without and with the control variables used in the main analysis.

\begin{table}[htbp]
\centering
\caption{Separate associations of Perspective Taking, Cognitive Reflection, and Response Time with Distribution Game choices}
\label{tab:separate_associations}
\begin{threeparttable}
\begin{tabular}{lcccccc}
\toprule
& (1) & (2) & (3) & (4) & (5) & (6) \\
\midrule

Perspective Taking (PT)
& 0.093
& 0.091
&
&
&
\\
& (0.056)
& (0.057)
&
&
&
\\

Cognitive Reflection (CRT)
&
&
& -0.020
& -0.020
&
\\
&
&
& (0.027)
& (0.029)
&
\\

Response Time (DG)
&
&
&
&
& -0.017
& -0.016
\\
&
&
&
&
& (0.022)
& (0.022)
\\

Constant
& 1.742***
& 2.084***
& 2.127***
& 2.438***
& 2.080***
& 2.526***
\\
& (0.202)
& (0.297)
& (0.079)
& (0.248)
& (0.056)
& (0.250)
\\

\midrule

Controls
& No
& Yes
& No
& Yes
& No
& Yes
\\

Observations
& 649
& 649
& 649
& 649
& 525
& 525
\\

$R^2$
& 0.004
& 0.020
& 0.001
& 0.017
& 0.001
& 0.022
\\

\bottomrule
\end{tabular}

\begin{tablenotes}[flushleft]
\footnotesize
\item Notes: The dependent variable is the ordered prosociality score in the Distribution Game, ranging from 1 to 3. Each pair of columns reports the association between the outcome and one focal variable separately. Columns (1), (3), and (5) include no additional controls. Columns (2), (4), and (6) include age, gender, generalized trust, association membership, voting status, and blood-donation status, as in the main specifications. Response Time coefficients refer to a 10-second increase. Robust standard errors are reported in parentheses. Coefficients are rounded to three decimal places. *** $p < .001$, ** $p < .01$, * $p < .05$.
\end{tablenotes}

\end{threeparttable}
\end{table}

\subsection{Multicollinearity and centered Response Time $\times$ Perspective Taking models}

Because the interaction between Response Time and Perspective Taking involves two continuous variables, the non-centered specifications may generate substantial collinearity between the main effects and their product term. Table~\ref{tab:s3_vif_pt_noncentered} reports the variance inflation factors from the non-centered Response Time models. The VIFs are high for both Response Time and the interaction term, indicating strong mechanical collinearity.

To address the substantial non-essential multicollinearity in the uncentered interaction models, we re-estimate the same specifications after mean-centering Response Time and Perspective Taking. Mean-centering does not alter the interaction coefficient, fitted values, or overall model fit, but it substantially reduces the mechanical collinearity between the constituent terms and their interaction and makes the lower-order coefficients interpretable at the sample mean of the interacting variable. As shown in Tables~\ref{tab:s3_centered_models_pt} and~\ref{tab:s3_vif_pt_centered}, the interaction coefficient is unchanged, while the VIF values fall sharply and become close to one. This indicates that the high VIFs observed in the uncentered specifications reflect non-essential collinearity induced by the interaction parameterization rather than a substantive instability of the model.

\begin{center}
\begin{longtable}{@{}L{6.0cm}*{4}{C{1.75cm}}@{}}
\caption{Variance inflation factors for the uncentered interaction models with Perspective Taking}
\label{tab:s3_vif_pt_noncentered}\\
\toprule
Variable & (1) & (2) & (3) & (4) \\
\midrule
\endfirsthead

\toprule
Variable & (1) & (2) & (3) & (4) \\
\midrule
\endhead

\midrule
\multicolumn{5}{r}{\textit{Continued on next page}}\\
\endfoot

\bottomrule
\endlastfoot

Cognitive Reflection Test (CRT)
    & 32.81 & 34.22 &  &  \\

Perspective Taking (PT)
    & 5.04 & 5.20 & 2.18 & 2.25 \\

Response Time (DG)
    &  &  & 39.24 & 40.40 \\

Cognitive Reflection (CRT) $\times$ PT
    & 35.96 & 37.40 & & \\

Response Time (DG) $\times$ PT
    & & & 40.82 & 42.04 \\

Age
    &  & 1.07 &  & 1.06 \\

Gender: Male
    &  & 1.17 &  & 1.10 \\

Gender: Prefer not to answer
    &  & 1.02 &  & 1.02 \\

Gender: Not identified
    &  & 1.02 &  & 1.02 \\

Trust in people
    &  & 1.03 &  & 1.04 \\

Association membership
    &  & 1.04 &  & 1.05 \\

Voting behavior: Did not vote
    &  & 1.04 &  & 1.06 \\

Voting behavior: Not eligible
    &  & 1.05 &  & 1.06 \\

Blood donation: Non-donor
    &  & 1.85 &  & 1.79 \\

Blood donation: Not eligible
    &  & 1.94 &  & 1.82 \\

\midrule
Controls & No & Yes & No & Yes \\
Mean VIF & 24.61 & 6.85 & 27.41 & 7.44 \\

\midrule

\multicolumn{5}{p{0.95\textwidth}}{\footnotesize
\textit{Notes:} The table reports variance inflation factors for the uncentered specifications reported in Table~\ref{tab:s3_main_models_pt}. Columns~(1)--(2) refer to the CRT $\times$ Perspective Taking models, while columns~(3)--(4) refer to the Response Time $\times$ Perspective Taking models. The mean VIF is computed across all regressors included in each specification.
}\\

\end{longtable}
\end{center}

\begin{center}
\begin{longtable}{@{}L{6.0cm}*{4}{C{1.75cm}}@{}}
\caption{Extended centered interaction models with Perspective Taking}
\label{tab:s3_centered_models_pt}\\
\toprule
& (1) & (2) & (3) & (4) \\
\midrule
\endfirsthead

\toprule
& (1) & (2) & (3) & (4) \\
\midrule
\endhead

\midrule
\multicolumn{5}{r}{\textit{Continued on next page}}\\
\endfoot

\bottomrule
\endlastfoot

Cognitive Reflection Test (CRT)
    & -0.018 & -0.019 &  &  \\
    & (0.028) & (0.029) &  &  \\

Perspective Taking (PT)
    & 0.092 & 0.092 & 0.097 & 0.093 \\
    & (0.056) & (0.057) & (0.061) & (0.063) \\

Response Time (DG)
    &  &  & 0.000 & 0.004 \\
    &  &  & (0.026) & (0.026) \\

Cognitive Reflection (CRT) $\times$ PT
    & 0.010 & 0.015 & & \\
    & (0.045) & (0.046) & & \\

Response Time (DG) $\times$ PT
    & & & -0.064* & -0.072** \\
    & & & (0.027) & (0.028) \\

Age
    &  & -0.010 &  & -0.010 \\
    &  & (0.009) &  & (0.009) \\

Gender: Male
    &  & -0.031 &  & -0.132 \\
    &  & (0.076) &  & (0.082) \\

Gender: Prefer not to answer
    &  & 0.536* &  & 0.351 \\
    &  & (0.253) &  & (0.411) \\

Gender: Not identified
    &  & -0.057 &  & -0.090 \\
    &  & (0.499) &  & (0.476) \\

Trust in people
    &  & 0.017 &  & 0.005 \\
    &  & (0.030) &  & (0.034) \\

Association membership
    &  & -0.041 &  & -0.073 \\
    &  & (0.080) &  & (0.090) \\

Voting behavior: Did not vote
    &  & 0.061 &  & 0.012 \\
    &  & (0.110) &  & (0.124) \\

Voting behavior: Not eligible
    &  & 0.017 &  & 0.116 \\
    &  & (0.129) &  & (0.144) \\

Blood donation: Non-donor
    &  & -0.206* &  & -0.238* \\
    &  & (0.096) &  & (0.104) \\

Blood donation: Not eligible
    &  & -0.136 &  & -0.228 \\
    &  & (0.112) &  & (0.122) \\

Constant
    & 2.074*** & 2.422*** & 2.049*** & 2.533*** \\
    & (0.033) & (0.236) & (0.037) & (0.249) \\

\midrule

Controls & No & Yes & No & Yes \\
Observations & 649 & 649 & 525 & 525 \\
R-squared & 0.005 & 0.021 & 0.012 & 0.034 \\
Log likelihood & -810.638 & -805.364 & -655.441 & -649.590 \\
AIC & 1629.276 & 1638.728 & 1318.881 & 1327.180 \\
BIC & 1647.178 & 1701.384 & 1335.935 & 1386.868 \\

\midrule

\multicolumn{5}{p{0.95\textwidth}}{\footnotesize
\textit{Notes:} The dependent variable is the Distribution Game outcome, coded so that higher values indicate more prosocial choices. Robust standard errors are reported in parentheses. The continuous variables entering the interaction terms are mean-centered within the relevant estimation sample. Perspective Taking (PT) is measured as the mean score of the corresponding Interpersonal Reactivity Index subscale. Gender, Voting behavior, and Blood donation are entered as categorical variables; the omitted categories are Female, Voted, and Donor, respectively. Association membership is coded as a binary indicator. Columns (1)--(2) use CRT performance, whereas columns (3)--(4) use choice-stage Response Time. *** $p<.001$, ** $p<.01$, * $p<.05$.
}\\

\end{longtable}
\end{center}

\begin{center}
\begin{longtable}{@{}L{6.0cm}*{4}{C{1.75cm}}@{}}
\caption{Variance inflation factors for the centered interaction models with Perspective Taking}
\label{tab:s3_vif_pt_centered}\\
\toprule
Variable & (1) & (2) & (3) & (4) \\
\midrule
\endfirsthead

\toprule
Variable & (1) & (2) & (3) & (4) \\
\midrule
\endhead

\midrule
\multicolumn{5}{r}{\textit{Continued on next page}}\\
\endfoot

\bottomrule
\endlastfoot

Cognitive Reflection Test (CRT)
    & 1.01 & 1.12 &  &  \\

Perspective Taking (PT)
    & 1.00 & 1.04 & 1.00 & 1.04 \\

Response Time (DG)
    &  &  & 1.22 & 1.24 \\

Cognitive Reflection (CRT) $\times$ PT
    & 1.01 & 1.05 & & \\

Response Time (DG) $\times$ PT
    & & & 1.21 & 1.25 \\

Age
    &  & 1.07 &  & 1.06 \\

Gender: Male
    &  & 1.17 &  & 1.10 \\

Gender: Prefer not to answer
    &  & 1.02 &  & 1.02 \\

Gender: Not identified
    &  & 1.02 &  & 1.02 \\

Trust in people
    &  & 1.03 &  & 1.04 \\

Association membership
    &  & 1.04 &  & 1.05 \\

Voting behavior: Did not vote
    &  & 1.04 &  & 1.06 \\

Voting behavior: Not eligible
    &  & 1.05 &  & 1.06 \\

Blood donation: Non-donor
    &  & 1.85 &  & 1.79 \\

Blood donation: Not eligible
    &  & 1.94 &  & 1.82 \\

\midrule
Controls & No & Yes & No & Yes \\
Mean VIF & 1.01 & 1.19 & 1.14 & 1.20 \\

\midrule

\multicolumn{5}{p{0.95\textwidth}}{\footnotesize
\textit{Notes:} The table reports variance inflation factors for the centered specifications reported in Table~\ref{tab:s3_centered_models_pt}. Columns~(1)--(2) refer to the CRT $\times$ Perspective Taking models, while columns~(3)--(4) refer to the Response Time $\times$ Perspective Taking models. The mean VIF is computed across all regressors included in each specification.
}\\

\end{longtable}
\end{center}

\subsection{Marginal effects of the Response Time $\times$ Perspective Taking interaction}

To clarify the substantive meaning of the negative interaction between Response Time and Perspective Taking, we compute the marginal effect of Response Time at different levels of Perspective Taking using the centered specifications. Table~\ref{tab:s3_margins_pt} reports the conditional marginal effects of Response Time when Perspective Taking is evaluated at one standard deviation below the mean, at the mean, and at one standard deviation above the mean.

The estimates show that the marginal effect of Response Time becomes more negative as Perspective Taking increases. In the baseline centered specification, the effect of Response Time is close to zero at the sample mean and becomes negative at higher levels of Perspective Taking. The same pattern emerges in the specification with controls. This is consistent with the interpretation of the interaction coefficient reported above: longer choice-stage response times are associated with fewer prosocial choices among participants with stronger perspective-taking dispositions.

Figure 1 in the main text illustrates this relationship using the controlled specification, with marginal effects expressed per 10 seconds of response time and Perspective Taking measured in standard deviation units.

\begin{center}
\begin{longtable}{lcc}
\caption{Marginal effect of Response Time at different levels of Perspective Taking}
\label{tab:s3_margins_pt}\\
\toprule
& Without controls & With controls \\
\midrule
\endfirsthead

\toprule
& Without controls & With controls \\
\midrule
\endhead

\midrule
\multicolumn{3}{r}{\textit{Continued on next page}}\\
\endfoot

\bottomrule
\endlastfoot

Response Time effect at PT = $-1$ SD
    & 0.039 & 0.047 \\
    & (0.040) & (0.040) \\

Response Time effect at PT = mean
    & 0.000 & 0.004 \\
    & (0.026) & (0.026) \\

Response Time effect at PT = $+1$ SD
    & -0.038* & -0.040* \\
    & (0.018) & (0.018) \\

\midrule

\multicolumn{3}{p{1\textwidth}}{\footnotesize
\textit{Notes:} The table reports conditional marginal effects of centered Response Time, expressed per 10-second increase, in the centered Response Time $\times$ Perspective Taking models. Perspective Taking is evaluated at one standard deviation below the mean, at the mean, and at one standard deviation above the mean. The standard deviation of Perspective Taking in the Response Time estimation sample is 0.603. Standard errors, reported in parentheses, are obtained by the delta method. Values are rounded to three decimal places. *** $p<.001$, ** $p<.01$, * $p<.05$.
}\\

\end{longtable}
\end{center}

\subsection{Difference-in-slopes tests for the centered Response Time models}

A statistically significant interaction coefficient implies that the effect of Response Time varies with Perspective Taking, but it is also useful to test directly whether the slope of Response Time differs between participants with relatively low and relatively high Perspective Taking. We therefore estimate the marginal effect of Response Time at one standard deviation below and above the mean of Perspective Taking and test whether these two slopes differ significantly.

Table~\ref{tab:s3_slopes_pt} reports the results. In both the baseline and the controlled specification, the difference between the low-PT and high-PT slopes is statistically significant. This confirms that the moderating role of Perspective Taking is not merely a feature of coefficient interpretation, but corresponds to a statistically meaningful difference in the conditional association between response time and prosocial behavior.

\begin{center}
\begin{longtable}{@{}L{7.5cm}*{2}{C{3.0cm}}@{}}
\caption{Difference-in-slopes tests for the centered Response Time $\times$ Perspective Taking models}
\label{tab:s3_slopes_pt}\\
\toprule
& Without controls & With controls \\
\midrule
\endfirsthead

\toprule
& Without controls & With controls \\
\midrule
\endhead

\midrule
\multicolumn{3}{r}{\textit{Continued on next page}}\\
\endfoot

\bottomrule
\endlastfoot

Slope of Response Time at low PT ($-1$ SD)
    & 0.039 & 0.047 \\
    & (0.040) & (0.040) \\

Slope of Response Time at high PT ($+1$ SD)
    & -0.038* & -0.040* \\
    & (0.018) & (0.018) \\

Difference: high PT slope -- low PT slope
    & -0.077* & -0.087** \\
    & (0.033) & (0.033) \\

\midrule

\multicolumn{3}{p{0.95\textwidth}}{\footnotesize
\textit{Notes:} The reported slopes are conditional marginal effects of centered Response Time, expressed per 10-second increase, evaluated at one standard deviation below and above the mean of centered Perspective Taking. Standard errors are obtained by the delta method. The last row reports the linear combination comparing the slope at high Perspective Taking with the slope at low Perspective Taking. *** $p<.001$, ** $p<.01$, * $p<.05$.
}\\

\end{longtable}
\end{center}

\subsection{Centered CRT $\times$ Perspective Taking models estimated on the Response Time subsample}

Because Response Time is available only for a reduced subsample, one possible concern is that the contrast between the CRT-based and Response Time-based interaction models may partly reflect differences in sample composition. To address this issue, we re-estimate the CRT $\times$ Perspective Taking models on the same subsample of observations for which Response Time is observed.

Table~\ref{tab:s3_crt_same_sample_pt} shows that the interaction between CRT and Perspective Taking remains statistically insignificant both without and with controls. The null result for CRT therefore does not depend on the larger sample used in the main CRT specifications, but also holds when the CRT models are estimated on the same subsample used in the Response Time analysis.

\begin{center}
\begin{longtable}{lcc}
\caption{Centered CRT $\times$ Perspective Taking models estimated on the Response Time subsample}
\label{tab:s3_crt_same_sample_pt}\\
\toprule
& (1) & (2) \\
\midrule
\endfirsthead

\toprule
& (1) & (2) \\
\midrule
\endhead

\midrule
\multicolumn{3}{r}{\textit{Continued on next page}}\\
\endfoot

\bottomrule
\endlastfoot

Cognitive Reflection Test (CRT)
    & -0.049 & -0.046 \\
    & (0.031) & (0.032) \\

Perspective Taking (PT)
    & 0.090 & 0.090 \\
    & (0.062) & (0.064) \\

CRT $\times$ Perspective Taking
    & -0.010 & 0.006 \\
    & (0.052) & (0.053) \\

Age
    &  & -0.011 \\
    &  & (0.009) \\

Gender: Male
    &  & -0.087 \\
    &  & (0.085) \\

Gender: Prefer not to answer
    &  & 0.369 \\
    &  & (0.411) \\

Gender: Not identified
    &  & -0.070 \\
    &  & (0.504) \\

Trust in people
    &  & 0.008 \\
    &  & (0.034) \\

Association membership
    &  & -0.061 \\
    &  & (0.090) \\

Voting behavior: Did not vote
    &  & 0.010 \\
    &  & (0.127) \\

Voting behavior: Not eligible
    &  & 0.079 \\
    &  & (0.145) \\

Blood donation: Non-donor
    &  & -0.249* \\
    &  & (0.104) \\

Blood donation: Not eligible
    &  & -0.211 \\
    &  & (0.123) \\

Constant
    & 2.047*** & 2.544*** \\
    & (0.037) & (0.248) \\

\midrule
Controls & No & Yes \\
Observations & 525 & 525 \\
R-squared & 0.010 & 0.029 \\
Log likelihood & -656.055 & -650.805 \\
AIC & 1320.111 & 1329.611 \\
BIC & 1337.164 & 1389.298 \\

\midrule

\multicolumn{3}{p{0.88\textwidth}}{\footnotesize
\textit{Notes:} The dependent variable is the Distribution Game outcome, coded so that higher values indicate more prosocial choices. Robust standard errors are reported in parentheses. The sample is restricted to the observations used in the Response Time specifications. CRT and Perspective Taking are mean-centered within this subsample before constructing the interaction term. *** $p<.001$, ** $p<.01$, * $p<.05$.
}\\

\end{longtable}
\end{center}

\subsection{Joint CRT and Response Time interaction models}

To compare the two cognitive-processing indicators directly, we estimate joint specifications on the common Response Time sample (\(N=525\)). These models include CRT performance, choice-stage Response Time, Perspective Taking, and both the CRT $\times$ PT and Response Time $\times$ PT interaction terms. All continuous variables entering the interactions are mean-centered within the common estimation sample.

Table~\ref{tab:s3_joint_crt_rt} reports the coefficients of interest. The Response Time $\times$ PT interaction remains negative and statistically significant both without and with controls. By contrast, the CRT $\times$ PT interaction remains close to zero and statistically insignificant. The same qualitative contrast therefore emerges when both indicators and both interaction terms are estimated simultaneously.

\begin{center}
\begin{longtable}{lcc}
\caption{Joint CRT and Response Time interaction models}
\label{tab:s3_joint_crt_rt}\\
\toprule
& Without controls & With controls \\
\midrule
\endfirsthead

\toprule
& Without controls & With controls \\
\midrule
\endhead

\midrule
\multicolumn{3}{r}{\textit{Continued on next page}}\\
\endfoot

\bottomrule
\endlastfoot

CRT $\times$ Perspective Taking
    & 0.003 & 0.020 \\
    & [$p=.955$] & [$p=.705$] \\

Response Time $\times$ Perspective Taking
    & -0.066* & -0.075** \\
    & [$p=.018$] & [$p=.008$] \\

\midrule
Controls & No & Yes \\
Observations & 525 & 525 \\
R-squared & 0.017 & 0.038 \\
Joint test of both interactions: $p$-value & 0.057 & 0.031 \\

\midrule

\multicolumn{3}{p{0.92\textwidth}}{\footnotesize
\textit{Notes:} The dependent variable is the Distribution Game outcome, coded so that higher values indicate more prosocial choices. The table reports the two interaction coefficients from joint OLS specifications that also include the centered lower-order terms for CRT, Response Time, and Perspective Taking. Response Time is expressed in units of 10 seconds. Column~(2) additionally includes the full set of demographic and survey-based controls used in the main analysis. Robust-inference $p$-values for the interaction coefficients are reported in square brackets. The last row reports the joint test that both interaction coefficients equal zero. *** $p<.001$, ** $p<.01$, * $p<.05$.
}\\

\end{longtable}
\end{center}

\subsection{Ordered logit robustness checks}
\label{sec:s3_ordered_logit}

As a robustness check, we re-estimate the main specifications using ordered logit models, which account for the ordinal nature of the Distribution Game outcome. Table~\ref{tab:s3_ordered_logit_pt} shows the same qualitative interaction pattern as the linear specifications: the CRT $\times$ Perspective Taking interaction remains statistically insignificant, whereas the Response Time $\times$ Perspective Taking interaction is negative and statistically significant. For the Response Time model, the average marginal effect of an additional 10 seconds on the probability of choosing the most prosocial allocation (Choice C) is positive but statistically insignificant at one standard deviation below mean Perspective Taking (0.0253, $p=.248$), approximately zero at the mean (0.0019, $p=.898$), and negative at one standard deviation above the mean ($-0.0231$, $p=.045$). The difference between the marginal effects at high and low Perspective Taking is statistically significant ($-0.0484$, $p=.019$).

\begin{center}
\begin{longtable}{@{}L{7.0cm}*{2}{C{3.0cm}}@{}}
\caption{Ordered logit robustness checks}
\label{tab:s3_ordered_logit_pt}\\
\toprule
& CRT model & Response Time model \\
\midrule
\endfirsthead

\toprule
& CRT model & Response Time model \\
\midrule
\endhead

\midrule
\multicolumn{3}{r}{\textit{Continued on next page}}\\
\endfoot

\bottomrule
\endlastfoot

Cognitive Reflection Test (CRT)
    & -0.040 &  \\
    & (0.064) &  \\

Response Time (DG)
    &  & 0.008 \\
    &  & (0.063) \\

Perspective Taking (PT)
    & 0.214 & 0.223 \\
    & (0.129) & (0.144) \\

Cognitive Reflection (CRT) $\times$ PT
    & 0.031 & \\
    & (0.107) & \\

Response Time (DG) $\times$ PT
    & & -0.175* \\
    & & (0.078) \\

\midrule
Controls & Yes & Yes \\
Observations & 649 & 525 \\
Pseudo R-squared & 0.011 & 0.017 \\

\midrule

\multicolumn{3}{p{0.95\textwidth}}{\footnotesize
\textit{Notes:} The dependent variable is the ordered Distribution Game outcome. Both specifications include the full set of controls used in the main analysis. Continuous variables entering the interactions are mean-centered. Response Time is expressed in units of 10 seconds. Robust standard errors are reported in parentheses. *** $p<.001$, ** $p<.01$, * $p<.05$.
}\\

\end{longtable}
\end{center}

\subsection{Response-time patterns across Distribution Game choices}
\label{sec:s3_rt_choices}

As discussed in the main text, choice-stage response time may reflect several features of the decision process, including decision conflict, difficulty, and preference strength, in addition to the amount of processing devoted to the choice. We therefore document how response times vary across the three Distribution Game alternatives and examine whether the main moderation result depends on the inclusion of the intermediate allocation.

Table~\ref{tab:s3_rt_by_choice} reports response-time descriptives separately for Choices A, B, and C in the Response Time estimation sample. Mean response times are 19.1 seconds for Choice A, 18.8 seconds for Choice B, and 17.9 seconds for Choice C, while the corresponding medians are 13, 15, and 15 seconds.

\begin{center}
\begin{table}[H]
\centering
\caption{Response Time by Distribution Game choice}
\label{tab:s3_rt_by_choice}
\begin{threeparttable}
\begin{tabular}{lcccc}
\toprule
Distribution Game choice & N & Mean & Median & SD \\
\midrule
Choice A (least prosocial) & 177 & 19.136 & 13 & 20.634 \\
Choice B (intermediate) & 146 & 18.753 & 15 & 14.116 \\
Choice C (most prosocial) & 202 & 17.851 & 15 & 13.402 \\
\bottomrule
\end{tabular}
\begin{tablenotes}[flushleft]
\footnotesize
\item \textit{Notes:} Response Time is measured in seconds on the final Distribution Game choice page. The table is restricted to the 525 observations used in the Response Time estimation sample.
\end{tablenotes}
\end{threeparttable}
\end{table}
\end{center}

Because Choice B is intermediate both in the decision-maker's payoff and in the distributional characteristics of the allocation, we also re-estimate the controlled Response Time $\times$ Perspective Taking specification after restricting the sample to the two endpoint allocations, Choices A and C. To preserve direct comparability with the main analysis, we retain the same OLS specification and the original coding of the outcome. In this restricted sample (\(N=379\)), the Response Time $\times$ Perspective Taking interaction remains negative and statistically significant (\(\beta=-0.0892\), SE \(=0.0329\), \(p=.007\)).

This endpoint-only specification shows that the estimated moderation pattern does not depend on the inclusion of the intermediate Distribution Game category. It does not identify the cognitive processes underlying response-time differences across choices, but indicates that the main interaction is also present when the comparison is restricted to the least and most prosocial allocations.

\subsection{Sensitivity to extreme Response Time values and session heterogeneity}
\label{sec:s3_rt_sensitivity}

Because the Response Time distribution is right-skewed, we assess whether the main interaction is driven by a small number of extreme decision times. We winsorize Response Time at the 1st and 99th percentiles, replacing observations beyond these thresholds with the corresponding percentile values. This procedure alters only 7 of the 525 observations in the Response Time estimation sample. The Response Time $\times$ Perspective Taking interaction remains negative and statistically significant both without controls (\(\beta=-0.0903\), \(p=.022\)) and with controls (\(\beta=-0.0964\), \(p=.017\)).

We also examine whether the interaction is sensitive to persistent differences across experimental sessions. Adding fixed effects for the 17 sessions in which Response Time is observed leaves the interaction negative and statistically significant (\(\beta=-0.0670\), SE \(=0.0280\), \(p=.017\)). The result is also robust to clustering standard errors at the session level (\(\beta=-0.0719\), SE \(=0.0236\), \(p=.008\)). These checks indicate that the negative interaction is not driven solely by extreme response-time observations or by between-session heterogeneity. As an additional inferential robustness check, we employ wild-cluster bootstrap tests with 9,999 replications to account for the limited number of sessions. The resulting $p$-values ($.081$ with Rademacher weights and $.072$ with Webb weights) indicate that the interaction remains statistically significant at the 10\% level under this alternative inference procedure.

We also examine whether the Response Time $\times$ Perspective Taking interaction is sensitive to features of the preceding bribery game. The interaction remains negative and statistically significant when controlling for the prior treatment ($\beta=-0.0658$, $p=.018$) and when additionally controlling for the number of rounds ($\beta=-0.0665$, $p=.016$), with no statistically significant differences in the interaction across treatments (joint test, $p=.798$).

\subsection{Response Time availability and sample comparability}
\label{sec:s3_missing_rt}

Among the 649 participants with a non-missing Distribution Game outcome, Response Time is observed for 525 participants and unavailable for 124. The unavailable observations are concentrated in five experimental sessions, indicating that response-time availability is a session-level feature of the dataset rather than individual nonresponse.

We compare participants with observed and missing Response Time on the variables used in the main specifications. As shown in Table~\ref{tab:s3_missing_rt_desc}, none of the differences in Distribution Game outcomes, Cognitive Reflection Test scores, Perspective Taking, age, or generalized trust is statistically significant. The same holds for the categorical controls: Pearson chi-squared tests yield $p=.114$ for gender, $p=.284$ for association membership, $p=.816$ for voting behavior, and $p=.517$ for blood-donation status. These checks reveal no statistically significant observable differences between the two groups on the characteristics considered.

\begin{center}
\begin{longtable}{lccc}
\caption{Comparison between participants with missing and observed Response Time (DG)}
\label{tab:s3_missing_rt_desc}\\
\toprule
 & Response Time missing & Response Time observed & $p$-value \\
\midrule
\endfirsthead

\toprule
 & Response Time missing & Response Time observed & $p$-value \\
\midrule
\endhead

\midrule
\multicolumn{4}{r}{\textit{Continued on next page}} \\
\endfoot

\bottomrule
\endlastfoot

Distribution Game outcome & 2.185 & 2.048 & .103 \\
 & (0.830) & (0.849) & \\

Cognitive Reflection Test (CRT) & 2.613 & 2.621 & .948 \\
 & (1.273) & (1.215) & \\

Perspective Taking (PT) & 3.558 & 3.589 & .608 \\
 & (0.611) & (0.603) & \\

Age & 22.153 & 22.141 & .976 \\
 & (4.985) & (3.829) & \\

Trust in people & 3.137 & 2.954 & .109 \\
 & (1.150) & (1.137) & \\

Observations & 124 & 525 & \\

\midrule
\multicolumn{4}{p{0.92\textwidth}}{\footnotesize \textit{Notes:} Entries report group means, with standard deviations in parentheses. The comparison is based on participants with a non-missing Distribution Game outcome ($N=649$). The last column reports two-sided two-sample t-tests of equality of means between participants with missing and observed Response Time.} \\

\end{longtable}
\end{center}

Because the Response Time models use a smaller set of sessions and participants, we also compare CRT and Response Time using both their maximum available samples and the common Response Time sample. As shown in Table~\ref{tab:s3_crt_same_sample_pt}, restricting the CRT analysis to the common sample does not generate a CRT $\times$ Perspective Taking interaction, while the joint specifications in Table~\ref{tab:s3_joint_crt_rt} preserve the negative Response Time $\times$ Perspective Taking interaction.

\subsection{Sensitivity of the interaction tests}

Because the sample size was determined by the broader experimental project rather than specifically for the present interaction analyses, we report a conventional sensitivity calculation for the incremental explanatory power of an additional interaction term. Under standard independent-observation assumptions, with a two-sided significance level of 5\% and 80\% power, the Response Time estimation sample (\(N=525\)) is approximately sensitive to an incremental \(R^2\) of 0.014, while the CRT estimation sample (\(N=649\)) is approximately sensitive to an incremental \(R^2\) of 0.012. These calculations describe the approximate magnitude of interaction effects that the available samples are well positioned to detect; they are not post-hoc estimates of the probability that the observed effects are true. In particular, the null CRT interaction should not be interpreted as ruling out smaller moderation effects.

\section{Empathic Concern as a Distinct Empathy Dimension}

In the main analysis, empathy is operationalized primarily through \textit{Perspective Taking} (PT), which captures the tendency to adopt the psychological point of view of others. PT is the empathy dimension most closely aligned with the cognitive component of social understanding and, conceptually, with the tendency to represent another person's perspective and the consequences of one's choices for others.

At the same time, the Interpersonal Reactivity Index also includes \textit{Empathic Concern} (EC), which reflects a different facet of empathy. While PT captures a more cognitive and perspective-based dimension, EC captures a more affective and compassion-based orientation, namely the tendency to feel sympathy, warmth, and concern for others. These two constructs are related, but they are not interchangeable. PT is closer to a mentalizing-oriented and cognitively mediated form of empathy, whereas EC reflects a more emotional and other-regarding form of empathic engagement.

For this reason, it is substantively important to examine whether the patterns documented for PT also emerge when empathy is instead measured through EC. This appendix section treats EC as a distinct empathy dimension and studies its relationship with prosocial behavior in the same empirical framework used for PT. The goal is not to identify which empathy dimension is ``better'' in general, but rather to assess whether different empathy traits display distinct associations with distributive behavior within the same empirical framework.

Overall, the results show that EC does not simply replicate the PT pattern. In the Response Time specifications, the interaction between Response Time and EC is positive, whereas the corresponding interaction with PT is negative. This suggests that the two empathy dimensions are linked to prosocial behavior in different ways. Moreover, when PT and EC are included jointly, the two interaction terms move in opposite directions and their difference is statistically significant, further supporting the idea that PT and EC should be treated as distinct moderators rather than as alternative proxies for the same latent trait.

\subsection*{S4.1 Response Time $\times$ Empathic Concern models}
\addcontentsline{toc}{subsection}{S4.1 Response Time \texorpdfstring{$\times$}{x} Empathic Concern models}

Before turning to the Response Time specifications, we briefly report the analogous models in which Empathic Concern interacts with the Cognitive Reflection Test. As with Perspective Taking in the main analysis, these CRT-based specifications do not yield statistically significant results: neither the main effect of Empathic Concern nor the interaction between CRT and Empathic Concern is significant, both without and with controls (Table~\ref{tab:s4_crt_ec}). We report these models for completeness and then apply the same Response Time framework used in the main PT analysis to Empathic Concern, allowing us to assess whether the moderation pattern differs across the cognitive and affective empathy dimensions.

\begin{center}
\footnotesize
\begin{longtable}{lcc}
\caption{Cognitive Reflection Test $\times$ Empathic Concern models}
\label{tab:s4_crt_ec}\\
\toprule
& (1) & (2) \\
\midrule
\endfirsthead

\toprule
& (1) & (2) \\
\midrule
\endhead

\midrule
\multicolumn{3}{r}{\textit{Continued on next page}}
\endfoot

\bottomrule
\endlastfoot

Cognitive Reflection Test (CRT)
    & -0.108 & -0.117 \\
    & (0.166) & (0.167) \\

Empathic Concern (EC)
    & 0.002 & -0.011 \\
    & (0.130) & (0.129) \\

CRT $\times$ Empathic Concern
    & 0.025 & 0.027 \\
    & (0.044) & (0.044) \\

Age
    &  & -0.009 \\
    &  & (0.009) \\

Male
    &  & -0.027 \\
    &  & (0.077) \\

Gender: prefer not to answer
    &  & 0.542* \\
    &  & (0.243) \\

Gender: not identified
    &  & -0.006 \\
    &  & (0.528) \\

Trust in people
    &  & 0.015 \\
    &  & (0.031) \\

Association membership
    &  & -0.041 \\
    &  & (0.079) \\

Did not vote
    &  & 0.070 \\
    &  & (0.111) \\

Not eligible to vote
    &  & 0.022 \\
    &  & (0.130) \\

Blood donor: no
    &  & -0.201* \\
    &  & (0.096) \\

Blood donor: not eligible
    &  & -0.133 \\
    &  & (0.111) \\

Constant
    & 2.108*** & 2.495*** \\
    & (0.496) & (0.543) \\

\midrule
Controls & No & Yes \\
Observations & 649 & 649 \\
R-squared & 0.004 & 0.019 \\
Log likelihood & -811.072 & -806.001 \\
AIC & 1630.144 & 1640.002 \\
BIC & 1648.046 & 1702.658 \\ \hline

\multicolumn{3}{p{1\textwidth}}{\footnotesize \textit{Notes:} The dependent variable is the Distribution Game outcome, coded so that higher values indicate more prosocial choices. Robust standard errors are reported in parentheses. Controls correspond to the same demographic and survey-based variables used in the main specifications. In both specifications, neither the main effect of Empathic Concern nor the interaction between CRT and Empathic Concern is statistically significant. *** $p<.001$, ** $p<.01$, * $p<.05$.} \\

\end{longtable}
\end{center}

We then replace Perspective Taking with Empathic Concern in the Response Time interaction models. Table~\ref{tab:s4_rt_ec_raw} reports the uncentered specifications. In contrast with the PT models, the interaction between Response Time and Empathic Concern is positive and statistically significant across all specifications. In the uncentered models, the coefficient on Response Time is negative, while the interaction term is positive. Substantively, this indicates that longer choice-stage response times are associated with fewer prosocial choices at lower levels of EC, whereas this negative association becomes weaker---and may even reverse---as EC increases.

The pattern remains stable after adding the standard demographic and survey-based controls and after additionally controlling for CRT. However, as in the uncentered PT models, the variance inflation factors are relatively high, especially for Response Time and the interaction term, indicating that multicollinearity affects the direct interpretation of lower-order coefficients in the raw specifications. These diagnostics are reported directly in Table~\ref{tab:s4_rt_ec_raw}. For this reason, we next turn to centered models.

\begin{center}
\footnotesize
\begin{longtable}{lcccc}
\caption{Response Time $\times$ Empathic Concern models (uncentered)}
\label{tab:s4_rt_ec_raw}\\
\toprule
& (1) & (2) & (3) & (4) \\
\midrule
\endfirsthead

\toprule
& (1) & (2) & (3) & (4) \\
\midrule
\endhead

\midrule
\multicolumn{5}{r}{\textit{Continued on next page}}
\endfoot

\bottomrule
\endlastfoot

Response Time
    & -0.144* & -0.154** & -0.148** & -0.156** \\
    & (0.056) & (0.055) & (0.057) & (0.056) \\

Empathic Concern
    & 0.027 & 0.017 & 0.009 & 0.004 \\
    & (0.077) & (0.079) & (0.078) & (0.080) \\

Response Time $\times$ Empathic Concern
    & 0.039* & 0.042* & 0.041* & 0.043* \\
    & (0.018) & (0.017) & (0.018) & (0.017) \\

CRT
    &  &  & -0.045 & -0.042 \\
    &  &  & (0.031) & (0.033) \\

Age
    &  & -0.010 &  & -0.011 \\
    &  & (0.009) &  & (0.009) \\

Male
    &  & -0.115 &  & -0.087 \\
    &  & (0.083) &  & (0.086) \\

Gender: prefer not to answer
    &  & 0.449 &  & 0.450 \\
    &  & (0.396) &  & (0.388) \\

Gender: not identified
    &  & -0.093 &  & -0.075 \\
    &  & (0.507) &  & (0.532) \\

Trust in people
    &  & -0.001 &  & 0.002 \\
    &  & (0.034) &  & (0.034) \\

Association membership
    &  & -0.076 &  & -0.077 \\
    &  & (0.090) &  & (0.090) \\

Did not vote
    &  & 0.045 &  & 0.032 \\
    &  & (0.125) &  & (0.128) \\

Not eligible to vote
    &  & 0.113 &  & 0.113 \\
    &  & (0.147) &  & (0.146) \\

Blood donor: no
    &  & -0.229* &  & -0.233* \\
    &  & (0.105) &  & (0.104) \\

Blood donor: not eligible
    &  & -0.213 &  & -0.197 \\
    &  & (0.122) &  & (0.122) \\

Constant
    & 1.952*** & 2.470*** & 2.130*** & 2.637*** \\
    & (0.285) & (0.368) & (0.313) & (0.392) \\

\midrule
Controls & No & Yes & No & Yes \\
CRT control & No & No & Yes & Yes \\
Observations & 525 & 525 & 525 & 525 \\
R-squared & 0.011 & 0.032 & 0.015 & 0.035 \\
Log likelihood & -655.642 & -650.085 & -654.583 & -649.233 \\
AIC & 1319.285 & 1328.171 & 1319.166 & 1328.465 \\
BIC & 1336.338 & 1387.858 & 1340.483 & 1392.416 \\
Mean VIF & 11.19 & 3.61 & 8.68 & 3.45 \\ \hline

\multicolumn{5}{p{0.90\textwidth}}{\footnotesize \textit{Notes:} The dependent variable is the Distribution Game outcome, coded so that higher values indicate more prosocial choices. Robust standard errors are reported in parentheses. Response Time is expressed in units of 10 seconds. Controls correspond to the same demographic and survey-based variables used in the main specifications. Uncentered models are reported here to show the raw interaction pattern; multicollinearity is addressed explicitly in the centered specifications below. *** $p<.001$, ** $p<.01$, * $p<.05$.} \\

\end{longtable}
\end{center}

\subsection*{S4.2 Centered Response Time $\times$ Empathic Concern models}
\addcontentsline{toc}{subsection}{S4.2 Centered Response Time \texorpdfstring{$\times$}{x} Empathic Concern models}

To make the lower-order coefficients interpretable and to reduce non-essential multicollinearity, we re-estimate the same models after mean-centering both Response Time and Empathic Concern. Table~\ref{tab:s4_rt_ec_centered} reports the centered specifications. The interaction between Response Time and EC remains positive and statistically significant in all four models, while the VIF values fall dramatically and become uniformly low.

The centered specifications show that the high VIFs in the uncentered models reflect non-essential collinearity induced by the interaction parameterization rather than changes in the underlying fitted model. The positive Response Time $\times$ Empathic Concern interaction remains statistically significant when demographic and survey-based controls are included and when CRT is added as a separate measure of analytic reflection. The corresponding fit statistics and multicollinearity diagnostics are reported in Table~\ref{tab:s4_rt_ec_centered}.

\begin{center}
\footnotesize
\begin{longtable}{lcccc}
\caption{Centered Response Time $\times$ Empathic Concern models}
\label{tab:s4_rt_ec_centered}\\
\toprule
& (1) & (2) & (3) & (4) \\
\midrule
\endfirsthead

\toprule
& (1) & (2) & (3) & (4) \\
\midrule
\endhead

\midrule
\multicolumn{5}{r}{\textit{Continued on next page}}
\endfoot

\bottomrule
\endlastfoot

Response Time, centered
    & 0.000 & 0.001 & 0.004 & 0.004 \\
    & (0.024) & (0.024) & (0.024) & (0.024) \\

Empathic Concern, centered
    & 0.099 & 0.095 & 0.085 & 0.084 \\
    & (0.063) & (0.065) & (0.064) & (0.066) \\

Response Time, centered $\times$ Empathic Concern, centered
    & 0.039* & 0.042* & 0.041* & 0.043* \\
    & (0.018) & (0.017) & (0.018) & (0.017) \\

CRT
    &  &  & -0.045 & -0.042 \\
    &  &  & (0.031) & (0.033) \\

Age
    &  & -0.010 &  & -0.011 \\
    &  & (0.009) &  & (0.009) \\

Male
    &  & -0.115 &  & -0.087 \\
    &  & (0.083) &  & (0.086) \\

Gender: prefer not to answer
    &  & 0.449 &  & 0.450 \\
    &  & (0.396) &  & (0.388) \\

Gender: not identified
    &  & -0.093 &  & -0.075 \\
    &  & (0.507) &  & (0.532) \\

Trust in people
    &  & -0.001 &  & 0.002 \\
    &  & (0.034) &  & (0.034) \\

Association membership
    &  & -0.076 &  & -0.077 \\
    &  & (0.090) &  & (0.090) \\

Did not vote
    &  & 0.045 &  & 0.032 \\
    &  & (0.125) &  & (0.128) \\

Not eligible to vote
    &  & 0.113 &  & 0.113 \\
    &  & (0.147) &  & (0.146) \\

Blood donor: no
    &  & -0.229* &  & -0.233* \\
    &  & (0.105) &  & (0.104) \\

Blood donor: not eligible
    &  & -0.213 &  & -0.197 \\
    &  & (0.122) &  & (0.122) \\

Constant
    & 2.052*** & 2.537*** & 2.170*** & 2.659*** \\
    & (0.037) & (0.247) & (0.090) & (0.264) \\

\midrule
Controls & No & Yes & No & Yes \\
CRT control & No & No & Yes & Yes \\
Observations & 525 & 525 & 525 & 525 \\
R-squared & 0.011 & 0.032 & 0.015 & 0.035 \\
Log likelihood & -655.642 & -650.085 & -654.583 & -649.233 \\
AIC & 1319.285 & 1328.171 & 1319.166 & 1328.465 \\
BIC & 1336.338 & 1387.858 & 1340.483 & 1392.416 \\
Mean VIF & 1.11 & 1.20 & 1.11 & 1.21 \\ \hline

\multicolumn{5}{p{0.90\textwidth}}{\footnotesize \textit{Notes:} The dependent variable is the Distribution Game outcome, coded so that higher values indicate more prosocial choices. Robust standard errors are reported in parentheses. Response Time and Empathic Concern are mean-centered before constructing the interaction term. Response Time is expressed in units of 10 seconds. Controls correspond to the same demographic and survey-based variables used in the main specifications. *** $p<.001$, ** $p<.01$, * $p<.05$.} \\

\end{longtable}
\end{center}

\subsection*{S4.3 Marginal effects of Response Time across levels of Empathic Concern}
\addcontentsline{toc}{subsection}{S4.3 Marginal effects of Response Time across levels of Empathic Concern}

To make the interaction easier to interpret, Table~\ref{tab:s4_rt_ec_margins} reports the marginal effect of Response Time evaluated at low EC ($-1$ SD), average EC, and high EC ($+1$ SD) in the centered models. Across all specifications, the simple slopes move monotonically upward as EC increases: the marginal effect of Response Time is negative at low EC, approximately zero around the mean, and positive at high EC. Although the simple slopes at low and high EC are not individually significant, the difference between them is statistically significant in every specification.

This is the key result of the EC moderation analysis. It implies that the association between choice-stage response time and prosociality differs significantly across levels of Empathic Concern, even if the slope within each endpoint is estimated imprecisely. In other words, EC significantly shifts the \textit{direction} of the Response Time gradient. Table~\ref{tab:s4_rt_ec_margins} reports these simple slopes and the corresponding difference tests.

\begin{center}
\footnotesize
\begin{longtable}{lcccc}
\caption{Marginal effect of Response Time across levels of Empathic Concern}
\label{tab:s4_rt_ec_margins}\\
\toprule
& (1) & (2) & (3) & (4) \\
\midrule
\endfirsthead

\toprule
& (1) & (2) & (3) & (4) \\
\midrule
\endhead

\midrule
\multicolumn{5}{r}{\textit{Continued on next page}}
\endfoot

\bottomrule
\endlastfoot

Slope at EC = $-1$ SD
    & -0.024 & -0.024 & -0.021 & -0.022 \\
    & (0.021) & (0.020) & (0.021) & (0.020) \\

Slope at EC = mean
    & 0.000 & 0.001 & 0.004 & 0.004 \\
    & (0.024) & (0.024) & (0.024) & (0.024) \\

Slope at EC = $+1$ SD
    & 0.024 & 0.027 & 0.029 & 0.031 \\
    & (0.031) & (0.031) & (0.031) & (0.031) \\

Difference: high EC $-$ low EC
    & 0.047* & 0.051* & 0.050* & 0.053* \\
    & (0.021) & (0.021) & (0.022) & (0.021) \\

\midrule
Controls & No & Yes & No & Yes \\
CRT control & No & No & Yes & Yes \\
Observations & 525 & 525 & 525 & 525 \\ \hline

\multicolumn{5}{p{0.90\textwidth}}{\footnotesize \textit{Notes:} Marginal effects are computed from the centered models reported in Table~\ref{tab:s4_rt_ec_centered} and refer to a 10-second increase in Response Time. Low and high EC are defined as one standard deviation below and above the sample mean of centered Empathic Concern ($SD = 0.611$). The last row reports the linear combination test for the difference between the high-EC and low-EC slopes. *** $p<.001$, ** $p<.01$, * $p<.05$.} \\

\end{longtable}
\end{center}

\begin{figure}[H]
\centering
\caption{Marginal effect of Response Time across levels of Empathic Concern}
\includegraphics[width=0.80\textwidth]{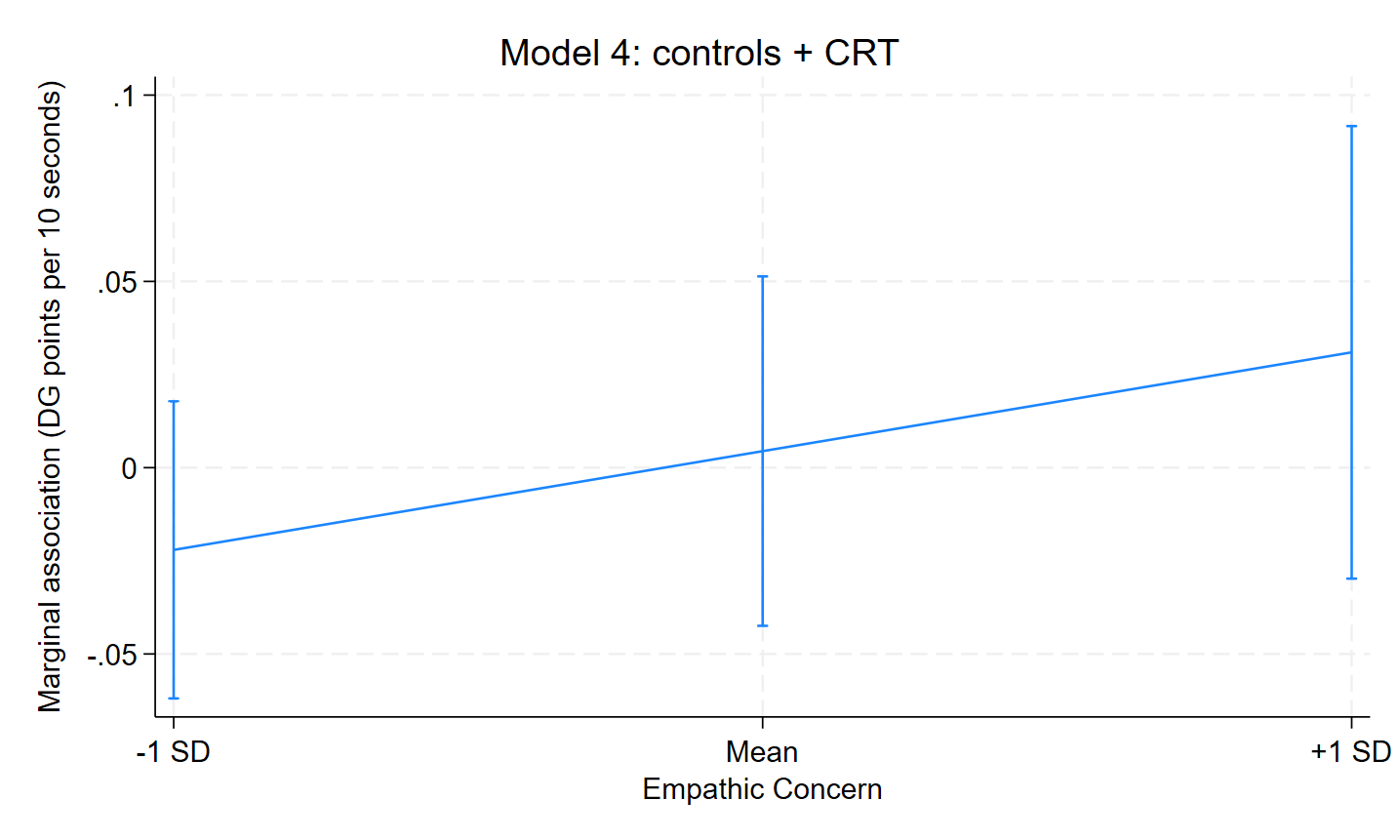}
\label{fig:s4_ec_margins}
\end{figure}

Figure~\ref{fig:s4_ec_margins} shows the marginal effect of Response Time across levels of Empathic Concern in the centered specification with controls and CRT. The main text also presents a graphical comparison of predicted Distribution Game outcomes across Response Time for both PT and EC.

\subsection*{S4.4 Joint model with Perspective Taking and Empathic Concern}
\addcontentsline{toc}{subsection}{S4.4 Joint model with Perspective Taking and Empathic Concern}

We estimate two joint models that include Perspective Taking (PT), Empathic Concern (EC), centered Response Time, and the interaction of Response Time with each empathy measure. Both models use the same 525 observations and the demographic and social controls used in the main analysis. Column (2) of Table~\ref{tab:s4_joint_pt_ec} additionally controls for CRT performance.

The estimated interaction with PT is negative, while the interaction with EC is positive. In both models, neither interaction is statistically significant on its own, but the tests reject both the null that the two coefficients are jointly zero and the null that they are equal (Table~\ref{tab:s4_joint_pt_ec_tests}). The estimates are similar when CRT performance is included.

\begin{center}
\footnotesize
\begin{longtable}{lcc}
\caption{Joint models including Perspective Taking and Empathic Concern}
\label{tab:s4_joint_pt_ec}\\
\toprule
& (1) & (2) \\
\midrule
\endfirsthead

\toprule
& (1) & (2) \\
\midrule
\endhead

\midrule
\multicolumn{3}{r}{\textit{Continued on next page}} \\
\endfoot

\bottomrule
\endlastfoot

Response Time, centered & 0.011 & 0.015 \\
& (0.026) & (0.026) \\

Perspective Taking, centered & 0.080 & 0.080 \\
& (0.065) & (0.065) \\

Empathic Concern, centered & 0.065 & 0.054 \\
& (0.066) & (0.067) \\

Response Time $\times$ Perspective Taking & -0.053 & -0.054 \\
& (0.034) & (0.034) \\

Response Time $\times$ Empathic Concern & 0.032 & 0.034 \\
& (0.022) & (0.022) \\

CRT & & -0.043 \\
& & (0.033) \\

Age & -0.010 & -0.011 \\
& (0.009) & (0.009) \\

Male & -0.119 & -0.091 \\
& (0.083) & (0.086) \\

Gender: prefer not to answer & 0.394 & 0.395 \\
& (0.395) & (0.388) \\

Gender: not identified & -0.087 & -0.069 \\
& (0.493) & (0.518) \\

Trust in people & -0.002 & 0.001 \\
& (0.035) & (0.035) \\

Association membership & -0.080 & -0.082 \\
& (0.090) & (0.090) \\

Did not vote & 0.029 & 0.016 \\
& (0.124) & (0.127) \\

Not eligible to vote & 0.131 & 0.131 \\
& (0.146) & (0.146) \\

Blood donor: no & -0.233* & -0.237* \\
& (0.105) & (0.104) \\

Blood donor: not eligible & -0.226 & -0.210 \\
& (0.122) & (0.122) \\

Constant & 2.555*** & 2.680*** \\
& (0.249) & (0.266) \\

\midrule
Controls & Yes & Yes \\
CRT control & No & Yes \\
Observations & 525 & 525 \\
$R^2$ & 0.038 & 0.042 \\
\addlinespace
\multicolumn{3}{p{0.90\textwidth}}{\footnotesize \textit{Notes:} The dependent variable is the Distribution Game prosociality score, ranging from 1 to 3. Robust standard errors are reported in parentheses. Response Time and both empathy scores are mean-centered within the estimation sample. Response Time is expressed in units of 10 seconds. Both specifications include the same demographic and social controls. Column (2) additionally controls for CRT performance. Coefficients and standard errors are rounded to three decimal places. *** $p<.001$, ** $p<.01$, * $p<.05$.} \\

\end{longtable}
\end{center}

\begin{center}
\footnotesize
\begin{longtable}{lcc}
\caption{Post-estimation tests from the joint PT--EC models}
\label{tab:s4_joint_pt_ec_tests}\\
\toprule
& (1) & (2) \\
\midrule
\endfirsthead

\toprule
& (1) & (2) \\
\midrule
\endhead

\midrule
\multicolumn{3}{r}{\textit{Continued on next page}} \\
\endfoot

\bottomrule
\endlastfoot

$H_0$: Response Time $\times$ PT $=0$ & .1177 & .1150 \\

$H_0$: Response Time $\times$ EC $=0$ & .1469 & .1342 \\

$H_0$: both interaction terms jointly $=0$ & .0274 & .0237 \\

$H_0$: Response Time $\times$ PT $=$ Response Time $\times$ EC & .0099 & .0086 \\

\addlinespace
\multicolumn{3}{p{0.90\textwidth}}{\footnotesize \textit{Notes:} Entries are $p$-values from robust Wald tests based on the corresponding specifications in Table~\ref{tab:s4_joint_pt_ec}. Both models use the same estimation sample ($N=525$).} \\

\end{longtable}
\end{center}

\subsection*{S4.5 Interpretation and theoretical implications}
\addcontentsline{toc}{subsection}{S4.5 Interpretation and theoretical implications}

The supplementary EC analysis differs from the main PT analysis. In the separate Response Time specifications, the Response Time $\times$ PT interaction is negative, whereas the Response Time $\times$ EC interaction is positive. In both joint models, the two interactions retain opposite signs; although neither is individually statistically significant, they are jointly significant and the hypothesis that they are equal is rejected. PT and EC therefore exhibit statistically distinguishable moderation patterns rather than acting as interchangeable indicators of a single empathy construct.

This contrast is consistent with the two IRI dimensions capturing different components of empathy: PT concerns the cognitive representation of others' perspectives, whereas EC captures affective concern for others. The predicted profiles presented in the main text further illustrate this distinction: the difference between high and low PT narrows as response time increases, whereas the corresponding difference for EC widens. Although the present design does not identify the underlying psychological processes directly, these results show that PT and EC are associated with different moderation patterns.

\clearpage
\section{Experimental Instructions}

This section reproduces the experimental instructions used in the study. For transparency and completeness, we report the instructions for the different Bribe Game treatments, as well as the materials for the Distribution Game, the Cognitive Reflection Test, the Interpersonal Reactivity Index, and the exit survey. Both the English version and the original Italian version are included.

\subsection{Experimental Instructions (English)}

\textbf{Welcome to VERA Lab!}

Hello, welcome to the VERA Lab. We thank you for participating in the study.

Before we begin, we kindly ask you to turn off your mobile phone. From this moment onward, no form of communication with the participants present in the session or with the outside is allowed.

If you have any questions during the course of the study, do not hesitate to call us; we will clarify any doubts you may have. You are allowed to use paper and pen for notes or calculations.

It is very important to follow the instructions carefully.

During the study, you will be asked to make decisions: you can change each decision as many times as you wish before pressing the "Next" button at the bottom of the page. After clicking the "Next" button, your last selected decision will become final.

Please note that the choices you make are anonymous. During the study, you will be part of a group whose composition will remain anonymous: you will not know who is in your group, and others will not know who is in theirs.

At the end of the study, an economic compensation is provided. You will receive 3 euros for your participation. Additionally, you will earn an extra payment based on the choices you make, which will be communicated to you at the end of the study.

For the study, you will be given experimental points. The points you earn during the study will be converted into euros.  
100 experimental points are equivalent to 3 euros.

In this study, you can earn up to 25 euros.

Click on the "Next" button when you are ready to begin.
\newpage
On the next page, you will be presented with the instructions for the first part of the study. After reading the instructions, you will need to answer some comprehension questions. These questions will not affect your final compensation but will help you ensure that you have understood the structure of the game.
\newpage

\subsubsection{Bribe Game Full - Buffat and Senn Replication}
The study will last between 12 and 15 periods. At the beginning of each period, every participant receives \textbf{20 points}. Each group is composed of 3 \textbf{Participant A's} and 1 \textbf{Participant B}. Roles are assigned randomly and remain the same throughout all periods. Each period is divided into \textbf{3 phases}.
\bigskip

\textbf{Phase 1}

\bigskip
If you are a \textbf{Participant A}, you must decide how to allocate the 20 points between two options: a \textbf{private account} and a \textbf{common project}. Any points not allocated to the common project are automatically credited to your private account.

\bigskip
\textbf{Your income from the private account}.

Every point credited to your private account provides you with one point of income.  
For example, if you allocate all 20 points to your private account (and contribute nothing to the common project), you earn exactly 20 points. If you leave 6 points in your private account, your income from the private account is 6 points.

\bigskip
\textbf{Your income from the common project}.
Every member of the group (both Participant A's and Participant B) equally benefits from the contributions made to the common project. Income from the common project is calculated as follows:

\begin{mdframed}
\begin{center}
    Income from the common project
\end{center}
\par
 \begin{center}
     $=$
 \end{center}
\par
\begin{center}
    (1.6 × sum of all contributions to the common project) / number of participants
\end{center}
\end{mdframed}

For example, if the total contributions to the common project amount to 60 points (i.e., all Participant A's contribute 20 points each), you and the other group members receive: (60 × 1.6) / 4 = 24 points.

If the total contributions to the common project amount to 20 points, you and the other group members receive: (20 × 1.6) / 4 = 8 points.

If no one contributes to the common project, no one receives any points from it.

The income from the common project is rounded down for decimal values less than 0.5, and rounded up for decimal values equal to or greater than 0.5.

\bigskip
\textbf{Income for Participant A after Phase 1}

\begin{mdframed}
\begin{center}
    Income from private account + Income from common project
\end{center}
 
\par
 \begin{center}
     $=$
 \end{center}
\par
\begin{center}
    20 - your contribution to the common project + [(1.6 × sum of all contributions to the common project) / 4]
\end{center}
\end{mdframed}
\bigskip

\textbf{If you are a Participant B}
You do not need to take any action during this Phase. 
\par
\bigskip

Click the "Next" button to proceed to the next part of the instructions. 

\textit{Note: The "Next" button will become available after 60 seconds.}

\newpage
\textbf{Phase 2}
\bigskip

\textbf{If you are a Participant A}. After Phase 1, you receive an additional allocation of 5 points.  

You can decide whether and how many of these points to transfer to Participant B. Any points not transferred to Participant B are added to your income.  

\bigskip
Each point you transfer to Participant B costs you 1 point and increases Participant B’s income by 1 point. For example, if you decide to transfer 3 points to Participant B, it will cost you 3 points, and Participant B’s income will increase by 3 points.

\bigskip
\textbf{Income for Participant A after Phase 2}. 
Your total income is calculated as:
\begin{mdframed}
\begin{center}
    Income from private account + Income from common project + 5 points - points transferred to Participant B
\end{center}
 
\par
 \begin{center}
     $=$
 \end{center}
\par
\begin{center}
    20 - contribution to the common project + income from the common project + 5 - points transferred to Participant B
\end{center}
\end{mdframed}
\bigskip

\textbf{If you are a Participant B}. You do not need to take any action during Phase 2.
\par
\bigskip

Click the "Next" button to proceed to the next part of the instructions.  

\textit{Note: The "Next" button will become available after 30 seconds}
\newpage

\textbf{Phase 3}

\bigskip
\textbf{If you are a Participant A}. You do not need to take any action during Phase 3. 

\bigskip
\textbf{If you are a Participant B}. You will be shown the contributions made by each Participant A to the common project, along with the number of points each Participant A has transferred to you.
\textbf{You must decide whether to accept or reject the points transferred to you by each Participant A}.

\bigskip
If you accept, the points will be credited to your account. If you reject, the points will be returned to the respective Participant A.

\bigskip
Note: The order in which Participant A's are displayed is random. This means that the decisions shown in different periods do not necessarily correspond to the same Participant A.

\bigskip
In addition, if you are a Participant B,\textbf{you receive another 15 points (which we will call deduction points)}. These points can be used to \textbf{reduce the earnings of the Participant As in your group}.

\bigskip

You must decide whether and how many \textbf{deduction points} to assign to the A participants in your group.

For each deduction point you assign, you pay \textbf{1 point}, and the income of the corresponding Participant A is reduced by \textbf{3 points}. 

\bigskip
Example: If you assign 3 deduction points to a Participant A, their income will decrease by \(3 \times 3 = 9\) points, while your income will decrease by 3 points.

\bigskip
Although you have 15 deduction points, \textbf{you can subtract a maximum of 10 points from each participant}. 

\bigskip 
Any unused deduction points are credited back to your income for the period.

\bigskip 
\textbf{Income for Participant B after Phase 3}.

\begin{mdframed}
\begin{center}
    Income for Participant B
\end{center}
\par
 \begin{center}
     $=$
 \end{center}
\par

\begin{center}
    20 + income from the common project + points transferred by Participant A's + unused deduction points - total deduction points assigned to Participant As
\end{center}
\end{mdframed}
\bigskip

\newpage
\textbf{Summary of Participant A and Participant B Income}

\bigskip
\textbf{For Participant A.}
Your income for the period is calculated as follows:

\begin{mdframed}
\begin{center}
Income of Participant A for the period\end{center}
\par
 \begin{center}
     $=$
 \end{center}
\par
\begin{center}
    20 - contribution to the common project + income from the common project + 5 - points transferred to Participant B - 3 × (deduction points assigned by Participant B)
\end{center}

\end{mdframed}

\textbf{For Participant B.} Your income for the period is calculated as follows:
\begin{mdframed}
\begin{center}
Income of Participant B for the period
\end{center}
\par
 \begin{center}
     $=$
 \end{center}
\par
\begin{center}20 + income from the common project + sum of accepted transfers + 15 - total deduction points assigned to Participant As
\end{center}
 \end{mdframed}

\bigskip
At the end of each round, you will be shown a summary of your decisions and those made by the other Participant As in the group (your income for the round, the contributions to the common project by the other Participant As, the deduction points assigned to Participant As, and any points accepted or rejected by Participant B). The order in which the Participant As are shown is random. This means that the decisions shown in different rounds do not necessarily correspond to the same Participant A.

\bigskip
You will now be presented with a quiz. Please answer the quiz questions by entering your responses on the computer. You will always have access to a summary of the instructions by clicking on the \textbf{"Instructions"} button.

\newpage

\subsubsection{Control Treatment}

The study will last between 12 and 15 periods. At the beginning of each period, every participant receives \textbf{20 points}. Each group is composed of 3 \textbf{Participant A's} and 1 \textbf{Participant B}. Roles are assigned randomly and remain the same throughout all periods. Each period is divided into \textbf{3 phases}.
\bigskip

\textbf{Phase 1}

\bigskip
If you are a \textbf{Participant A}, you must decide how to allocate the 20 points between two options: a \textbf{private account} and a \textbf{common project}. Any points not allocated to the common project are automatically credited to your private account.

\bigskip
\textbf{Your income from the private account}.

Every point credited to your private account provides you with one point of income.  
For example, if you allocate all 20 points to your private account (and contribute nothing to the common project), you earn exactly 20 points. If you leave 6 points in your private account, your income from the private account is 6 points.

\bigskip
\textbf{Your income from the common project}.
Every member of the group (both Participant A's and Participant B) equally benefits from the contributions made to the common project. Income from the common project is calculated as follows:

\begin{mdframed}
\begin{center}
    Income from the common project
\end{center}
\par
 \begin{center}
     $=$
 \end{center}
\par
\begin{center}
    (1.6 × sum of all contributions to the common project) / number of participants
\end{center}
\end{mdframed}

For example, if the total contributions to the common project amount to 60 points (i.e., all Participant A's contribute 20 points each), you and the other group members receive: (60 × 1.6) / 4 = 24 points.

If the total contributions to the common project amount to 20 points, you and the other group members receive: (20 × 1.6) / 4 = 8 points.

If no one contributes to the common project, no one receives any points from it.

The income from the common project is rounded down for decimal values less than 0.5, and rounded up for decimal values equal to or greater than 0.5.

\bigskip
\textbf{Income for Participant A after Phase 1}

\begin{mdframed}
\begin{center}
    Income from private account + Income from common project
\end{center}
 
\par
 \begin{center}
     $=$
 \end{center}
\par
\begin{center}
    20 - your contribution to the common project + [(1.6 × sum of all contributions to the common project) / 4]
\end{center}
\end{mdframed}
\bigskip

\textbf{If you are a Participant B}
You do not need to take any action during this Phase. 
\par
\bigskip

Click the "Next" button to proceed to the next part of the instructions. 

\textit{Note: The "Next" button will become available after 60 seconds.}

\newpage
\textbf{Phase 2}
\bigskip

\textbf{If you are a Participant A}. After Phase 1, you receive an additional allocation of 5 points.  

You can decide whether and how many of these points to transfer to Participant B. Any points not transferred to Participant B are added to your income.  

\bigskip
Each point you transfer to Participant B costs you 1 point and increases Participant B’s income by 1 point. For example, if you decide to transfer 3 points to Participant B, it will cost you 3 points, and Participant B’s income will increase by 3 points.

\bigskip
\textbf{Income for Participant A after Phase 2}. 
Your total income is calculated as:
\begin{mdframed}
\begin{center}
    Income from private account + Income from common project + 5 points - points transferred to Participant B
\end{center}
 
\par
 \begin{center}
     $=$
 \end{center}
\par
\begin{center}
    20 - contribution to the common project + income from the common project + 5 - points transferred to Participant B
\end{center}
\end{mdframed}
\bigskip

\textbf{If you are a Participant B}. You do not need to take any action during Phase 2.
\par
\bigskip

Click the "Next" button to proceed to the next part of the instructions.  

\textit{Note: The "Next" button will become available after 30 seconds}
\newpage

\textbf{Phase 3}

\bigskip
\textbf{If you are a Participant A}. You do not need to take any action during Phase 3. 

\bigskip
\textbf{If you are a Participant B}. You will be shown the contributions made by each Participant A to the common project, along with the number of points each Participant A has transferred to you.
\textbf{You must decide whether to accept or reject the points transferred to you by each Participant A}.

\bigskip
If you accept, the points will be credited to your account. If you reject, the points will be returned to the respective Participant A.

\bigskip
Note: The order in which Participant A's are displayed is random. This means that the decisions shown in different periods do not necessarily correspond to the same Participant A.

\bigskip
In addition, if you are a Participant B,\textbf{you receive another 15 points (which we will call deduction points)}. These points can be used to \textbf{reduce the earnings of the Participant As in your group}.

\bigskip

You must decide whether and how many \textbf{deduction points} to assign to the A participants in your group.

For each deduction point you assign, you pay \textbf{1 point}, and the income of the corresponding Participant A is reduced by \textbf{3 points}. 

\bigskip
Example: If you assign 3 deduction points to a Participant A, their income will decrease by \(3 \times 3 = 9\) points, while your income will decrease by 3 points.

\bigskip
Although you have 15 deduction points, \textbf{you can subtract a maximum of 10 points from each participant}. 

\bigskip 
Any unused deduction points are credited back to your income for the period.

\bigskip 
\textbf{Income for Participant B after Phase 3}.

\begin{mdframed}
\begin{center}
    Income for Participant B
\end{center}
\par
 \begin{center}
     $=$
 \end{center}
\par

\begin{center}
    20 + income from the common project + points transferred by Participant A's + unused deduction points - total deduction points assigned to Participant As
\end{center}
\end{mdframed}
\bigskip

\newpage
\textbf{Summary of Participant A and Participant B Income}

\bigskip
\textbf{For Participant A.}
Your income for the period is calculated as follows:

\begin{mdframed}
\begin{center}
Income of Participant A for the period\end{center}
\par
 \begin{center}
     $=$
 \end{center}
\par
\begin{center}
    20 - contribution to the common project + income from the common project + 5 - points transferred to Participant B - 3 × (deduction points assigned by Participant B)
\end{center}

\end{mdframed}

\textbf{For Participant B.} Your income for the period is calculated as follows:
\begin{mdframed}
\begin{center}
Income of Participant B for the period
\end{center}
\par
 \begin{center}
     $=$
 \end{center}
\par
\begin{center}20 + income from the common project + sum of accepted transfers + 15 - total deduction points assigned to Participant As
\end{center}
 \end{mdframed}

\bigskip
At the end of each round, Participant As will be shown a summary of their own decisions and those made by the other Participant As in the group (their income for the round, the contributions to the common project by the other Participant As, and the deduction points assigned to Participant As). The order in which Participant As are displayed is random. This means that the decisions shown in different rounds do not necessarily correspond to the same Participant A. Participant B will also be shown a summary of their own decisions and those made by the other Participant As in the group (their income for the round, the contributions to the common project by the other Participant As, the deduction points assigned to Participant As, and the points transferred by the Participant As).

\bigskip
You will now be presented with a quiz. Please answer the quiz questions by entering your responses on the computer. You will always have access to a summary of the instructions by clicking on the \textbf{"Instructions"} button.

\newpage
\textbf{Quiz}

\bigskip
Imagine that the other Participants A in the group have contributed 10 points each to the common project. You are also a Participant A and have decided to contribute 10 points.
\begin{itemize}
    \item[1] How many points do you get if Participant B does not deduct any points?
    \begin{todolist}
        \item 30 x 1.6 = 48. 48 : 4 = 12. 20 + 12 = 32
        \item 30 x 1.6 = 48. 48 : 4 = 12. 20 - 10 + 12 + 5 = 27
    \end{todolist}
    \item[2] How many points do you get in total if Participant B deducts 9 points from you?
    \begin{todolist}
        \item 20 + 12 - 9 = 23
        \item 320 - 10 = 10. 10 + 12 + 5 - 9 = 18
    \end{todolist}
    \item[3] How many points do you get if you transfer 2 points to Participant B?

    \begin{todolist}
        \item 20 - 10 + 12 + 5 - 2 = 25
        \item 20 - 10 + 12 - 2 = 20
    \end{todolist}
\end{itemize}

\newpage

\textbf{Role}

\bigskip
Your role is \textbf{Participant A}

\bigskip
\bigskip

\textbf{Role}

\bigskip
Your role is \textbf{Participant B}
\newpage

\textbf{Let’s start} 

\bigskip
Now you are ready to start! The decisions you will make in the next pages will have an impact on your final income.
\newpage

\textbf{Phase 1 (Participant A)}
\bigskip

\rightline{\textbf{Period 1}}
\bigskip
How many points do you want to allocate to the common project?
  \begin{todolist}
        \item 0
        \item 1
        \item 2
        \item 3
        \item 4
        \item 5
        \item 6
        \item 7
        \item 8
        \item 9
        \item 10
        \item 11
        \item 12
        \item 13
        \item 14
        \item 15
        \item 16
        \item 17
        \item 18
        \item 19
        \item 20
    \end{todolist}

\newpage

\textbf{Phase 1 Results}
{\hfill \textbf{Period 1}}

\bigskip
You have made a contribution of 4 points to the joint project. The common project is now worth 13 points.

\bigskip
Your income from the private account is 16 points and the one from the common project is 3 points.

\bigskip
So, in Phase 1 you get 19 points.
\newpage

\textbf{Phase 2 (Participant A)}

\rightline{\textbf{Period 1}}
\bigskip
You have 5 points available. Indicate if and how many points you want to transfer to Participant B.
  \begin{todolist}
        \item 0
        \item 1
        \item 2
        \item 3
        \item 4
        \item 5
    \end{todolist}

\newpage

\textbf{Phase 3 (Participant B)}

\rightline{\textbf{Period 1}}
\bigskip
Here is the decision made by the following Participant A: 1 point placed into the joint project and 2 points transferred to you. 

\bigskip
Do you want to accept the points transferred to you by this Participant A?
  \begin{todolist}
        \item Yes
        \item No
    \end{todolist}
Please indicate whether and how many points you want to deduct from Participant A.
  \begin{todolist}
        \item 0
        \item 1
        \item 2
        \item 3
        \item 4
        \item 5
        \item 6
        \item 7
        \item 8
        \item 9
        \item 10
    \end{todolist}

\bigskip
Here is the decision made by the following Participant A: 4 points placed into the joint project and 1 point transferred to you. 

\bigskip
Do you want to accept the points transferred to you by this Participant A?
  \begin{todolist}
        \item Yes
        \item No
    \end{todolist}
Please indicate whether and how many points you want to deduct from Participant A.
  \begin{todolist}
        \item 0
        \item 1
        \item 2
        \item 3
        \item 4
        \item 5
        \item 6
        \item 7
        \item 8
        \item 9
        \item 10
    \end{todolist}

\bigskip
Here is the decision made by the following Participant A: 3 points placed into the joint project and 1 point transferred to you. 

\bigskip
Do you want to accept the points transferred to you by this Participant A?
  \begin{todolist}
        \item Yes
        \item No
    \end{todolist}
Please indicate whether and how many points you want to deduct from Participant A.
  \begin{todolist}
        \item 0 
        \item 1
        \item 2
        \item 3
        \item 4
        \item 5
        \item 6
        \item 7
        \item 8
        \item 9
        \item 10
    \end{todolist}

\newpage

\textbf{Results (Participant B)}

\rightline{\textbf{Period 1}}
\bigskip

Participants A in the group made the following choices:
\begin{itemize}
    \item participate in the common project by contributing 1 point and transfer 2 points to you. You deducted 3 points from this participant.
    \item participate in the common project by contributing 4 points and transfer 1 point to you. You deducted 1 point from this participant.
    \item participate in the common project by contributing 3 points and transfer 1 point to you. You deducted 2 points from this participant.
\end{itemize}

\bigskip
You got 3 points from the joint project. Participant As transferred a total of 4 points to you and you accepted 4 points.
So, you got 26 points in this period.
\newpage

\textbf{Results (Participant A)}

\rightline{\textbf{Period 1}}
\bigskip

You contributed 3 points to the common project and transferred 1 point to Participant B. Participant B accepted the points you transferred and deducted 6 points. 

\bigskip
Other Participants A in the group made the following choices:
\begin{itemize}
    \item participate in the common project by contributing 1 point. Participant B deducted 9 points.
    \item participate in the common project by contributing 4 points. Participant B deducted 3 points.
\end{itemize}

\bigskip
You got 3 points from the common project. So, you got 18 points in this period.
\newpage

\subsubsection{Transparency Treatment}
The study will last between 12 and 15 periods. At the beginning of each period, every participant receives \textbf{20 points}. Each group is composed of 3 \textbf{Participant A's} and 1 \textbf{Participant B}. Roles are assigned randomly and remain the same throughout all periods. Each period is divided into \textbf{3 phases}.
\bigskip

\textbf{Phase 1}

\bigskip
If you are a \textbf{Participant A}, you must decide how to allocate the 20 points between two options: a \textbf{private account} and a \textbf{common project}. Any points not allocated to the common project are automatically credited to your private account.

\bigskip
\textbf{Your income from the private account}.
Every point credited to your private account provides you with one point of income.  
For example, if you allocate all 20 points to your private account (and contribute nothing to the common project), you earn exactly 20 points.  If you leave 6 points in your private account, your income from the private account is 6 points.

\bigskip
\textbf{Your income from the common project}.

Every member of the group (both Participant As and Participant B) equally benefits from the contributions made to the common project. Income from the common project is calculated as follows:

\begin{mdframed}
\begin{center}
  Income from the common project  
\end{center}
\par
 \begin{center}
     $=$
 \end{center}
\par
\begin{center}
    (1.6 × sum of all contributions to the common project) / number of participants
\end{center}
\end{mdframed}

For example, if the total contributions to the common project amount to 60 points (i.e., all Participant As contribute 20 points each), you and the other group members receive: (60 × 1.6) / 4 = 24 points.

If the total contributions to the common project amount to 20 points, you and the other group members receive: (20 × 1.6) / 4 = 8 points.

If no one contributes to the common project, no one receives any points from it.

\bigskip
The income from the common project is rounded down for decimal values less than 0.5, and rounded up for decimal values equal to or greater than 0.5.

\bigskip
\textbf{Income for Participant A after Phase 1}

\bigskip
\begin{mdframed}
\begin{center}
    Income from private account + Income from common project
\end{center}
 
\par
 \begin{center}
     $=$
 \end{center}
\par
\begin{center}
    20 - your contribution to the common project + [(1.6 × sum of all contributions to the common project) / 4]
\end{center}
\end{mdframed}
\bigskip

\textbf{If you are a Participant B}
You do not need to take any action during this Phase. 
\par
\bigskip

Click the "Next" button to proceed to the next part of the instructions. 

\textit{Note: The "Next" button will become available after 60 seconds.}

\newpage
\textbf{Phase 2}
\bigskip

\textbf{If you are a Participant A}. After Phase 1, you receive an additional allocation of 5 points.  

You can decide whether and how many of these points to transfer to Participant B. Any points not transferred to Participant B are added to your income.  

\bigskip
Each point you transfer to Participant B costs you 1 point and increases Participant B’s income by 1 point. For example, if you decide to transfer 3 points to Participant B, it will cost you 3 points, and Participant B’s income will increase by 3 points.

\bigskip
\textbf{Income for Participant A after Phase 2}. 

\begin{mdframed}
\begin{center}
    Income from private account + Income from common project + 5 points - points transferred to Participant B
\end{center}
 
\par
 \begin{center}
     $=$
 \end{center}
\par
\begin{center}
    20 - contribution to the common project + income from the common project + 5 - points transferred to Participant B
\end{center}
\end{mdframed}
\bigskip

\textbf{If you are a Participant B}. You do not need to take any action during Phase 2.
\par
\bigskip

Click the "Next" button to proceed to the next part of the instructions.  

\textit{Note: The "Next" button will become available after 30 seconds}
\newpage

\textbf{Phase 3}

\bigskip
\textbf{If you are a Participant A}. You do not need to take any action during Phase 3.

\bigskip
\textbf{If you are a Participant B}. You will be shown the contributions made by each Participant A to the common project, along with the number of points each Participant A has transferred to you.  

\textbf{You must decide whether to accept or reject the points transferred to you by each Participant A}.

\bigskip
If you accept, the points will be credited to your account. If you reject, the points will be returned to the respective Participant A.

\bigskip
Note: The order in which Participant A's are displayed is random. This means that the decisions shown in different periods do not necessarily correspond to the same Participant A.
\bigskip

In addition, if you are a Participant B, \textbf{you receive another 15 points
(which we will call deduction points)}. These points can be used to \textbf{reduce
the earnings of the Participant As in your group}.

\bigskip
You must decide whether and how many \textbf{deduction points} to assign to the
A participants in your group.

For each deduction point you assign, you pay \textbf{1 point}, and the income of
the corresponding Participant A is reduced by \textbf{3 points}.

\bigskip
Example: If you assign 3 deduction points to a Participant A, their income will decrease by \(3 \times 3 = 9\) points, while your income will decrease by 3 points.

\bigskip
Although you have 15 deduction points, \textbf{you can subtract a maximum of 10 points from each participant}.

\bigskip
Any unused deduction points will be added to your income for the round.

\bigskip
\textbf{Income for Participant B after Phase 3}. 

\begin{mdframed}
\begin{center}
    Income for Participant B
\end{center}
 
\par
 \begin{center}
     $=$
 \end{center}
\par
\begin{center}
    20 + income from the common project + points transferred by Participant As + unused deduction points - total deduction points assigned to Participant As
\end{center}
\end{mdframed}
\bigskip

\textbf{Information}

\bigskip
\textbf{If you are a Participant A}, before making your decision in Phase 1, you can purchase \textbf{information} about Participant B, specifically, the decisions Participant B made in the previous period regarding the other Participants A in the group (e.g., whether any points transferred by Participant A members were accepted or rejected, deduction points assigned, and contributions from Participant A members to the common project). The cost of this information is \textbf{3 points} per period. You can start purchasing this information from the second period onward.

\bigskip
Click the "Next" button to proceed to the next part of the instructions.  

\textit{Note: The "Next" button will become available after 30 seconds.}

\newpage
\textbf{Summary of Participant A and Participant B Income }

\bigskip
\textbf{For Participant A}, your income for the period is calculated as follows:

\begin{mdframed}
\begin{center}
Income of Participant A for the period
\end{center}
\par
 \begin{center}
     $=$
 \end{center}
\par
\begin{center} 20 - cost of information - contribution to the common project + income from the common project + 5 - points transferred to Participant B - 3 × (deduction points assigned by Participant B)
\end{center}
 
\end{mdframed}

\textbf{For Participant B}, your income for the period is calculated as follows:
\begin{mdframed}
\begin{center}
Income of Participant B for the period \end{center}
\par
 \begin{center}
     $=$
 \end{center}
\par
\begin{center}
20 + income from the common project + sum of accepted transfers + 15 - total deduction points assigned to Participant A's
\end{center}
 \end{mdframed}

\bigskip

At the end of each round, you will be shown a summary of your decisions and those made by the other Participant As in the group (your income for the round, the contributions to the common project by the other Participant As, and the deduction points assigned to Participant As). The order in which the Participant As are shown is random. This means that the decisions shown in different rounds do not necessarily correspond to the same Participant A. Participant B will also be shown a summary of their own decisions and those made by the other Participant As in the group (their income for the round, the contributions to the common project by the other Participant As, the deduction points assigned to Participant As, and the points transferred by the Participant As).

\bigskip
You will now be presented with a quiz. Please answer the quiz questions by entering your responses on the computer. You will always have access to a summary of the instructions by clicking on the \textbf{"Instructions"} button.

\newpage
\subsubsection{Exit Treatment}
The study will last between 12 and 15 periods. At the beginning of each period, every participant receives \textbf{20 points}. Each group is composed of 3 \textbf{Participant A's} and 1 \textbf{Participant B}. Roles are assigned randomly and remain the same throughout all periods. Each period is divided into \textbf{3 phases}.
\bigskip

\textbf{Phase 1}

\bigskip
If you are a \textbf{Participant A}, you must decide how to allocate the 20 points between two options: a \textbf{private account} and a \textbf{common project}. Any points not allocated to the common project are automatically credited to your private account.

\bigskip
\textbf{Your income from the private account}.

Every point credited to your private account provides you with one point of income.  
For example, if you allocate all 20 points to your private account (and contribute nothing to the common project), you earn exactly 20 points. If you leave 6 points in your private account, your income from the private account is 6 points.

\bigskip
\textbf{Your income from the common project}.

Every member of the group (both Participant As and Participant B) equally benefits from the contributions made to the common project. Income from the common project is calculated as follows:

\begin{mdframed}
\begin{center}
    Income from the common project
\end{center}
\par
 \begin{center}
     $=$
 \end{center}
\par
\begin{center}
    (1.6 × sum of all contributions to the common project) / number of participants
\end{center}
\end{mdframed}

\bigskip
For example, if the total contributions to the common project amount to 60 points (i.e., all Participant As contribute 20 points each), you and the other group members receive: (60 × 1.6) / 4 = 24 points.

If the total contributions to the common project amount to 20 points, you and the other group members receive: (20 × 1.6) / 4 = 8 points.

If no one contributes to the common project, no one receives any points from it.

\bigskip

The income from the common project is rounded down for decimal values less than 0.5, and rounded up for decimal values equal to or greater than 0.5

\bigskip
\textbf{Income for Participant A after Phase 1}

\begin{mdframed}
\begin{center}
    Income from private account + Income from common project
\end{center}
 
\par
 \begin{center}
     $=$
 \end{center}
\par
\begin{center}
    20 - your contribution to the common project + [(1.6 × sum of all contributions to the common project) / 4]
\end{center}
\end{mdframed}
\bigskip

\textbf{If you are a Participant B}
You do not need to take any action during this Phase. 
\par
\bigskip

Click the "Next" button to proceed to the next part of the instructions. 

\textit{Note: The "Next" button will become available after 60 seconds.}

\newpage
\textbf{Phase 2}
\bigskip

\textbf{If you are a Participant A}. After Phase 1, you receive an additional allocation of \textbf{5 points}.  

You can decide whether and how many of these points to transfer to Participant B. Any points not transferred to Participant B are added to your income.  

\bigskip
Each point you transfer to Participant B costs you 1 point and increases Participant B’s income by 1 point.  For example, if you decide to transfer 3 points to Participant B, it will cost you 3 points, and Participant B’s income will increase by 3 points.

\bigskip
\textbf{Income for Participant A after Phase 2}. 

\begin{mdframed}
\begin{center}
    Income from private account + Income from common project + 5 points - points transferred to Participant B
\end{center}
 
\par
 \begin{center}
     $=$
 \end{center}
\par
\begin{center}
    20 - contribution to the common project + income from the common project + 5 - points transferred to Participant B
\end{center}
\end{mdframed}
\bigskip

\textbf{If you are a Participant B}. You do not need to take any action during Phase 2.
\par
\bigskip

Click the "Next" button to proceed to the next part of the instructions.  

\textit{Note: The "Next" button will become available after 30 seconds}
\newpage

\textbf{Phase 3}

\bigskip
\textbf{If you are a Participant A}. You do not need to take any action during Phase 3.

\bigskip
\textbf{If you are a Participant B}. You will be shown the contributions made by each Participant A to the common project, along with the number of points each Participant A has transferred to you.  

\textbf{You must decide whether to accept or reject the points transferred to you by each Participant A}.

\bigskip
If you accept, the points will be credited to your account. If you reject, the points will be returned to the respective Participant A.

\bigskip
Note: The order in which Participant A's are displayed is random. This means that the decisions shown in different periods do not necessarily correspond to the same Participant A.

\bigskip

In addition, if you are a Participant B, \textbf{you receive another 15 points (which we will call deduction points)}. These points can be used to \textbf{reduce the earnings of the Participant As in your group}.

\bigskip
You must decide whether and how many \textbf{deduction points} to assign to the
A participants in your group.

For each deduction point you assign, you pay \textbf{1 point}, and the income of
the corresponding Participant A is reduced by\textbf{3 points}.

\bigskip
Example: if you assign 3 deduction points to a Participant A, their income will decrease by \(3 \times 3 = 9\) points, while your income will decrease by 3 points.

\bigskip
Although you have 15 deduction points, \textbf{you can subtract a maximum
of 10 points from each participant}.

\bigskip
Any unused deduction points will be added to your income for the round.

\bigskip
\textbf{Income for Participant B after Phase 3}. 

\begin{mdframed}
\begin{center}
Income of Participant B
\end{center}
\par
 \begin{center}
     $=$
 \end{center}
\par
\begin{center}
     20 + income from the common project + points transferred by Participant As + unused deduction points - total deduction points assigned to Participant As
\end{center}
   
\end{mdframed}
\bigskip

\textbf{Information} 

\bigskip
\textbf{If you are a Participant A}, before making your decision in Phase 1, you can purchase \textbf{information} about Participant B, specifically, the decisions Participant B made in the previous period regarding the other Participants A in the group (e.g., whether any points transferred by Participant A members were accepted or rejected, deduction points assigned, and contributions from Participant A members to the common project). The cost of this information is \textbf{3 points} per period. You can start purchasing this information from the second period onward.
\bigskip

Additionally, in Phase 1, you can decide to \textbf{not participate in the common project}. If you do not participate in the common project, you will not receive income from the common project, and the common project will be divided among the Participant A members who do participate and Participant B. If a Participant A leaves the group, Participant B has 10 deduction points available instead of 15. 

If two or more Participant A members do not participate in the common project, the common project will not be carried out, and all Participant A members will retain their initial endowment (20 + 5). If a Participant A had transferred points to Participant B, the transfer will not occur, and the points will be returned to Participant A. Participant B will retain their initial endowment (20).

\bigskip
Click the "Next" button to proceed to the next part of the instructions.  

\textit{Note: The "Next" button will become available after 30 seconds.}

\newpage
\textbf{Summary of Participant A and Participant B Income }

\bigskip
\textbf{For Participant A who participates in the common project.}
Your income for the period is calculated as follows:

\begin{mdframed}
\begin{center}
Income of Participant A for the period
 \end{center}
\par
 \begin{center}
     $=$
 \end{center}
\par
\begin{center}
20 - cost of information - contribution to the common project + income from the common project + 5 - points transferred to Participant B - 3 × (deduction points assigned by Participant B)
\end{center}
 
\end{mdframed}

\textbf{For Participant A who doesn't participate in the common project.}
Your income for the period is calculated as follows:

\begin{mdframed}
\begin{center}
Income of Participant A for the period 
\end{center}
\par
 \begin{center}
     $=$
 \end{center}
\par
\begin{center}
20 - cost of information + 5
\end{center}
 
\end{mdframed}

\textbf{For Participant B.} Your income for the period is calculated as follows:
\begin{mdframed}
\begin{center}
Income of Participant B for the period 
\end{center}
\par
 \begin{center}
     $=$
 \end{center}
\par
\begin{center}
20 + income from the common project + sum of accepted transfers + 15 - total deduction points assigned to Participant A's
\end{center}
 \end{mdframed}

\textbf{For Participant B if the common project is not realized.} Your income for the period is calculated as follows:
\begin{mdframed}
\begin{center}
Income of Participant B for the period 
\end{center}
\par
 \begin{center}
     $=$
 \end{center}
\par
\begin{center}
20 
\end{center}
 \end{mdframed}

\bigskip
At the end of each round, Participant As will be shown a summary of their own decisions and those made by the other Participant As in the group (their income for the round, the contributions to the common project by the other Participant As, and the deduction points assigned to Participant As). The order in which the Participant As are shown is random. This means that the decisions displayed in different rounds do not necessarily correspond to the same Participant A. Participant B will also be shown a summary of their own decisions and those made by the other Participant As in the group (their income for the round, the contributions to the common project by the other Participant As, the deduction points assigned to Participant As, and the points transferred by Participant As).

\bigskip
You will now be presented with a quiz. Please answer the quiz questions by entering your responses on the computer. You will always have access to a summary of the instructions by clicking on the \textbf{"Instructions"} button.

\newpage
\subsubsection{Distribution Game}
\textbf{Instructions}

In this part of the study, new groups are created. You will find three choices indicating the possible points to allocate to the four members of your group.

Please note that the total number of points to be allocated to the group does not change across the three choices. Only the number of points assigned to each member varies.

The possible choices are the following:

\begin{figure}[h]
\centering
    \includegraphics[width=0.49\linewidth]{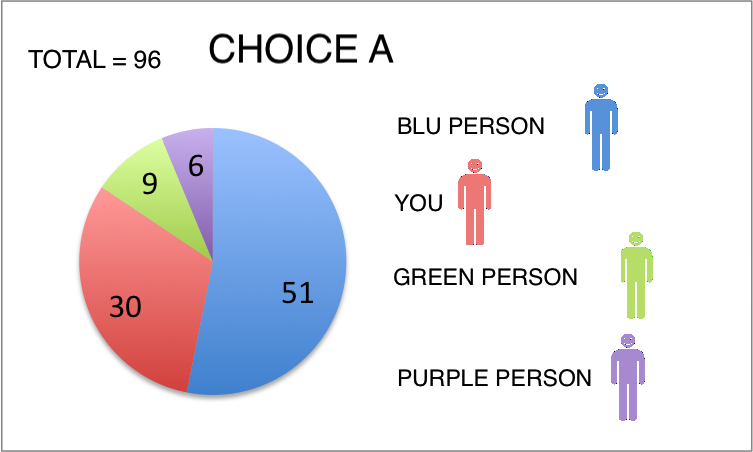}
    \includegraphics[width=0.49\linewidth]{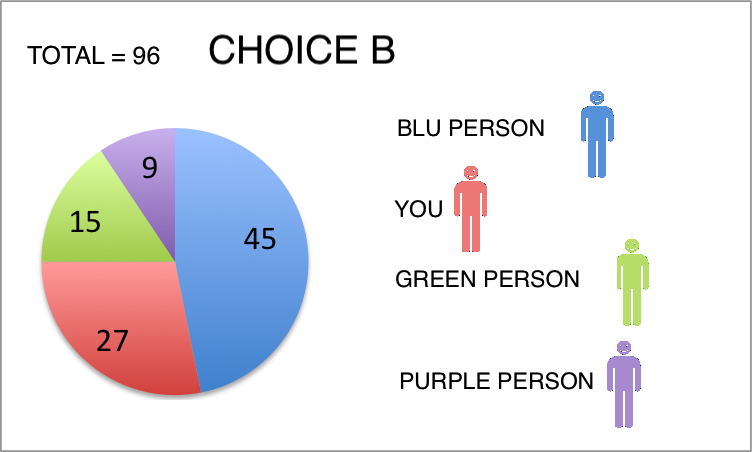}\\
    \includegraphics[width=0.49\linewidth]{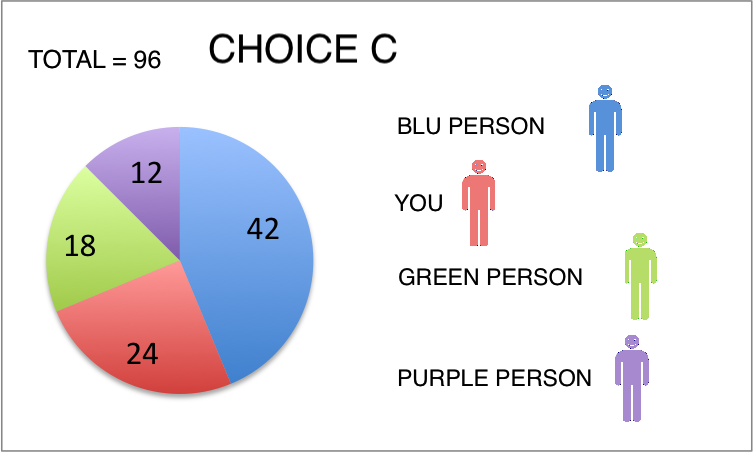}
\end{figure}

\textbf{What do you need to do?} You need to indicate which choice you prefer.

\bigskip
\textbf{How is your compensation determined?} Once everyone has made their choice, the computer will randomly select one person to decide the compensation for all the members of your group.
\bigskip

\textbf{For example, if you are selected, your choices will determine the compensation for all the members of your group.} 

\bigskip
The selected person, in this case, you, will receive the compensation marked in RED with the label YOU, while the other members of the group will receive the number of points indicated in the chart, depending on whether they are assigned as the BLUE, GREEN, or PURPLE person.

For instance, if you choose Option A and are selected to decide the compensation for your group, you will receive 30 points, and the others will receive 51, 9, and 6 points, depending on whether the computer assigns them as the BLUE, GREEN, or PURPLE person.

\bigskip
\textbf{If, for example, you are NOT selected, someone else's choices will determine your compensation and that of all other group members.}

\bigskip
Your earnings will correspond to the points allocated to the BLUE, GREEN, or PURPLE person, depending on the computer's selection.

For example, if the person selected by the computer to decide the compensation for the group chooses Option C, that person receives 24 points. If the computer assigns you as the PURPLE person, you will receive 12 points, while the other two group members receive 18 and 42 points.

\bigskip
The instructions will remain available on the screen throughout this part of the study.

\bigskip
Before making your choice, we ask you to answer two questions that will not affect your final compensation but will help you confirm your understanding of the instructions.

\bigskip
Click "Next" when you are ready to continue.

\textit{Note: The "Next" button will become available after 60 seconds.}
\newpage
\textbf{Control Questions}

\begin{itemize}
    \item[1] Imagine you chose B. If you are selected by the computer to decide the compensation for everyone, you will earn: 
     \begin{todolist}
        \item 15 points
        \item 27 points
    \end{todolist}
    \item[2] Imagine you chose A. Another member of your group is selected to decide the compensation for everyone, and this person has chosen C. In this case, you will earn:
    \begin{todolist}
        \item 42, 18, or 12 points, depending on the color the computer randomly selects and assigns to you
        \item 45, 15, or 9 points, depending on the color the computer randomly selects and assigns to you

    \end{todolist}
\end{itemize}
\begin{figure}[h]
\centering
\includegraphics[width=0.49\linewidth]{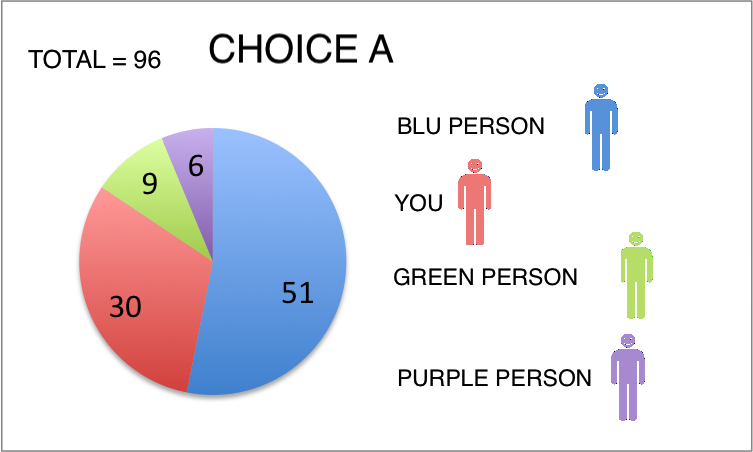}
    \includegraphics[width=0.49\linewidth]{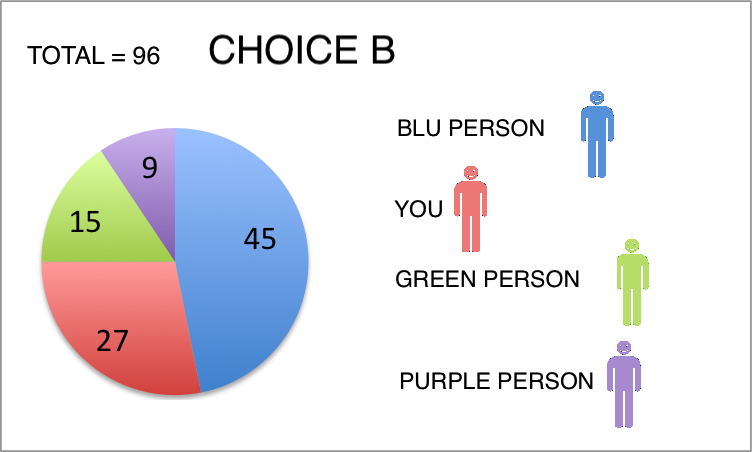}\\
    \includegraphics[width=0.49\linewidth]{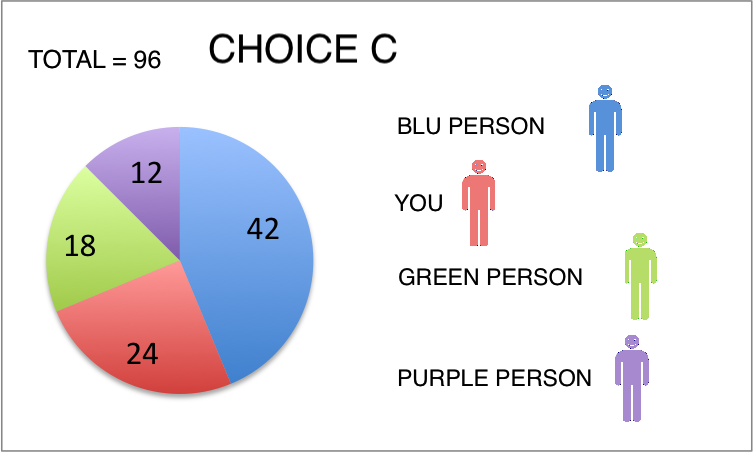}
\end{figure}

\newpage
\textbf{Let’s start} 

\bigskip
Now you are ready to start! The decisions you will make in the next pages will have an impact on your final income.
\newpage

\textbf{Decision}
Now you can make your choice:
\begin{figure}[h]
 \centering   \includegraphics[width=0.49\linewidth]{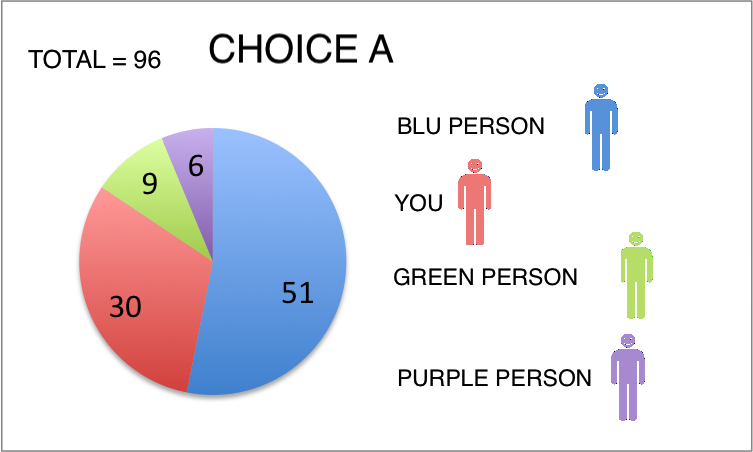}
    \includegraphics[width=0.49\linewidth]{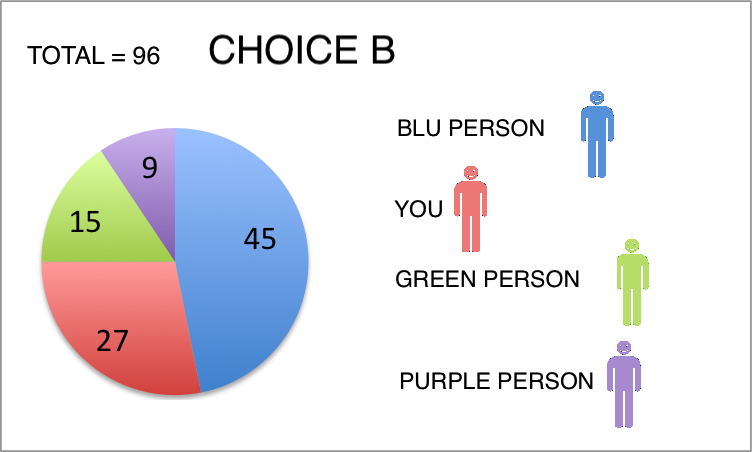}\\
    \includegraphics[width=0.49\linewidth]{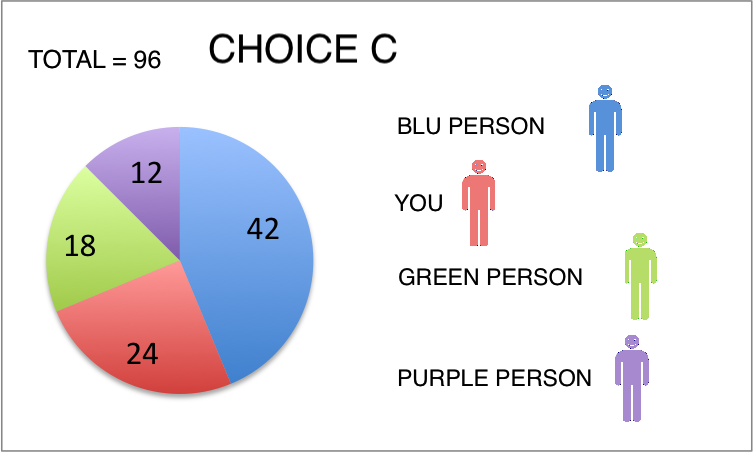}
\end{figure}

 \begin{todolist}
        \item Choice A
        \item Choice B
        \item Choice C

    \end{todolist}

\newpage
\subsubsection{Cognitive Reflection Test (CRT)}
In this part of the study, you will need to answer some questions. Your score will depend on the number of correct answers. For each correct answer, you will earn 7 experimental points (equivalent to €0.21). When you are ready, click the "Next" button.

\bigskip
\bigskip 

\begin{itemize}
    \item[1] If you flipped a fair coin three times, what is the probability that it would land ‘Heads’ at least once? The answer must be expressed as a PERCENTAGE
    \item[2] If three workers can wrap three toys in one hour, how many workers are needed to wrap six toys in two hours? The answer must be expressed as WORKERS
    \item[3] Giulia and Sara are running around a track. They run equally fast but Giulia started later. When Giulia has run 5 laps, Sara has run 15 laps. When Giulia has run 30 laps, how many has Sara run? The answer must be expressed as LAPS
    \item[4] Marco received both the 15th highest and the 15th lowest mark in the class. How many students are in the class?The answer must be expressed as STUDENTS
    \item[5] In an athletics team, tall members are three times more likely to win a medal than short members. This year, the team has won 60 medals so far. How many of these have been won by short athletes? The answer must be expressed as MEDALS
\newpage
\end{itemize}

\subsubsection{Interpersonal Reactivity Index (IRI)}
In this part of the study, you will need to answer some questions. When you are ready, click the "Next" button.
\bigskip
\bigskip

Please answer the following questions by indicating a value from 1 "never true" to 5 "always true”

\begin{itemize}
    \item [1] I daydream and fantasize quite regularly about things that might happen to me
     \begin{todolist}
        \item 1  
        \item 2
        \item 3
        \item 4 
        \item 5
    \end{todolist}
    \item [2] I often feel tenderness and concern for people less fortunate than I am
     \begin{todolist}
        \item 1  
        \item 2
        \item 3
        \item 4 
        \item 5
    \end{todolist}
    \item [3] Sometimes I find it difficult to see things from another person's point of view
         \begin{todolist}
        \item 1  
        \item 2
        \item 3
        \item 4 
        \item 5
    \end{todolist}
    \item [4] Sometimes I don't feel much sorry for other people who have problems
     \begin{todolist}
        \item 1  
        \item 2
        \item 3
        \item 4 
        \item 5
    \end{todolist}
    \item [5] I get truly involved in the moods of the main characters in a story
     \begin{todolist}
        \item 1  
        \item 2
        \item 3
        \item 4 
        \item 5
    \end{todolist}
    \item [6] In emergency situations, I feel apprehensive and uncomfortable
     \begin{todolist}
        \item 1  
        \item 2
        \item 3
        \item 4 
        \item 5
    \end{todolist}
    \item [7] I can usually be objective when watching a movie or a play and I rarely get completely involved
         \begin{todolist}
        \item 1  
        \item 2
        \item 3
        \item 4 
        \item 5
    \end{todolist}
    \item [8] In case of disagreement, I try to take everyone's point of view into account before making a decision
     \begin{todolist}
        \item 1  
        \item 2
        \item 3
        \item 4 
        \item 5
    \end{todolist}
    \item [9] When I see someone being exploited, I have protective feelings toward them
     \begin{todolist}
        \item 1  
        \item 2
        \item 3
        \item 4 
        \item 5
    \end{todolist}
    \item [10] Sometimes I feel helpless when I find myself in very emotionally-involving situations
     \begin{todolist}
        \item 1  
        \item 2
        \item 3
        \item 4 
        \item 5
    \end{todolist}
    \item [11] Sometimes I try to understand my friends better by imagining how things may appear from their perspective
     \begin{todolist}
        \item 1  
        \item 2
        \item 3
        \item 4 
        \item 5
    \end{todolist}
    \item [12] I rarely get really involved with a good book or a good movie
     \begin{todolist}
        \item 1  
        \item 2
        \item 3
        \item 4 
        \item 5
    \end{todolist}
    \item [13] When I see someone getting hurt, I tend to stay calm
     \begin{todolist}
        \item 1  
        \item 2
        \item 3
        \item 4 
        \item 5
    \end{todolist}
    \item [14] Sometimes other people's misfortunes don't upset me much
     \begin{todolist}
        \item 1  
        \item 2
        \item 3
        \item 4 
        \item 5
    \end{todolist}
    \item [15] If I'm sure I'm right about something, I don't waste much time listening to other people's arguments
     \begin{todolist}
        \item 1  
        \item 2
        \item 3
        \item 4 
        \item 5
    \end{todolist}
    \item [16] After watching a play or a movie, I feel as if I am one of the protagonists myself
     \begin{todolist}
        \item 1  
        \item 2
        \item 3
        \item 4 
        \item 5
    \end{todolist}
    \item [17] Finding myself in situations that cause emotional tension scares me
     \begin{todolist}
        \item 1  
        \item 2
        \item 3
        \item 4 
        \item 5
    \end{todolist}
    \item [18] When I see someone being treated unfairly, I sometimes don’t feel so much pity for them
         \begin{todolist}
        \item 1  
        \item 2
        \item 3
        \item 4 
        \item 5
    \end{todolist}
    \item [19] I am usually quite efficient in dealing with emergency situations
         \begin{todolist}
        \item 1  
        \item 2
        \item 3
        \item 4 
        \item 5
    \end{todolist}
    \item [20] Things that happen affect me quite often
     \begin{todolist}
        \item 1  
        \item 2
        \item 3
        \item 4 
        \item 5
    \end{todolist}
    \item [21] I believe that there are two opposite sides to every story and I try to take both into consideration
     \begin{todolist}
        \item 1  
        \item 2
        \item 3
        \item 4 
        \item 5
    \end{todolist}
    \item [22] I could describe myself as a person with a rather tender heart
         \begin{todolist}
        \item 1  
        \item 2
        \item 3
        \item 4 
        \item 5
    \end{todolist}
    \item [23] When I watch a good movie, I can very easily put myself in the shoes of a main character
     \begin{todolist}
        \item 1  
        \item 2
        \item 3
        \item 4 
        \item 5
    \end{todolist}
    \item [24] I tend to lose control in emergency cases
     \begin{todolist}
        \item 1  
        \item 2
        \item 3
        \item 4 
        \item 5
    \end{todolist}
    \item [25] When I am in conflict with someone, I usually try to put myself in their shoes for a moment
     \begin{todolist}
        \item 1  
        \item 2
        \item 3
        \item 4 
        \item 5
    \end{todolist}    
    \item [26] When I read an interesting story or tale, I imagine how I would feel if those events were happening to me
     \begin{todolist}
        \item 1  
        \item 2
        \item 3
        \item 4 
        \item 5
    \end{todolist}
    \item [27] When I see someone who urgently needs help in an emergency situation, I break down
     \begin{todolist}
        \item 1  
        \item 2
        \item 3
        \item 4 
        \item 5
    \end{todolist}
    \item [28] Before criticizing someone, I try to imagine how I would feel if I were in their place
     \begin{todolist}
        \item 1  
        \item 2
        \item 3
        \item 4 
        \item 5
    \end{todolist}

\end{itemize}
\newpage
\subsubsection{Exit Survey}
Please answer the following questions.
\begin{itemize}
    \item[1] Age
    \item[2] Gender
         \begin{todolist}
        \item Male  
        \item Female
        \item I don’t identify with either of them
        \item I prefer not to answer

    \end{todolist}
    \item[3] Current residence
    \item[4] Place of birth
    \item[5] Highest education level
     \begin{todolist}
         \item Phd  
        \item Master's Degree
        \item Bachelor's Degree
        \item High school diploma 
        \item Middle school
        \item Elementary school
    \end{todolist}
    \item[6] Employment status
     \begin{todolist}
        \item Employed  
        \item Unemployed
        \item Student
        \item Retired
        \item Not looking for a job
    \end{todolist}
    \item[7] Where did you attend elementary school?
    \item[8] In general, would you say people can be trusted or that you should be very careful?
Please indicate a value from 1 to 6, where 1 means you should be very careful and 6 means you can trust people
     \begin{todolist}
        \item 1  
        \item 2
        \item 3
        \item 4 
        \item 5
        \item 6
    \end{todolist}
    \item[9] Have you ever participated in a project of an association or your community without deriving any profit from it?
    \begin{todolist}
        \item Yes  
        \item No
    \end{todolist}
    \item[10]Did you vote in the last election?
    \begin{todolist}
        \item Yes  
        \item No
        \item I didn't meet the requirements or couldn't do it because of my age
    \end{todolist}
    \item[11] Are you a blood donor?
   \begin{todolist}
        \item Yes  
        \item No
        \item I can't
    \end{todolist}
    \item[12] Are you on the organ donation list?
   \begin{todolist}
        \item Yes  
        \item No
        \item I don't know
    \end{todolist}
    \item[13] For each of the following actions, indicate a value from 1 to 6, where 1 means people can never be justified and 6 means people can always be justified
    \begin{itemize}
        \item Applying for government benefits you are not entitled to
           \begin{todolist}
        \item 1  
        \item 2
        \item 3
        \item 4 
        \item 5
        \item 6
    \end{todolist}
        \item Avoiding paying the bus fare
          \begin{todolist}
        \item 1  
        \item 2
        \item 3
        \item 4 
        \item 5
        \item 6
    \end{todolist}
        \item Not paying taxes if you have the chance
          \begin{todolist}
        \item 1  
        \item 2
        \item 3
        \item 4 
        \item 5
        \item 6
    \end{todolist}
        \item Accepting a bribe in the work environment
          \begin{todolist}
        \item 1  
        \item 2
        \item 3
        \item 4 
        \item 5
        \item 6
    \end{todolist}
    \end{itemize}

    \item[14] How widespread do you think corruption is in this country?
      \begin{todolist}
        \item Hardly any of the public offices are corrupt
        \item Few public officials are corrupt
        \item Most public officials are corrupt
        \item Almost all public officials are corrupt
    \end{todolist}

\end{itemize}
\newpage
\textbf{Final Results}
\bigskip

In the first part of the study, you scored XXX points.
\bigskip

In the second part, you scored XXX points.
\bigskip

In the third part, you scored XXX points.

\bigskip

Converting these points into euros and adding € 3.00 that you get for participating, you earn € XX.00.
\bigskip

Thank you for participating!

\clearpage

\subsection{Experimental Instructions (Italian)}

\textbf{Benvenuto al VERA Lab!}
Ciao, benvenuto al VERA Lab. Ti ringraziamo per la tua partecipazione allo studio.

Prima di cominciare, ti chiediamo gentilmente di spegnere il telefono cellulare. Da questo momento in poi non è consentita alcuna forma di comunicazione con i partecipanti presenti alla sessione né con l'esterno. 

Se hai domande da fare durante lo svolgimento dello studio, non esitare a chiamarci: chiariremo qualsiasi tuo dubbio. Ti è consentito l'uso di carta e penna per eventuali annotazioni o calcoli. 

È molto importante seguire attentamente le istruzioni.  

Durante lo studio ti verrà richiesto di prendere delle decisioni: ogni decisione può essere cambiata quante volte si vuole, prima di premere il pulsante "Successivo" in fondo alla pagina. Dopo aver cliccato sul pulsante "Successivo", l'ultima decisione selezionata diventa definitiva.

Ti ricordiamo che le scelte che effettui sono anonime. Durante lo studio farai parte di un gruppo, la cui composizione resterà anonima: non sarai in grado di sapere chi fa parte del tuo gruppo, così come gli altri non sapranno chi è nel loro. 

Alla fine dello studio, è previsto un compenso economico. Riceverai 3 euro per la tua partecipazione. Inoltre, riceverai un guadagno aggiuntivo, che dipenderà dalle scelte che effettuerai e ti sarà comunicato al termine dello studio.   

Per lo studio avrai a disposizione dei punti sperimentali. I punti che ottieni durante lo studio verranno convertiti in euro. 100 punti sperimentali equivalgono a 3 euro.  

In questo studio puoi guadagnare fino a 25 euro.

Clicca sul bottone "Successivo" quando sei pronto per iniziare. 

\newpage

Nella prossima pagina ti verranno presentate le istruzioni della prima parte dello studio. Dopo aver letto le istruzioni dovrai rispondere a delle domande di comprensione, che non avranno effetto sul tuo compenso finale, ma ti serviranno per verificare di aver compreso la struttura del gioco.
\newpage

\subsubsection{Bribe Game Full - Buffat and Senn Replication}
Lo studio durerà  tra i 12 e i 15 periodi. All'inizio di ogni periodo ogni partecipante riceve \textbf{20 punti}. Ogni gruppo è composto da 3 \textbf{Partecipanti A} e 1 \textbf{Partecipante B}. L'assegnazione dei ruoli avviene in maniera casuale e rimane la stessa per tutti i periodi. Ogni periodo è diviso in \textbf{3 fasi}.
\bigskip

\textbf{Fase 1}

\bigskip
Se sei un \textbf{Partecipante A}, devi decidere come allocare i 20 punti tra le diverse opzioni: un \textbf{conto privato} e un \textbf{progetto comune}. Ogni punto non allocato al progetto comune è accreditato automaticamente al tuo conto privato. 

\bigskip
\textbf{Il tuo reddito dal conto privato.} 
Ogni punto che viene accreditato nel tuo conto privato ti darà un punto. Per esempio, se lasci 20 punti nel tuo conto privato (quindi non contribuisci al progetto comune), ottieni esattamente 20 punti. Se lasci 6 punti nel conto privato, il tuo reddito dal conto privato è di 6 punti. 

\bigskip    
\textbf{Il tuo reddito dal progetto comune}. Ogni membro del gruppo (Partecipante A e Partecipante B) beneficia in misura eguale delle contribuzioni derivanti dal progetto comune. Il reddito dal progetto comune è determinato come segue:
\begin{mdframed}
\begin{center}
  Reddito da progetto comune  
\end{center}\par
 \begin{center}
     $=$
 \end{center}
\par
\begin{center}
   (1.6 x somma di tutte le contribuzioni al progetto comune) / numero dei partecipanti 
\end{center} 
\end{mdframed}

Per esempio, se la somma di tutte le contribuzioni al progetto è di 60 punti (cioè tutti i partecipanti A contribuiscono 20 punti), tu e gli altri partecipanti del gruppo ricevete: (60 x 1.6) / 4 = 24 punti. 

Se la somma delle contribuzioni al progetto comune è di 20 punti,  tu e gli altri partecipanti del gruppo ricevete: (20 x 1.6) / 4 = 8 punti.

Se nessuno contribuisce al progetto comune, nessuno riceve punti dal progetto comune.

Il reddito da progetto comune viene automaticamente arrotondato per difetto per valori decimali minori di 0.5 e arrotondato per eccesso per valori decimali maggiori o uguali a 0.5. 

\bigskip
\textbf{Il reddito dei Partecipanti A dopo la Fase 1}
\begin{mdframed}
\begin{center}
    Reddito dal conto privato + reddito dal progetto comune  
\end{center}
\par
 \begin{center}
     $=$
 \end{center}
\par
\begin{center}
     20 - il tuo contributo al progetto comune + [(1.6 x somma di tutte le contribuzioni al progetto comune) / 4 ]
\end{center}
       
\end{mdframed}

\textbf{Se sei un Partecipante B,} non devi fare nulla nella Fase 1.  

\bigskip
Clicca il pulsante "Successivo" per accedere alla prossima parte delle istruzioni.

\textit{Nota: il pulsante "Successivo" apparirà dopo 60 secondi.}

\newpage
\textbf{Fase 2}

\bigskip
\textbf{Se sei un Partecipante A}, dopo la Fase 1 ottieni una dotazione addizionale di 5 punti. 

Puoi decidere se e quanti punti trasferire al Partecipante B. I punti che non vengono trasferiti al Partecipante B, sono aggiunti al tuo reddito.

\bigskip
Ogni punto trasferito al Partecipante B, costa a te 1 punto e accresce il reddito del Partecipante B di 1 punto. Se, per esempio, decidi di trasferire 3 punti al Partecipante B, ti costerà 3 punti e il reddito del Partecipante B aumenterà di 3 punti.

\bigskip
\textbf{Il reddito dei Partecipanti A dopo la Fase 2}
\begin{mdframed}
\begin{center}
    Reddito dal conto privato + reddito dal progetto comune + 5 punti - punti trasferiti al Partecipante B 
\end{center}
 
\par
 \begin{center}
     $=$
 \end{center}
\par
\begin{center}
    20 - contribuzione al progetto comune + reddito dal progetto comune + 5 - punti trasferiti al Partecipante B 
\end{center}
\end{mdframed}
\bigskip

\textbf{Se sei un Partecipante B,} non devi fare nulla nella Fase 2. 

\bigskip
Clicca il pulsante "Successivo" per accedere alla prossima parte delle istruzioni.

\textit{Nota: il pulsante "Successivo" apparirà dopo 30 secondi.}

\newpage
\textbf{Fase 3}

\bigskip
 \textbf{Se sei un Partecipante A,} non devi fare nulla nella Fase 3. 

 \bigskip
 \textbf{Se sei un Partecipante B,} ti verranno mostrate le contribuzioni al progetto comune dei Partecipanti A insieme al numero di punti che ogni Partecipante A ti ha trasferito. \textbf{Devi decidere se accettare o rifiutare i punti trasferiti a te da ogni Partecipante A}.

\bigskip
Se accetti un trasferimento, i punti verranno accreditati sul tuo conto. Se rifiuti, i punti verranno restituiti al Partecipante A.   

\bigskip
Importante: l'ordine in cui i Partecipanti A sono mostrati è casuale. Questo significa che le decisioni mostrate nei diversi periodi non coincidono necessariamente con lo stesso Partecipante A. 
    
\bigskip
Inoltre, se sei un Partecipante B, \textbf{ottieni altri 15 punti (che chiameremo punti di deduzione)}. Questi punti possono essere usati per \textbf{ridurre i redditi dei Partecipanti A nel tuo gruppo}. 

   \bigskip
Devi decidere se e quanti \textbf{punti di deduzione} assegnare ai partecipanti A del tuo gruppo. 

Per ogni punto di deduzione che assegni, devi pagare \textbf{1 punto}. Per ogni punto di deduzione assegnato, il reddito del Partecipante A verrà ridotto di \textbf{3 punti}. 

\bigskip
Esempio: se assegni 3 punti di deduzione, il reddito del Partecipante A sarà ridotto di 3 x 3 = 9 punti, mentre il tuo reddito del sarà ridotto di 3 punti. 

\bigskip
 Sebbene hai 15 punti di deduzione, \textbf{a ogni partecipante puoi togliere al massimo 10 punti.}
 
\bigskip
I punti di deduzione non usati saranno accreditati al tuo reddito del periodo.

\bigskip
\textbf{Il reddito del Partecipante B dopo la Fase 3 }

\begin{mdframed}
\begin{center}
    Reddito del periodo Partecipante B
\end{center}
 
\par
 \begin{center}
     $=$
 \end{center}
\par
\centering
    20 + reddito dal progetto comune + punti trasferiti dai Partecipanti A + punti di deduzione - somma dei punti di deduzione assegnati ai Partecipanti A
\end{mdframed}

\bigskip
Clicca il pulsante "Successivo" per accedere alla prossima parte delle istruzioni.

\textit{Nota: il pulsante "Successivo" apparirà dopo 30 secondi.}

\newpage
\textbf{Riassunto del reddito dei Partecipanti A e Partecipante B}

    \bigskip
\textbf{Se sei un Partecipante A,} il tuo reddito del periodo è calcolato come segue: 

\begin{mdframed}
\begin{center}
    Reddito del Partecipante A
\end{center}
 
\par
 \begin{center}
     $=$
 \end{center}
\par
\begin{center}
     20 - contribuzione al progetto comune + reddito dal progetto comune + 5 - punti trasferiti al Partecipante B - 3 x (punti di deduzione assegnati dal Partecipante B)  
\end{center}
\end{mdframed}

\textbf{Se sei un Partecipante B,} il tuo reddito del periodo è calcolato come segue:
\begin{mdframed}
\begin{center}
    Reddito del Partecipante B
\end{center}
 
\par
 \begin{center}
     $=$
 \end{center}
\par
\begin{center}
     20 + reddito dal progetto comune + somma trasferimenti accettati + 15 - somma dei punti di deduzione assegnati ai Partecipanti A
\end{center}
\end{mdframed}

  \bigskip  
Alla fine di ogni periodo ti verrà mostrato il riassunto delle tue decisioni e di quelle prese dagli altri partecipanti A del gruppo (il tuo reddito del periodo, le contribuzioni al progetto comune degli altri partecipanti A, i punti di deduzione assegnati ai partecipanti A, ed eventuali punti accettati o rifiutati dal partecipante B). L'ordine con cui i Partecipanti A sono mostrati è casuale. Questo significa che le decisioni mostrate nei diversi periodi non coincidono necessariamente con lo stesso Partecipante A. 
 
\bigskip
Nelle prossime pagine ti saranno presentate delle domande di comprensione. Per favore rispondi a queste domande inserendo le risposte nella schermata del computer. Avrai sempre a disposizione un riassunto delle istruzioni, cliccando \textbf{"Istruzioni"}

\newpage

\subsubsection{Control Treatment}

Lo studio durerà  tra i 12 e i 15 periodi. All'inizio di ogni periodo ogni partecipante riceve \textbf{20 punti}. Ogni gruppo è composto da 3 \textbf{Partecipanti A} e 1 \textbf{Partecipante B}. L'assegnazione dei ruoli avviene in maniera casuale e rimane la stessa per tutti i periodi. Ogni periodo è diviso in \textbf{3 fasi}.
\bigskip

\textbf{Fase 1}

\bigskip
Se sei un \textbf{Partecipante A}, devi decidere come allocare i 20 punti tra le diverse opzioni: un \textbf{conto privato} e un \textbf{progetto comune}. Ogni punto non allocato al progetto comune è accreditato automaticamente al tuo conto privato. 

\bigskip
\textbf{Il tuo reddito dal conto privato.} 
Ogni punto che viene accreditato nel tuo conto privato ti darà un punto. Per esempio, se lasci 20 punti nel tuo conto privato (quindi non contribuisci al progetto comune), ottieni esattamente 20 punti. Se lasci 6 punti nel conto privato, il tuo reddito dal conto privato è di 6 punti. 

\bigskip    
\textbf{Il tuo reddito dal progetto comune}. Ogni membro del gruppo (Partecipante A e Partecipante B) beneficia in misura eguale delle contribuzioni derivanti dal progetto comune. Il reddito dal progetto comune è determinato come segue:
\begin{mdframed}
\begin{center}
  Reddito da progetto comune  
\end{center}\par
 \begin{center}
     $=$
 \end{center}
\par
\begin{center}
   (1.6 x somma di tutte le contribuzioni al progetto comune) / numero dei partecipanti 
\end{center} 
\end{mdframed}

Per esempio, se la somma di tutte le contribuzioni al progetto è di 60 punti (cioè tutti i partecipanti A contribuiscono 20 punti), tu e gli altri partecipanti del gruppo ricevete: (60 x 1.6) / 4 = 24 punti. 

Se la somma delle contribuzioni al progetto comune è di 20 punti,  tu e gli altri partecipanti del gruppo ricevete: (20 x 1.6) / 4 = 8 punti.

Se nessuno contribuisce al progetto comune, nessuno riceve punti dal progetto comune.

Il reddito da progetto comune viene automaticamente arrotondato per difetto per valori decimali minori di 0.5 e arrotondato per eccesso per valori decimali maggiori o uguali a 0.5. 

\bigskip
\textbf{Il reddito dei Partecipanti A dopo la Fase 1}
\begin{mdframed}
\begin{center}
    Reddito dal conto privato + reddito dal progetto comune  
\end{center}
\par
 \begin{center}
     $=$
 \end{center}
\par
\begin{center}
     20 - il tuo contributo al progetto comune + [(1.6 x somma di tutte le contribuzioni al progetto comune) / 4 ]
\end{center}
       
\end{mdframed}

\textbf{Se sei un Partecipante B,} non devi fare nulla nella Fase 1.  

\bigskip
Clicca il pulsante "Successivo" per accedere alla prossima parte delle istruzioni.

\textit{Nota: il pulsante "Successivo" apparirà dopo 60 secondi.}

\newpage
\textbf{Fase 2}

\bigskip
\textbf{Se sei un Partecipante A}, dopo la Fase 1 ottieni una dotazione addizionale di 5 punti. 

Puoi decidere se e quanti punti trasferire al Partecipante B. I punti che non vengono trasferiti al Partecipante B, sono aggiunti al tuo reddito.

\bigskip
Ogni punto trasferito al Partecipante B, costa a te 1 punto e accresce il reddito del Partecipante B di 1 punto. Se, per esempio, decidi di trasferire 3 punti al Partecipante B, ti costerà 3 punti e il reddito del Partecipante B aumenterà di 3 punti.

\bigskip
\textbf{Il reddito dei Partecipanti A dopo la Fase 2}
\begin{mdframed}
\begin{center}
    Reddito dal conto privato + reddito dal progetto comune + 5 punti - punti trasferiti al Partecipante B 
\end{center}
 
\par
 \begin{center}
     $=$
 \end{center}
\par
\begin{center}
    20 - contribuzione al progetto comune + reddito dal progetto comune + 5 - punti trasferiti al Partecipante B 
\end{center}
\end{mdframed}
\bigskip

\textbf{Se sei un Partecipante B,} non devi fare nulla nella Fase 2. 

\bigskip
Clicca il pulsante "Successivo" per accedere alla prossima parte delle istruzioni.

\textit{Nota: il pulsante "Successivo" apparirà dopo 30 secondi.}

\newpage
\textbf{Fase 3}

\bigskip
 \textbf{Se sei un Partecipante A,} non devi fare nulla nella Fase 3. 

 \bigskip
 \textbf{Se sei un Partecipante B,} ti verranno mostrate le contribuzioni al progetto comune dei Partecipanti A insieme al numero di punti che ogni Partecipante A ti ha trasferito. \textbf{Devi decidere se accettare o rifiutare i punti trasferiti a te da ogni Partecipante A}.

\bigskip
Se accetti un trasferimento, i punti verranno accreditati sul tuo conto. Se rifiuti, i punti verranno restituiti al Partecipante A.   

\bigskip
Importante: l'ordine in cui i Partecipanti A sono mostrati è casuale. Questo significa che le decisioni mostrate nei diversi periodi non coincidono necessariamente con lo stesso Partecipante A. 
    
\bigskip
Inoltre, se sei un Partecipante B, \textbf{ottieni altri 15 punti (che chiameremo punti di deduzione)}. Questi punti possono essere usati per \textbf{ridurre i redditi dei Partecipanti A nel tuo gruppo}. 

   \bigskip
Devi decidere se e quanti \textbf{punti di deduzione} assegnare ai partecipanti A del tuo gruppo. 

Per ogni punto di deduzione che assegni, devi pagare \textbf{1 punto}. Per ogni punto di deduzione assegnato, il reddito del Partecipante A verrà ridotto di \textbf{3 punti}. 

\bigskip
Esempio: se assegni 3 punti di deduzione, il reddito del Partecipante A sarà ridotto di 3 x 3 = 9 punti, mentre il tuo reddito del sarà ridotto di 3 punti. 

\bigskip
 Sebbene hai 15 punti di deduzione, \textbf{a ogni partecipante puoi togliere al massimo 10 punti.}
 
\bigskip
I punti di deduzione non usati saranno accreditati al tuo reddito del periodo.

\bigskip
\textbf{Il reddito del Partecipante B dopo la Fase 3 }

\begin{mdframed}
\begin{center}
    Reddito del periodo Partecipante B
\end{center}
 
\par
 \begin{center}
     $=$
 \end{center}
\par
\begin{center}
    20 + reddito dal progetto comune + punti trasferiti dai Partecipanti A + punti di deduzione - somma dei punti di deduzione assegnati ai Partecipanti A
\end{center}
\end{mdframed}

\bigskip
Clicca il pulsante "Successivo" per accedere alla prossima parte delle istruzioni.

\textit{Nota: il pulsante "Successivo" apparirà dopo 30 secondi.}

\newpage
\textbf{Riassunto del reddito dei Partecipanti A e Partecipante B}

    \bigskip
\textbf{Se sei un Partecipante A,} il tuo reddito del periodo è calcolato come segue: 

\begin{mdframed}
\begin{center}
    Reddito del Partecipante A
\end{center}
 
\par
 \begin{center}
     $=$
 \end{center}
\par
\begin{center}
     20 - contribuzione al progetto comune + reddito dal progetto comune + 5 - punti trasferiti al Partecipante B - 3 x (punti di deduzione assegnati dal Partecipante B)  
\end{center}
\end{mdframed}

\textbf{Se sei un Partecipante B,} il tuo reddito del periodo è calcolato come segue:
\begin{mdframed}
\begin{center}
    Reddito del Partecipante B
\end{center}
 
\par
 \begin{center}
     $=$
 \end{center}
\par
\begin{center}
     20 + reddito dal progetto comune + somma trasferimenti accettati + 15 - somma dei punti di deduzione assegnati ai Partecipanti A
\end{center}
\end{mdframed}

  \bigskip  
Alla fine di ogni periodo ai partecipanti A verrà mostrato il riassunto delle proprie decisioni e di quelle prese degli altri partecipanti A del gruppo (il suo reddito del periodo, le contribuzioni al progetto comune degli altri partecipanti A, i punti di deduzione assegnati ai partecipanti A). L'ordine con cui i Partecipanti A sono mostrati è casuale. Questo significa che le decisioni mostrate nei diversi periodi non coincidono necessariamente con lo stesso Partecipante A. Anche al partecipante B verrà mostrato il riassunto delle sue decisioni e di quelle prese dagli altri partecipanti A del gruppo (il proprio reddito del periodo, le contribuzioni al progetto comune degli altri partecipanti A, i punti di deduzione assegnati ai partecipanti A, i punti trasferiti dai partecipanti A).
 
\bigskip
Nelle prossime pagine ti saranno presentate delle domande di comprensione. Per favore rispondi a queste domande inserendo le risposte nella schermata del computer. Avrai sempre a disposizione un riassunto delle istruzioni, cliccando \textbf{"Istruzioni"}

\newpage
\textbf{Domande di comprensione}

\bigskip
Immagina che gli altri Partecipanti A del gruppo abbiano messo 10 punti ciascuno nel progetto comune. Anche tu sei un Partecipante A e hai deciso di mettere 10 punti.
\begin{itemize}
    \item[1] Quanti punti ottieni alla fine del periodo, se il partecipante B non ti toglie nessun punto?
    \begin{todolist}
        \item 30 x 1.6 = 48. 48 : 4 = 12. 20 + 12 = 32
        \item 30 x 1.6 = 48. 48 : 4 = 12. 20 - 10 + 12 + 5 = 27
    \end{todolist}
    \item[2] Quanti punti ottieni alla fine del periodo, se il partecipante B ti toglie 9 punti?
    \begin{todolist}
        \item 20 + 12 - 9 = 23
        \item 320 - 10 = 10. 10 + 12 + 5 - 9 = 18
    \end{todolist}
    \item[3] Quanti punti ottieni se trasferisci 2 punti al partecipante B?
    \begin{todolist}
        \item 20 - 10 + 12 + 5 - 2 = 25
        \item 20 - 10 + 12 - 2 = 20
    \end{todolist}
\end{itemize}

\newpage

\textbf{Ruolo}

\bigskip
Il tuo ruolo è \textbf{Partecipante A}

\bigskip
\bigskip

\textbf{Ruolo}

\bigskip
Il tuo ruolo è \textbf{Partecipante B}
\newpage

\textbf{Cominciamo} 

\bigskip
Ora sei pronto per cominciare! Le decisioni che prenderai nelle prossime pagine avranno effetto sul tuo compenso finale. 
\newpage

\textbf{Fase 1 (Partecipante A)}
\bigskip

\rightline{\textbf{Periodo 1}}
\bigskip
Quanto vuoi mettere nel progetto comune?
  \begin{todolist}
        \item 0
        \item 1
        \item 2
        \item 3
        \item 4
        \item 5
        \item 6
        \item 7
        \item 8
        \item 9
        \item 10
        \item 11
        \item 12
        \item 13
        \item 14
        \item 15
        \item 16
        \item 17
        \item 18
        \item 19
        \item 20
    \end{todolist}

\newpage

\noindent\textbf{Risultato Fase 1}\hfill\textbf{Periodo 1}

\bigskip
Hai contribuito 4 punti al progetto comune. Il progetto comune vale 13 punti.

\bigskip
Il tuo reddito dal conto privato è 16 punti e il tuo reddito dal progetto comune è 3 punti.

\bigskip
Perciò nella Fase 1 ottieni 19 punti.
\newpage

\textbf{Fase 2 (Partecipante A)}

\rightline{\textbf{Periodo 1}}
\bigskip
Hai a disposizione 5 punti. Indica se e quanti punti vuoi trasferire al Partecipante B
  \begin{todolist}
        \item 0
        \item 1
        \item 2
        \item 3
        \item 4
        \item 5
    \end{todolist}

\newpage

\textbf{Fase 3 (Partecipante B)}

\rightline{\textbf{Periodo 1}}
\bigskip
Ecco la decisione effettuata dal seguente partecipante A: 1 punto nel progetto comune e ti ha trasferito 2 punti

\bigskip
Vuoi accettare i punti che ti ha trasferito il partecipante A? 
  \begin{todolist}
        \item Sì
        \item No
    \end{todolist}
Indica se e quanti punti vuoi togliere al partecipante A
  \begin{todolist}
        \item 0
        \item 1
        \item 2
        \item 3
        \item 4
        \item 5
        \item 6
        \item 7
        \item 8
        \item 9
        \item 10
    \end{todolist}
    
\bigskip
Ecco la decisione effettuata dal seguente partecipante A: 4 punti nel progetto comune e ti ha trasferito 1 punto.

\bigskip
Vuoi accettare i punti che ti ha trasferito il partecipante A? 
  \begin{todolist}
        \item Sì
        \item No
    \end{todolist}
Indica se e quanti punti vuoi togliere al partecipante A
  \begin{todolist}
        \item 0
        \item 1
        \item 2
        \item 3
        \item 4
        \item 5
        \item 6
        \item 7
        \item 8
        \item 9
        \item 10
    \end{todolist}

\bigskip
Ecco la decisione effettuata dal seguente partecipante A: 3 punti nel progetto comune e ti ha trasferito 1 punto.

\bigskip
Vuoi accettare i punti che ti ha trasferito il partecipante A? 
  \begin{todolist}
        \item Sì
        \item No
    \end{todolist}
Indica se e quanti punti vuoi togliere al partecipante A
  \begin{todolist}
        \item 0 
        \item 1
        \item 2
        \item 3
        \item 4
        \item 5
        \item 6
        \item 7
        \item 8
        \item 9
        \item 10
    \end{todolist}

\newpage

\textbf{Risultati Periodo (Partecipante B)}

\rightline{\textbf{Periodo 1}}
\bigskip
I partecipanti A del gruppo hanno effettuato le seguenti scelte:
\begin{itemize}
    \item partecipare al progetto comune mettendo 1 punto e ti ha trasferito 2 punti. Al partecipante hai tolto 3 punti.
    \item partecipare al progetto comune mettendo 4 punti e ti ha trasferito 1 punto. Al partecipante hai tolto 0 punti.
    \item partecipare al progetto comune mettendo 3 punti e ti ha trasferito 1 punto. Al partecipante hai tolto 1 punto.
\end{itemize}

\bigskip
Hai ottenuto 3 punti dal progetto comune. I partecipanti A ti hanno trasferito complessivamente 4 punti e hai accettato 4 punti. Dunque questo periodo ottieni 26 punti.

\newpage

\textbf{Risultati Periodo  (Partecipante A)}

\rightline{\textbf{Periodo 1}}
\bigskip

Hai deciso di mettere 3 punti nel progetto comune e hai trasferito 1 punto al partecipante B.  
Il partecipante B ha accettato i punti da te trasferiti e ti ha tolto 6 punti. 
    
Gli altri partecipanti A del gruppo hanno effettuato le seguenti scelte:
 \begin{itemize}
     \item  partecipare al progetto comune mettendo 1 punto. Il Partecipante B gli ha tolto 9 punti.
 \item  partecipare al progetto comune mettendo 4 punti. Il Partecipante B gli ha tolto 3 punti.
 \end{itemize} 
Dal progetto comune ottieni 3 punti. Sommando la tua quota del progetto comune al tuo conto privato, questo periodo ottieni 18 punti.

\newpage

\subsubsection{Transparency Treatment}

Lo studio durerà  tra i 12 e i 15 periodi. All'inizio di ogni periodo ogni partecipante riceve \textbf{20 punti}. Ogni gruppo è composto da 3 \textbf{Partecipanti A} e 1 \textbf{Partecipante B}. L'assegnazione dei ruoli avviene in maniera casuale e rimane la stessa per tutti i periodi. Ogni periodo è diviso in \textbf{3 fasi}.
\bigskip

\textbf{Fase 1}

\bigskip
Se sei un \textbf{Partecipante A}, devi decidere come allocare i 20 punti tra le diverse opzioni: un \textbf{conto privato} e un \textbf{progetto comune}. Ogni punto non allocato al progetto comune è accreditato automaticamente al tuo conto privato. 

\bigskip
\textbf{Il tuo reddito dal conto privato.} 

Ogni punto che viene accreditato nel tuo conto privato ti darà un punto. Per esempio, se lasci 20 punti nel tuo conto privato (quindi non contribuisci al progetto comune), ottieni esattamente 20 punti. Se lasci 6 punti nel conto privato, il tuo reddito dal conto privato è di 6 punti. 

 \bigskip   
\textbf{Il tuo reddito dal progetto comune}. 

Ogni membro del gruppo (Partecipante A e Partecipante B) beneficia in misura eguale delle contribuzioni derivanti dal progetto comune. Il reddito dal progetto comune è determinato come segue:
\bigskip
\begin{mdframed}
\begin{center}
    Reddito da progetto comune
\end{center}
\par
 \begin{center}
     $=$
 \end{center}
\par
\begin{center}
    (1.6 x somma di tutte le contribuzioni al progetto comune) / numero dei partecipanti
\end{center}
\end{mdframed}

\bigskip
Per esempio, se la somma di tutte le contribuzioni al progetto è di 60 punti (cioè tutti i partecipanti A contribuiscono 20 punti), tu e gli altri partecipanti del gruppo ricevete: (60 x 1.6) / 4 = 24 punti. 

Se la somma delle contribuzioni al progetto comune è di 20 punti,  tu e gli altri partecipanti del gruppo ricevete: (20 x 1.6) / 4 = 8 punti.

Se nessuno contribuisce al progetto comune, nessuno riceve punti dal progetto comune.

\bigskip
Il reddito da progetto comune viene automaticamente arrotondato per difetto per valori decimali minori di 0.5 e arrotondato per eccesso per valori decimali maggiori o uguali a 0.5. 

\bigskip
\textbf{Il reddito dei Partecipanti A dopo la Fase 1}
\begin{mdframed}
\begin{center}
Reddito dal conto privato + reddito dal progetto comune
    \end{center}
\par
 \begin{center}
     $=$
 \end{center}
\par
\begin{center}
      20 - il tuo contributo al progetto comune + [(1.6 x somma di tutte le contribuzioni al progetto comune) / 4 ]    
\end{center}
  
\end{mdframed}

\bigskip        
\textbf{Se sei un Partecipante B,} non devi fare nulla nella Fase 1.  

\bigskip
Clicca il pulsante "Successivo" per accedere alla prossima parte delle istruzioni.

\textit{Nota: il pulsante "Successivo" apparirà dopo 60 secondi.}

\newpage
\textbf{Fase 2}

\bigskip
\textbf{Se sei un Partecipante A}, dopo la Fase 1 ottieni una dotazione addizionale di 5 punti. 

Puoi decidere se e quanti punti trasferire al Partecipante B. I punti che non vengono trasferiti al Partecipante B, sono aggiunti al tuo reddito. 

\bigskip
Ogni punto trasferito al Partecipante B, costa a te 1 punto e accresce il reddito del Partecipante B di 1 punto. Se, per esempio, decidi di trasferire 3 punti al Partecipante B, ti costerà 3 punti e il reddito del Partecipante B aumenterà di 3 punti.

\bigskip
\textbf{Il reddito dei Partecipanti A dopo la Fase 2}
\begin{mdframed}
\begin{center}
    Reddito dal conto privato + reddito dal progetto comune + 5 punti - punti trasferiti al Partecipante B 
\end{center}
 
\par
 \begin{center}
     $=$
 \end{center}
\par
\begin{center}
    20 - contribuzione al progetto comune + reddito dal progetto comune + 5 - punti trasferiti al Partecipante B 
\end{center}
\end{mdframed}
\bigskip

\textbf{Se sei un Partecipante B,} non devi fare nulla nella Fase 2. 

\bigskip
Clicca il pulsante "Successivo" per accedere alla prossima parte delle istruzioni.

\textit{Nota: il pulsante "Successivo" apparirà dopo 30 secondi.}

\newpage
\textbf{Fase 3}

\bigskip
 \textbf{Se sei un Partecipante A,} non devi fare nulla nella Fase 3. 
 
 \bigskip
 \textbf{Se sei un Partecipante B,} ti verranno mostrate le contribuzioni al progetto comune dei Partecipanti A insieme al numero di punti che ogni Partecipante A ti ha trasferito. \textbf{Devi decidere se accettare o rifiutare i punti trasferiti a te da ogni Partecipante A.}

\bigskip Se accetti un trasferimento, i punti verranno accreditati sul tuo conto. Se rifiuti, i punti verranno restituiti al Partecipante A.   

\bigskip
Importante: l'ordine in cui i Partecipanti A sono mostrati è casuale. Questo significa che le decisioni mostrate nei diversi periodi non coincidono necessariamente con lo stesso Partecipante A. 
    
\bigskip
Inoltre, se sei un Partecipante B, \textbf{ottieni altri 15 punti (che chiameremo punti di deduzione)}. Questi punti possono essere usati per \textbf{ridurre i redditi dei Partecipanti A nel tuo gruppo}. 

   \bigskip
Devi decidere se e quanti \textbf{punti di deduzione} assegnare ai partecipanti A del tuo gruppo. 

Per ogni punto di deduzione che assegni, devi pagare \textbf{1 punto}. Per ogni punto di deduzione assegnato, il reddito del Partecipante A verrà ridotto di \textbf{3 punti}. 

\bigskip
Esempio: se assegni 3 punti di deduzione, il reddito del Partecipante A sarà ridotto di 3 x 3 = 9 punti, mentre il tuo reddito del sarà ridotto di 3 punti. 

\bigskip
Sebbene hai 15 punti di deduzione, \textbf{a ogni partecipante puoi togliere al massimo 10 punti.}

\bigskip
I punti di deduzione non usati saranno accreditati al tuo reddito del periodo.

\bigskip
\textbf{Il reddito del Partecipante B dopo la Fase 3 }

\begin{mdframed}
\begin{center}
    Reddito del periodo Partecipante B
\end{center}
 
\par
 \begin{center}
     $=$
 \end{center}
\par
\begin{center}
    20 + reddito dal progetto comune + punti trasferiti dai Partecipanti A + punti di deduzione - somma dei punti di deduzione assegnati ai Partecipanti A
\end{center}
\end{mdframed}

\bigskip
\textbf{Informazione}

\bigskip
\textbf{Se sei un Partecipante A},prima di prendere la tua decisione nella Fase 1, puoi acquistare una \textbf{informazione} sul Partecipante B e cioè sapere le decisioni prese dal Partecipante B nel periodo precedente nei confronti degli altri Partecipanti A del gruppo (se ha accettato o rifiutato eventuali punti trasferiti dai partecipanti A, punti di deduzione assegnati, contribuzione dei partecipanti A al progetto comune). Il costo di questa informazione è di \textbf{3 punti} per ogni periodo. Puoi comprare questa informazione a partire dal secondo periodo.

\bigskip
Clicca il pulsante "Successivo" per accedere alla prossima parte delle istruzioni.

\textit{Nota: il pulsante "Successivo" apparirà dopo 30 secondi.}

\newpage
\textbf{Riassunto del reddito dei Partecipanti A e Partecipante B}

    \bigskip
    
\textbf{Se sei un Partecipante A,} il tuo reddito del periodo è calcolato come segue: 

\begin{mdframed}
\begin{center}
    Reddito del Partecipante A
\end{center}
 
\par
 \begin{center}
     $=$
 \end{center}
\par
\begin{center}
     20 - costo dell'informazione - contribuzione al progetto comune + reddito dal progetto comune + 5 - punti trasferiti al Partecipante B - 3 x (punti di deduzione assegnati dal Partecipante B)  
\end{center}
\end{mdframed}

\textbf{Se sei un Partecipante B,} il tuo reddito del periodo è calcolato come segue:
\begin{mdframed}
\begin{center}
    Reddito del Partecipante B
\end{center}
 
\par
 \begin{center}
     $=$
 \end{center}
\par
\begin{center}
     20 + reddito dal progetto comune + somma trasferimenti accettati + 15 - somma dei punti di deduzione assegnati ai Partecipanti A
\end{center}
\end{mdframed}

Alla fine di ogni periodo ti verrà mostrato il riassunto delle tue decisioni e di quelle prese dagli altri partecipanti A del gruppo (il tuo reddito del periodo, le contribuzioni al progetto comune degli altri partecipanti A, i punti di deduzione assegnati ai partecipanti A). L'ordine con cui i Partecipanti A sono mostrati è casuale. Questo significa che le decisioni mostrate nei diversi periodi non coincidono necessariamente con lo stesso Partecipante A. Anche al partecipante B verrà mostrato il riassunto delle sue decisioni e di quelle prese dagli altri partecipanti A del gruppo (il proprio reddito del periodo, le contribuzioni al progetto comune degli altri partecipanti A, i punti di deduzione assegnati ai partecipanti A, i punti trasferiti dai partecipanti A).
 
\bigskip
Nelle prossime pagine ti saranno presentate delle domande di comprensione. Per favore rispondi a queste domande inserendo le risposte nella schermata del computer. Avrai sempre a disposizione un riassunto delle istruzioni, cliccando \textbf{"Istruzioni"}

\newpage

\subsubsection{Exit}
Lo studio durerà  tra i 12 e i 15 periodi. All'inizio di ogni periodo ogni partecipante riceve \textbf{20 punti}. Ogni gruppo è composto da 3 \textbf{Partecipanti A} e 1 \textbf{Partecipante B}. L'assegnazione dei ruoli avviene in maniera casuale e rimane la stessa per tutti i periodi. Ogni periodo è diviso in \textbf{3 fasi}.
\bigskip

\textbf{Fase 1}

\bigskip
Se sei un \textbf{Partecipante A}, devi decidere come allocare i 20 punti tra le diverse opzioni: un \textbf{conto privato} e un \textbf{progetto comune}. Ogni punto non allocato al progetto comune è accreditato automaticamente al tuo conto privato. 

\bigskip
\textbf{Il tuo reddito dal conto privato.} 

Ogni punto che viene accreditato nel tuo conto privato ti darà un punto. Per esempio, se lasci 20 punti nel tuo conto privato (quindi non contribuisci al progetto comune), ottieni esattamente 20 punti. Se lasci 6 punti nel conto privato, il tuo reddito dal conto privato è di 6 punti. 

\bigskip
\textbf{Il tuo reddito dal progetto comune}. 

Ogni membro del gruppo (Partecipante A e Partecipante B) beneficia in misura eguale delle contribuzioni derivanti dal progetto comune. Il reddito dal progetto comune è determinato come segue:

\begin{mdframed}
\begin{center}
    Reddito da progetto comune
\end{center}
\par
 \begin{center}
     $=$
 \end{center}
\par
\begin{center}
    (1.6 x somma di tutte le contribuzioni al progetto comune) / numero dei partecipanti
\end{center} 
\end{mdframed}

\bigskip
Per esempio, se la somma di tutte le contribuzioni al progetto è di 60 punti (cioè tutti i partecipanti A contribuiscono 20 punti), tu e gli altri partecipanti del gruppo ricevete: (60 x 1.6) / 4 = 24 punti. 

Se la somma delle contribuzioni al progetto comune è di 20 punti,  tu e gli altri partecipanti del gruppo ricevete: (20 x 1.6) / 4 = 8 punti.

Se nessuno contribuisce al progetto comune, nessuno riceve punti dal progetto comune.

 \bigskip
Il reddito da progetto comune viene automaticamente arrotondato per difetto per valori decimali minori di 0.5 e arrotondato per eccesso per valori decimali maggiori o uguali a 0.5. 

\bigskip
\textbf{Il reddito dei Partecipanti A dopo la Fase 1}
\begin{mdframed}
\begin{center}
   Reddito dal conto privato + reddito dal progetto comune  
\end{center}
\par
 \begin{center}
     $=$
 \end{center}
\par
\begin{center}
  20 - il tuo contributo al progetto comune + [(1.6 x somma di tutte le contribuzioni al progetto comune) / numero dei partecipanti ]  
\end{center}
        
\end{mdframed}

\textbf{Se sei un Partecipante B,} non devi fare nulla nella Fase 1.  

\bigskip
Clicca il pulsante "Successivo" per accedere alla prossima parte delle istruzioni.

\textit{Nota: il pulsante "Successivo" apparirà dopo 60 secondi.}

\newpage
\textbf{Fase 2}

\bigskip
\textbf{Se sei un Partecipante A}, dopo la Fase 1 ottieni una dotazione addizionale di 5 punti. 

Puoi decidere se e quanti punti trasferire al Partecipante B. I punti che non vengono trasferiti al Partecipante B, sono aggiunti al tuo reddito. 

\bigskip
 Ogni punto trasferito al Partecipante B, costa a te 1 punto e accresce il reddito del Partecipante B di 1 punto. Se, per esempio, decidi di trasferire 3 punti al Partecipante B, ti costerà 3 punti e il reddito del Partecipante B aumenterà di 3 punti.

\bigskip
\textbf{Il reddito dei Partecipanti A dopo la Fase 2}
\begin{mdframed}
\begin{center}
    Reddito dal conto privato + reddito dal progetto comune + 5 punti - punti trasferiti al Partecipante B 
\end{center}
 
\par
 \begin{center}
     $=$
 \end{center}
\par
\begin{center}
    20 - contribuzione al progetto comune + reddito dal progetto comune + 5 - punti trasferiti al Partecipante B 
\end{center}
\end{mdframed}
\bigskip

\textbf{Se sei un Partecipante B,} non devi fare nulla nella Fase 2. 

\bigskip
Clicca il pulsante "Successivo" per accedere alla prossima parte delle istruzioni.

\textit{Nota: il pulsante "Successivo" apparirà dopo 30 secondi.}

\newpage
\textbf{Fase 3}

\bigskip
 \textbf{Se sei un Partecipante A,} non devi fare nulla nella Fase 3. 
 
 \bigskip
 \textbf{Se sei un Partecipante B,} ti verranno mostrate le contribuzioni al progetto comune dei Partecipanti A insieme al numero di punti che ogni Partecipante A ti ha trasferito. 
 
\textbf{Devi decidere se accettare o rifiutare i punti trasferiti a te da ogni Partecipante A.}

\bigskip
Se accetti un trasferimento, i punti verranno accreditati sul tuo conto. Se rifiuti, i punti verranno restituiti al Partecipante A.   

\bigskip
Importante: l'ordine in cui i Partecipanti A sono mostrati è casuale. Questo significa che le decisioni mostrate nei diversi periodi non coincidono necessariamente con lo stesso Partecipante A. 
    
\bigskip
Inoltre, se sei un Partecipante B, \textbf{ottieni altri 15 punti (che chiameremo punti di deduzione)}. Questi punti possono essere usati per \textbf{ridurre i redditi dei Partecipanti A nel tuo gruppo}. 

\bigskip
Devi decidere se e quanti \textbf{punti di deduzione} assegnare ai partecipanti A del tuo gruppo. 

Per ogni punto di deduzione che assegni, devi pagare \textbf{1 punto}. Per ogni punto di deduzione assegnato, il reddito del Partecipante A verrà ridotto di \textbf{3 punti}. 

\bigskip
Esempio: se assegni 3 punti di deduzione, il reddito del Partecipante A sarà ridotto di 3 x 3 = 9 punti, mentre il tuo reddito del sarà ridotto di 3 punti. 

\bigskip
 Sebbene hai 15 punti di deduzione, \textbf{a ogni partecipante puoi togliere al massimo 10 punti.}

\bigskip
I punti di deduzione non usati saranno accreditati al tuo reddito del periodo.

\bigskip
\textbf{Il reddito del Partecipante B dopo la Fase 3 }

\begin{mdframed}
\begin{center}
    Reddito del periodo Partecipante B
\end{center}
 
\par
 \begin{center}
     $=$
 \end{center}
\par
\begin{center}
    20 + reddito dal progetto comune + punti trasferiti dai Partecipanti A + punti di deduzione - somma dei punti di deduzione assegnati ai Partecipanti A
\end{center}
\end{mdframed}

\bigskip
\textbf{Informazione}

\bigskip
\textbf{Se sei un Partecipante A},prima di prendere la tua decisione nella Fase 1, puoi acquistare una \textbf{informazione} sul Partecipante B e cioè sapere le decisioni prese dal Partecipante B nel periodo precedente nei confronti degli altri Partecipanti A del gruppo (se ha accettato o rifiutato eventuali punti trasferiti dai partecipanti A, punti di deduzione assegnati, contribuzione dei partecipanti A al progetto comune). Il costo di questa informazione è di \textbf{3 punti} per ogni periodo. Puoi comprare questa informazione a partire dal secondo periodo.

\bigskip
Inoltre, nella Fase 1 puoi decidere di \textbf{non partecipare al progetto comune}. Se non partecipi al progetto comune, non ricevi reddito dal progetto comune e il progetto comune viene diviso tra i Partecipanti A che partecipano al progetto comune e il Partecipante B. 
Se un partecipante A esce del gruppo, il partecipante B ha a disposizione 10 punti di deduzione, invece di 15.

Se due o più partecipanti A non partecipano al progetto comune, il progetto comune non viene realizzato, e tutti i partecipanti A mantengono la dotazione iniziale (20 + 5). Se un partecipante A aveva trasferito dei punti al partecipante B, il trasferimento non avviene e i punti vengono restituiti al partecipante A. Il partecipante B mantiene la sua dotazione iniziale (20). 

\bigskip
Clicca il pulsante "Successivo" per accedere alla prossima parte delle istruzioni 

 \textit{Nota: il pulsante "Successivo" apparirà dopo 30 secondi.}

\newpage
\textbf{Riassunto del reddito dei Partecipanti A e Partecipante B}

\bigskip
\textbf{Se sei un Partecipante A, e partecipi al progetto comune,} il tuo reddito del periodo è calcolato come segue: 

\begin{mdframed}
\begin{center}
    Reddito del Partecipante A
\end{center}
 
\par
 \begin{center}
     $=$
 \end{center}
\par
\begin{center}
     20 - costo dell'informazione  - contribuzione al progetto comune + reddito dal progetto comune + 5 - punti trasferiti al Partecipante B - 3 x (punti di deduzione assegnati dal Partecipante B)  
\end{center}
\end{mdframed}

\textbf{Se sei un Partecipante A, e non partecipi al progetto comune,} il tuo reddito del periodo è calcolato come segue: 

\begin{mdframed}
\begin{center}
    Reddito del Partecipante A
\end{center}
 
\par
 \begin{center}
     $=$
 \end{center}
\par
\begin{center}
     20 - costo dell'informazione + 5
\end{center}
\end{mdframed}

\textbf{Se sei un Partecipante B,} il tuo reddito del periodo è calcolato come segue:
\begin{mdframed}
\begin{center}
    Reddito del Partecipante B
\end{center}
 
\par
 \begin{center}
     $=$
 \end{center}
\par
\begin{center}
     20 + reddito dal progetto comune + somma trasferimenti accettati + 15 - somma dei punti di deduzione assegnati ai Partecipanti A
\end{center}
\end{mdframed}

\textbf{Se sei un Partecipante B, e il progetto comune non viene realizzato,} il tuo reddito del periodo è calcolato come segue:
\begin{mdframed}
\begin{center}
    Reddito del Partecipante B
\end{center}
 
\par
 \begin{center}
     $=$
 \end{center}
\par
\begin{center}
     20 
\end{center}
\end{mdframed}

\bigskip
Alla fine di ogni periodo ai partecipanti A verrà mostrato il riassunto delle proprie decisioni e di quelle prese degli altri partecipanti A del gruppo (il proprio reddito del periodo, le contribuzioni al progetto comune degli altri partecipanti A, i punti di deduzione assegnati ai partecipanti A). L'ordine con cui i Partecipanti A sono mostrati è casuale. Questo significa che le decisioni mostrate nei diversi periodi non coincidono necessariamente con lo stesso Partecipante A. Anche al partecipante B verrà mostrato il riassunto delle sue decisioni e di quelle prese dagli altri partecipanti A del gruppo (il suo reddito del periodo, le contribuzioni al progetto comune degli altri partecipanti A, i punti di deduzione assegnati ai partecipanti A, i punti trasferiti dai partecipanti A).

\bigskip
Nelle prossime pagine ti saranno presentate delle domande di comprensione. Per favore rispondi a queste domande inserendo le risposte nella schermata del computer. Avrai sempre a disposizione un riassunto delle istruzioni, cliccando \textbf{"Istruzioni"}

\newpage
\subsubsection{Distribution Game}
\textbf{Istruzioni}

In questa parte dello studio trovi tre scelte che indicano i possibili punti da assegnare ai quattro componenti del tuo gruppo.

Nota bene che il totale dei punti da assegnare al gruppo non cambia per le tre scelte. Cambia solo il numero di punti assegnato a ciascun componente.

\begin{figure}[h]
 \centering   \includegraphics[width=0.49\linewidth]{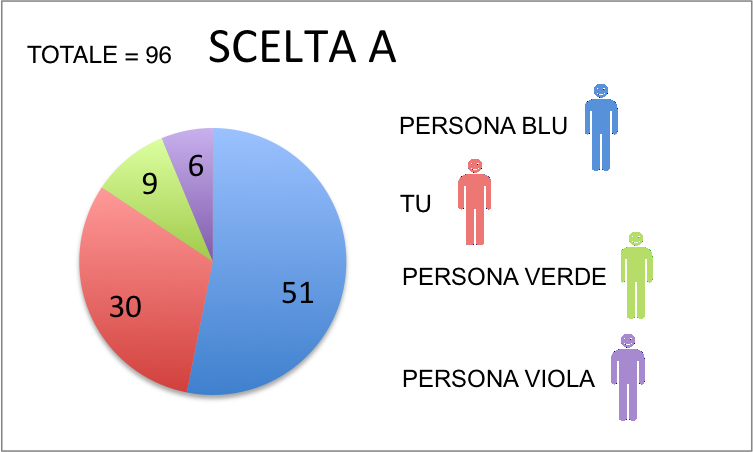}
    \includegraphics[width=0.49\linewidth]{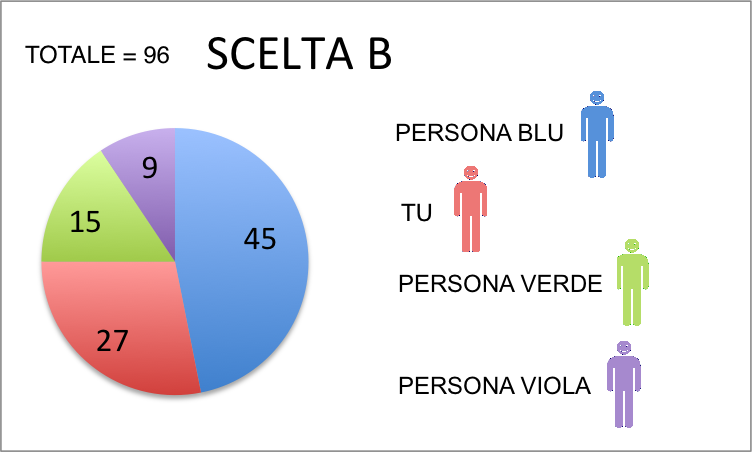}\\
    \includegraphics[width=0.49\linewidth]{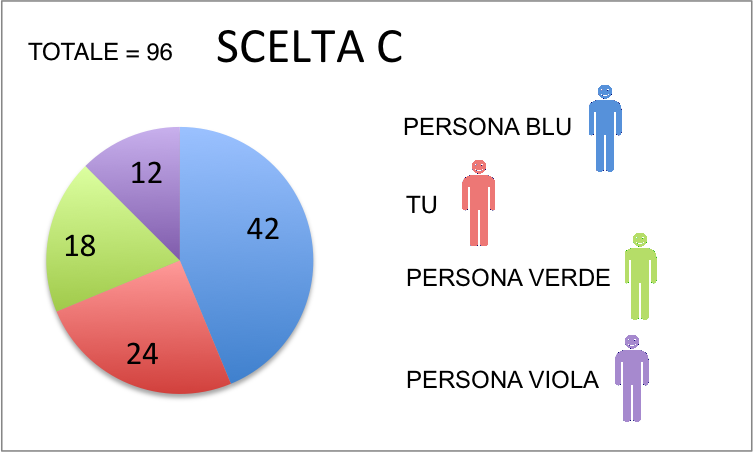}
\end{figure}

\textbf{Cosa devi fare?} Devi indicare quale scelta preferisci.

\bigskip
\textbf{Come viene determinato il tuo compenso?}Quando tutti avrete effettuato la vostra scelta, il computer estrarrà a caso chi deciderà i compensi di tutti i componenti del tuo gruppo.

\bigskip
\textbf{Se ad esempio vieni estratto tu saranno le tue scelte a determinare i compensi di tutti i componenti del tuo gruppo.}

\bigskip
Il soggetto estratto, in questo caso tu, prenderà il compenso indicato in ROSSO con la scritta TU, mentre gli altri componenti del gruppo prenderanno un numero di punti pari a quelli indicati nel disegno, a seconda che vengano estratti per essere Persona BLU, VERDE o VIOLA.

Se tu scegli A ad esempio, e vieni estratto per decidere i compensi di tutti i componenti del tuo gruppo, tu prendi 30 punti, gli altri ottengono 51, 9, e 6 punti a seconda che il computer decida che siano estratti come Persona BLU, VERDE o VIOLA.

\bigskip
\textbf{Se ad esempio NON vieni estratto tu saranno le scelte di un altro a determinare i compensi tuoi e di tutti gli altri componenti del tuo gruppo.}

\bigskip
Il tuo guadagno potrà corrispondere al guadagno indicato con Persona BLU, VERDE o VIOLA a seconda dell'estrazione effettuata dal computer.

Se ad esempio la persona estratta dal computer per decidere i compensi di tutti i componenti del gruppo ha indicato la Scelta C, questa persona ottiene 24 punti e tu sei stato estratto dal computer Persona VIOLA ottieni 12 punti, gli altri due componenti del gruppo ottengono 18 e 42 punti.

\bigskip
Le istruzioni resteranno disponibili sullo schermo per tutta la durata di questa parte dello studio.

\bigskip
Prima di effettuare la scelta, ti chiediamo di rispondere a due domande che non avranno effetto sul tuo compenso finale, ma che servono a te per verificare di aver compreso le istruzioni.

\bigskip
Clicca su Successivo quando sei pronto per continuare. 

 \textit{Nota: il pulsante "Successivo" apparirà dopo 60 secondi.}
\newpage
{Domande di Comprensione}

\begin{itemize}
    \item[1] Immagina di aver scelto B. Se tu vieni selezionato dal computer per decidere i compensi di tutti, tu guadagni:
     \begin{todolist}
        \item 15 punti
        \item 27 punti
    \end{todolist}
    \item[2] Immagina di aver scelto A. Un altro componente del tuo gruppo viene estratto per decidere i compensi di tutti e questa persona ha scelto C. In questo caso tu guadagni:
    \begin{todolist}
        \item 42 oppure 18 oppure 12 punti, a seconda del colore che il computer estrae a caso e ti assegna
        \item 45 oppure 15 oppure 9 punti, a seconda del colore che il computer estrae a caso e ti assegna

    \end{todolist}
\end{itemize}
\begin{figure}[h]
 \centering   \includegraphics[width=0.49\linewidth]{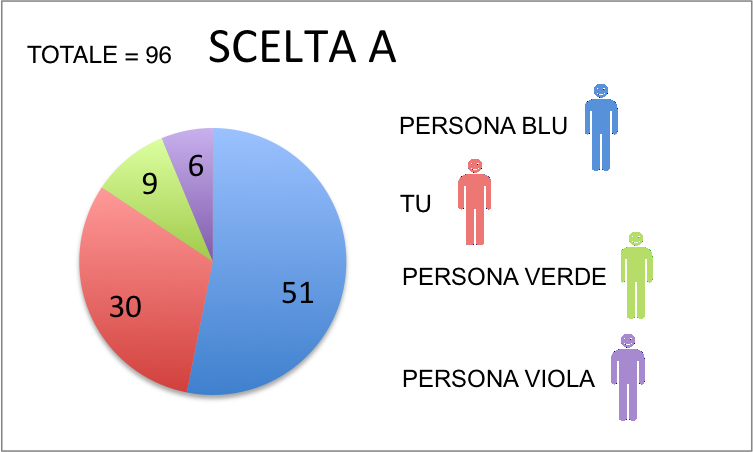}
    \includegraphics[width=0.49\linewidth]{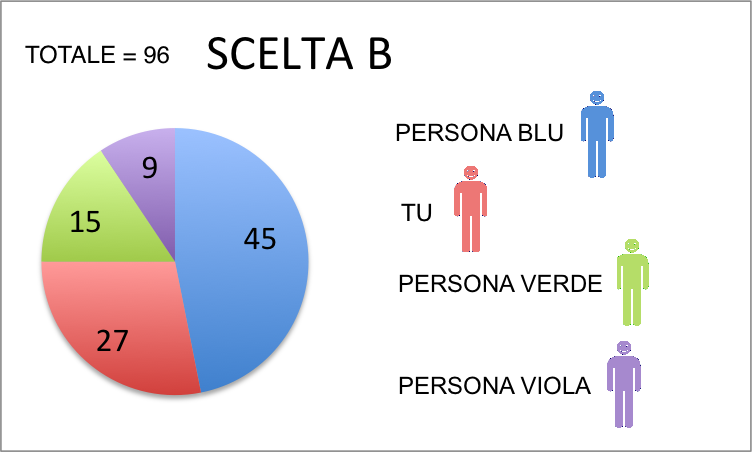}\\
    \includegraphics[width=0.49\linewidth]{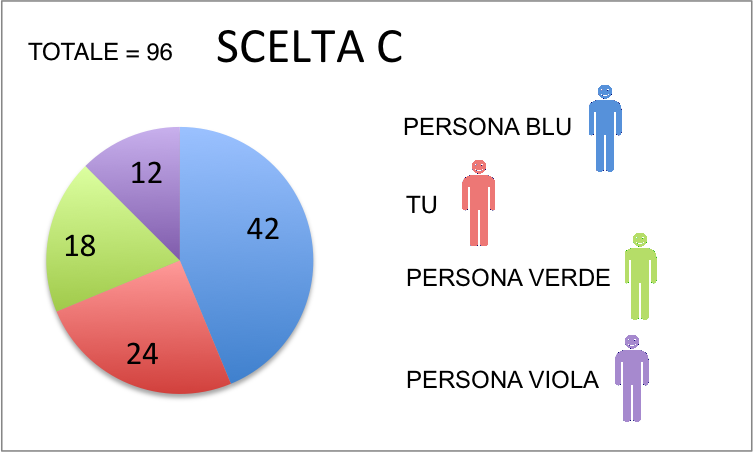}
\end{figure}

\newpage
\textbf{Cominciamo!}
 Ora sei pronto per cominciare. Le decisioni che prenderai nelle prossime pagine determineranno il tuo punteggio. 
 
\newpage
\textbf{Decisione}
Ora puoi effettuare la tua scelta:
\begin{figure}[h]
 \centering   \includegraphics[width=0.49\linewidth]{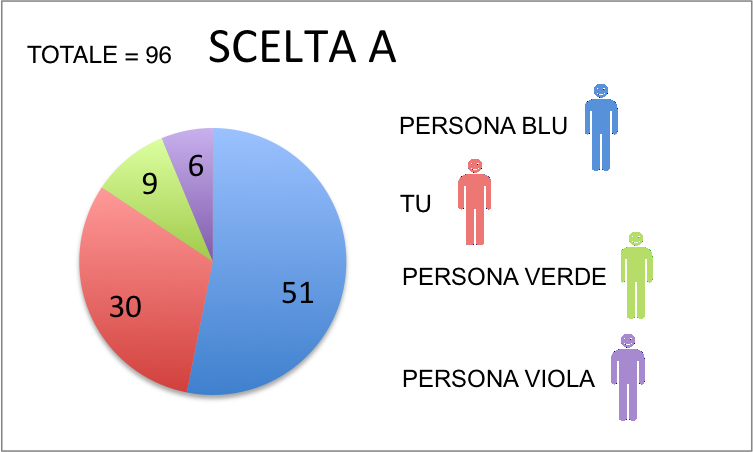}
    \includegraphics[width=0.49\linewidth]{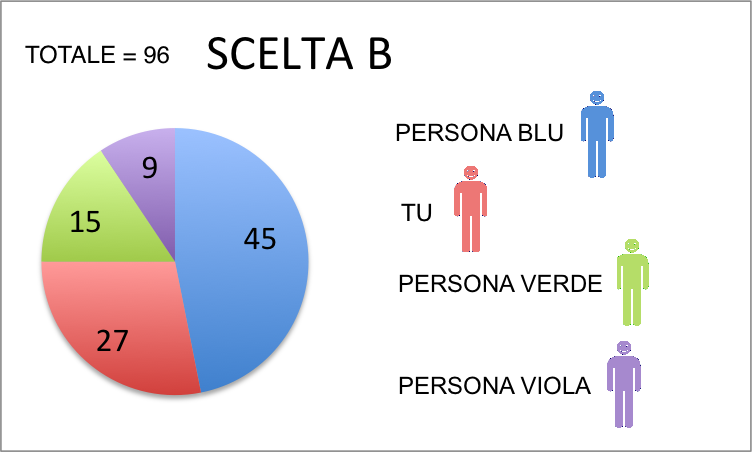}\\
    \includegraphics[width=0.49\linewidth]{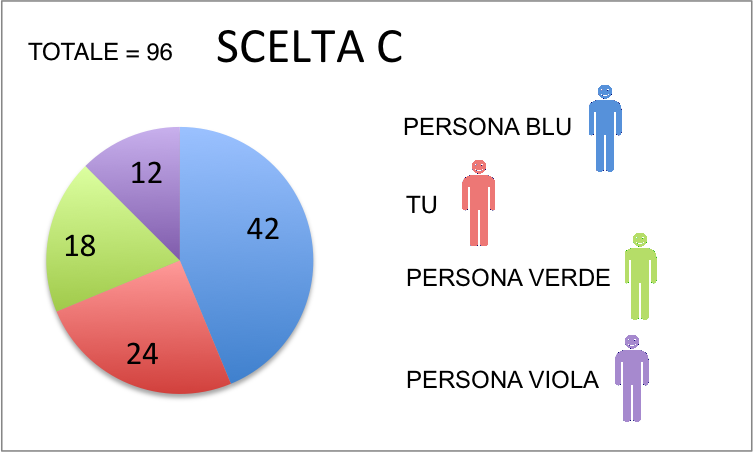}
\end{figure}
 \begin{todolist}
        \item Scelta A
        \item Scelta B
        \item Scelta C

    \end{todolist}

\newpage
\subsubsection{Cognitive Reflection Test (CRT)}
In questa parte dello studio dovrai rispondere a delle domande. Il tuo punteggio dipenderà dal numero di risposte corrette. Per ogni risposta corretta ottieni 7 punti sperimentali (quindi guadagni 0.21 euro). Quando sei pronto clicca sul pulsante "Successivo".

\bigskip
\bigskip 

\begin{itemize}
    \item[1] Se lanci una moneta non truccata tre volte, qual è la probabilità in percentuale che esca Testa almeno una volta? La risposta va espressa in PERCENTUALE:
    \item[2] Se 3 commessi possono incartare 3 giocattoli in 1 ora, quanti commessi sono necessari per incartare 6 giocattoli in 2 ore? La risposta va espressa in COMMESSI:
    \item[3] Giulia e Sara stanno correndo in una pista circolare. Corrono alla stessa velocità ma Giulia è partita dopo. Quando Giulia ha percorso 5 giri, Sara ne ha percorsi 15. Quando Giulia ne avrà percorsi 30, quanti ne avrà percorsi Sara? La risposta va espressa in GIRI:
    \item[4]Il voto di Marco è sia il 15esimo voto più alto che il 15esimo voto più basso della classe. Quanti studenti ci sono nella sua classe? La risposta va espressa in STUDENTI:
    \item[5] In una squadra di atletica gli atleti alti tendono a vincere 3 volte di più di quelli bassi. Quest'anno la squadra ha vinto 60 medaglie. Quante medaglie sono state vinte dagli atleti bassi? La risposta va espressa in MEDAGLIE:
\end{itemize}

\newpage
\subsubsection{Interpersonal Reactivity Index (IRI)}
 In questa parte dello studio dovrai rispondere a delle domande. Quando sei pronto clicca sul pulsante "Successivo".
 \bigskip
 \bigskip
 
Per piacere rispondi alle seguenti domande, indicando un valore da 1 "mai vero" a 5 "sempre vero".

\begin{itemize}
    \item [1] Sogno ad occhi aperti e fantastico, con una certa regolarità, sulle cose che potrebbero accadermi
     \begin{todolist}
        \item 1  
        \item 2
        \item 3
        \item 4 
        \item 5
    \end{todolist}
    \item [2] Provo spesso sentimenti di tenerezza e di preoccupazione per le persone meno fortunate di me
     \begin{todolist}
        \item 1  
        \item 2
        \item 3
        \item 4 
        \item 5
    \end{todolist}
    \item [3] A volte trovo difficile vedere le cose dal punto di vista di un'altra persona
         \begin{todolist}
        \item 1  
        \item 2
        \item 3
        \item 4 
        \item 5
    \end{todolist}
    \item [4] A volte non mi sento molto dispiaciuto per le altre persone che hanno problemi
     \begin{todolist}
        \item 1  
        \item 2
        \item 3
        \item 4 
        \item 5
    \end{todolist}
    \item [5] Resto veramente coinvolto dagli stati d'animo dei protagonisti di un racconto
     \begin{todolist}
        \item 1  
        \item 2
        \item 3
        \item 4 
        \item 5
    \end{todolist}
    \item [6] In situazioni di emergenza, mi sento apprensivo e a disagio
     \begin{todolist}
        \item 1  
        \item 2
        \item 3
        \item 4 
        \item 5
    \end{todolist}
    \item [7] Riesco solitamente ad essere obiettivo quando guardo un film o una rappresentazione teatrale e raramente mi lascio coinvolgere del tutto
         \begin{todolist}
        \item 1  
        \item 2
        \item 3
        \item 4 
        \item 5
    \end{todolist}
    \item [8] In caso di disaccordo, cerco di tener conto del punto di vista di ognuno prima di prendere una decisione
     \begin{todolist}
        \item 1  
        \item 2
        \item 3
        \item 4 
        \item 5
    \end{todolist}
    \item [9] Quando vedo qualcuno che viene sfruttato, provo sentimenti di protezione nei suoi confronti
     \begin{todolist}
        \item 1  
        \item 2
        \item 3
        \item 4 
        \item 5
    \end{todolist}
    \item [10] A volte mi sento indifeso quando mi trovo in situazioni emotivamente molto coinvolgenti
     \begin{todolist}
        \item 1  
        \item 2
        \item 3
        \item 4 
        \item 5
    \end{todolist}
    \item [11] A volte cerco di comprendere meglio i miei amici immaginando come le cose appaiano dalla loro prospettiva
     \begin{todolist}
        \item 1  
        \item 2
        \item 3
        \item 4 
        \item 5
    \end{todolist}
    \item [12] Mi accade raramente di essere molto coinvolto da un buon libero o da un bel film
     \begin{todolist}
        \item 1  
        \item 2
        \item 3
        \item 4 
        \item 5
    \end{todolist}
    \item [13] Quando vedo qualcuno farsi male, tendo a restare calmo
     \begin{todolist}
        \item 1  
        \item 2
        \item 3
        \item 4 
        \item 5
    \end{todolist}
    \item [14] Le sventure delle altre persone a volte non mi turbano molto
     \begin{todolist}
        \item 1  
        \item 2
        \item 3
        \item 4 
        \item 5
    \end{todolist}
    \item [15] Se sono sicuro di aver ragione riguardo a qualcosa, non spreco molto tempo ad ascoltare le argomentazioni degli altri 
     \begin{todolist}
        \item 1  
        \item 2
        \item 3
        \item 4 
        \item 5
    \end{todolist}
    \item [16] Dopo aver visto una rappresentazione teatrale o un film, mi sono sentito come se io stesso fossi uno dei protagonisti
     \begin{todolist}
        \item 1  
        \item 2
        \item 3
        \item 4 
        \item 5
    \end{todolist}
    \item [17] Trovarmi in situazioni che provocano tensione emotiva mi spaventa
     \begin{todolist}
        \item 1  
        \item 2
        \item 3
        \item 4 
        \item 5
    \end{todolist}
    \item [18] Quando vedo qualcuno che viene trattato ingiustamente, talvolta mi capita di non provare molta pietà per lui
         \begin{todolist}
        \item 1  
        \item 2
        \item 3
        \item 4 
        \item 5
    \end{todolist}
    \item [19] Sono di solito piuttosto efficiente nel far fronte alle situazioni di emergenza
         \begin{todolist}
        \item 1  
        \item 2
        \item 3
        \item 4 
        \item 5
    \end{todolist}
    \item [20] Le cose che accadono mi colpiscono molto spesso
     \begin{todolist}
        \item 1  
        \item 2
        \item 3
        \item 4 
        \item 5
    \end{todolist}
    \item [21] Credo che esistano due opposti aspetti in ogni vicenda e cerco di prenderli in considerazione entrambi
     \begin{todolist}
        \item 1  
        \item 2
        \item 3
        \item 4 
        \item 5
    \end{todolist}
    \item [22] Potrei descrivermi come una persona dal cuore piuttosto tenero
         \begin{todolist}
        \item 1  
        \item 2
        \item 3
        \item 4 
        \item 5
    \end{todolist}
    \item [23] Quando guardo un buon film, riesco molto facilmente a mettermi nei panni di un personaggio principale
     \begin{todolist}
        \item 1  
        \item 2
        \item 3
        \item 4 
        \item 5
    \end{todolist}
    \item [24] Tendo a perdere il controllo in casi di emergenza
     \begin{todolist}
        \item 1  
        \item 2
        \item 3
        \item 4 
        \item 5
    \end{todolist}
    \item [25] Quando sono in contrasto con qualcuno, di solito cerco di "mettermi nei suoi panni" per un attimo
     \begin{todolist}
        \item 1  
        \item 2
        \item 3
        \item 4 
        \item 5
    \end{todolist}    
    \item [26] Quando leggo una storia o un racconto interessante, immagino come mi sentirei se gli avvenimenti della storia stessero accadendo a me
     \begin{todolist}
        \item 1  
        \item 2
        \item 3
        \item 4 
        \item 5
    \end{todolist}
    \item [27] Quando vedo qualcuno che ha urgente bisogno di aiuto in una situazione di emergenza, crollo
     \begin{todolist}
        \item 1  
        \item 2
        \item 3
        \item 4 
        \item 5
    \end{todolist}
    \item [28] Prima di criticare qualcuno, cerco di immaginare cosa proverei se fossi al suo posto
     \begin{todolist}
        \item 1  
        \item 2
        \item 3
        \item 4 
        \item 5
    \end{todolist}

\newpage
\subsubsection{Exit Survey}
Per piacere rispondi alle seguenti domande.
\begin{itemize}
    \item[1] Età
    \item[2] Genere
         \begin{todolist}
        \item Maschio  
        \item Femmina
        \item Non mi identifico con i precedenti
        \item Preferisco non rispondere
        
    \end{todolist}
    \item[3] Residenza attuale
    \item[4] Luogo di nascita 
    \item[5] Livello d'istruzione superiore
     \begin{todolist}
         \item Dottorato di ricerca 
        \item Laurea specialistica/magistrale
        \item Laurea triennale
        \item Diploma scuola superiore
        \item Licenza media
        \item Licenza elementare
    \end{todolist}
    \item[6] Situazione lavorativa
     \begin{todolist}
        \item Occupato
        \item Disoccupato
        \item Studente
        \item Pensionato
        \item Non in cerca di lavoro
    \end{todolist}
    \item[7]  Dove hai frequentato le scuole elementari? 
    \item[8] In generale, diresti che ci si possa fidare delle persone o che bisogna stare molto attenti? Indica un valore da 1 a 6, dove 1 significa che bisogna stare molto attenti e 6 che ci si può fidare
     \begin{todolist}
        \item 1  
        \item 2
        \item 3
        \item 4 
        \item 5
        \item 6
    \end{todolist}
    \item[9]  Hai mai partecipato ad un progetto di un'associazione o della tua comunità, senza derivarne alcun guadagno?
    \begin{todolist}
        \item Sì
        \item No
    \end{todolist}
    \item[10] Hai votato alle ultime elezioni?
    \begin{todolist}
        \item Sì
        \item No
        \item Non avevo i requisiti o non potevo per via della mia età 
    \end{todolist}
    \item[11] Sei donatore di sangue?
   \begin{todolist}
        \item Sì  
        \item No
        \item Non posso donare 
    \end{todolist}
    \item[12] Sei iscritto nella lista donazione di organi?
   \begin{todolist}
        \item Sì 
        \item No
        \item Non lo so
    \end{todolist}
    \item[13] Per ciascuna delle seguenti azioni, indica un valore da 1 a 6, dove 1 significa che le persone non possono mai essere giustificate e 6 che le persone possono sempre essere giustificate
    \begin{itemize}
        \item Richiedere sussidi statali che non ti spettano
           \begin{todolist}
        \item 1  
        \item 2
        \item 3
        \item 4 
        \item 5
        \item 6
    \end{todolist}
        \item Evitare di pagare il biglietto dell'autobus
          \begin{todolist}
        \item 1  
        \item 2
        \item 3
        \item 4 
        \item 5
        \item 6
    \end{todolist}
        \item Non pagare le tasse se ne hai la possibilità
          \begin{todolist}
        \item 1  
        \item 2
        \item 3
        \item 4 
        \item 5
        \item 6
    \end{todolist}
        \item Accettare una bustarella nel contesto lavorativo
          \begin{todolist}
        \item 1  
        \item 2
        \item 3
        \item 4 
        \item 5
        \item 6
    \end{todolist}
    \end{itemize}

    \item[14] Quanto pensi sia diffusa la corruzione in questo paese?
      \begin{todolist}
        \item Quasi nessuno dei pubblici uffici è corrotto
        \item Pochi funzionari pubblici sono corrotti
        \item La maggior parte dei funzionari è corrotta
        \item Quasi tutti i funzionari pubblici sono corrotti
    \end{todolist}
\end{itemize}

\end{itemize}
\newpage